\documentclass[a4paper, amsfonts, amssymb, amsmath, reprint, showkeys, nofootinbib, twoside, superscriptaddress, footnotes]{revtex4-1}
\usepackage[english]{babel}
\usepackage[toc,page]{appendix}
\usepackage{float}
\usepackage[utf8]{inputenc}
\usepackage{tikz}
\usepackage{subcaption} 
\usetikzlibrary{decorations.pathreplacing} % extra needed for decorating quantum circuits!
\usepackage[colorinlistoftodos, color=green!40, prependcaption]{todonotes}
\usepackage{amsthm}
\usepackage{mathtools}
\usepackage{physics}
\usepackage{xcolor}
\usepackage{graphicx}
\usepackage[left=23mm,right=13mm,top=35mm,columnsep=15pt]{geometry} 
\usepackage{adjustbox}
\usepackage{placeins}
\usepackage[T1]{fontenc}
\usepackage{lipsum}
\usepackage{csquotes}
\usepackage{tikz}
\usepackage{float}
\usetikzlibrary{quantikz}

\usepackage[unicode=true, colorlinks=true]{hyperref} % For hyperlinks in the PDF
\newcommand{\prep}{\mbox{P{\scriptsize REP}}}
\newcommand{\select}{\mbox{S\scriptsize ELECT}}

\begin{document}
\title{Calculating the Single-Particle Many-body Green's Functions via the Quantum Singular Value Transform Algorithm}
% \title{Application of the Quantum Singular Value Transform Algorithm to calculate the single-particle many body Green's Functions}

\author{Alexis Ralli}
    \email{alexis.ralli.18@ucl.ac.uk}% Your name
    \affiliation{Quantinuum\\ 13-15 Hills Road, CB2 1NL Cambridge\\ United Kingdom}
    \affiliation{Centre for Computational Science \\ Department of Chemistry, University College London, WC1H 0AJ \\ United Kingdom}

\author{Gabriel Greene-Diniz}
    \affiliation{Quantinuum\\ 13-15 Hills Road, CB2 1NL Cambridge\\ United Kingdom}

\author{David Mu\~{n}oz Ramo}
    \affiliation{Quantinuum\\ 13-15 Hills Road, CB2 1NL Cambridge\\ United Kingdom}

\author{Nathan Fitzpatrick}
    \email{nathan.fitzpatrick@quantinuum.com}% Your name
    \affiliation{Quantinuum\\ 13-15 Hills Road, CB2 1NL Cambridge\\ United Kingdom}
\date{\today} % Leave empty to omit a date

\begin{abstract}
The Quantum Singular Value Transformation (QSVT) is a technique that provides a unified framework for describing many of the quantum algorithms discovered to date. We implement a noise-free simulation of the technique to investigate how it can be used to perform matrix inversion, which is an important step in calculating the single-particle Green's function in the Lehmann representation. Due to the inverse function not being defined at zero, we explore the effect of approximating $f(x)=1/x$ with a polynomial. This is carried out by calculating the single-particle Green's function of the two-site single-impurity Anderson model. We also propose a new circuit construction for the linear combination of unitaries block encoding technique, that reduces the number of single and two-qubit gates required. 
\end{abstract}

\maketitle

\section{Introduction} \label{sec:introduction}
% https://mathworld.wolfram.com/InhomogeneousLinearOrdinaryDifferentialEquationwithConstantCoefficients.html
In physics, many systems are described using linear inhomogeneous differential equations. In the 1820s George Green developed tools to deal with such problems \cite{green1889essay, duffy2015green}.
Unfortunately, his work remained largely undiscovered during his lifetime, but was luckily rediscovered by William Thomson (later Lord Kelvin) \cite{challis2003green}. The mathematical methods Green proposed are used throughout physics now and were a major component in the discovery of quantum field theory. Julian Schwinger, who shared the 1965 Nobel prize in physics with Sin-Itiro Tomonaga and Richard P. Feynman \cite{Nobel1965}, actually acknowledged Green's contribution to his work in \cite{schwinger1993greening}. %The function Green introduced was coined ``Green's funciton'' by Bernhard Riemann \cite{kline1990mathematical}.

Since then, the Green's function technique has become one of the most important tools in many-body theories \cite{szabo2012modern}.
The method allows one to calculate many properties of a system, for example: excitation and ionization energies, ground-state energies, transition matrix elements, absorption coefficients, and dynamical polarizabilities, as well as elastic and inelastic electron cross sections \cite{onida2002electronic}. In fact, self-consistent perturbation theories can be formulated in terms of the Green’s function \cite{onida2002electronic}. More details on Green's function theory may be found in standard textbooks on many-body theory \cite{farid1999electron}.

In this paper, we restrict our work to describing how to obtain the single-particle many-body Green's function using a quantum computer. There have been several different proposals utilising quantum devices to calculate the matrix elements of the Green's function. Some focus on variational approaches \cite{rungger2019dynamical, cai2020quantum, endo2020calculation, chen2021variational, jamet2021krylov, sakurai2022hybrid, zhu2022calculating} and other near-term methods \cite{steckmann2021simulating}. Alternatively, some proposals utilize quantum phase estimation \cite{bauer2016hybrid, kosugi2020construction}. We follow a similar approach to Tong \textit{et al.} in \cite{tong2021fast}, where the quantum singular value transform (QSVT) algorithm is used \cite{gilyen2019quantum}. However, we do not apply their fast inversion strategy, that can reduce the query complexity to the block encoding by preconditioning the linear problem. At a high level, this method assumes the matrix to block encode $H$ can be split as $H = A + B$, under the assumption the spectral norm of $A$ is much greater than $B$: $\| A\| >> \| B\|$. By preconditioning the problem using $A^{-1}$, the new query complexity depends on $\| B\|$ rather than $\| H\|$, thereby reducing the overall cost \cite{tong2021fast}. 

% d ˆA can be efficiently unitarily diagonalized as in
% This was because we did not perform circuit sampling and instead did a full classical unitary simulation of the quantum circuits and classically used projectors on the resulting matrices to deterministically project the state into the required subspace. % As this algorithm will require a fault tolerant quantum computer, amplitude amplification will likely be used to boost the probability of success. % \cite{rungger2019dynamical, cai2020quantum, chen2021variational, jamet2021krylov,  kosugi2020construction, sakurai2022hybrid, zhu2022calculating, steckmann2021simulating}. 

We tested our approach on the two-site single-impurity Anderson model (SIAM) \cite{anderson1961localized}, which is a four-qubit problem. Despite its simplicity, it captures some very interesting physics, for example it can be used to approximate the Mott insulator phase transition \cite{potthoff2001two}. We calculate the single-particle Green's function for this system, via the QSVT, and use the results to plot this phase transition. We show how approximating the inverse function can lead to singular values not being inverted properly, which causes errors in a given calculation - due to the matrix not being inverted properly. We then discuss possible ways to mitigate against this.

The outline of this paper is as follows. Section \ref{sec:background} introduces all the necessary background material for the paper. The single particle Green's function, the QSVT algorithm and the single-impurity Anderson model are reviewed. Our numerical study is then presented in Section  \ref{sec:discussion}. Finally, in Section \ref{sec:LCU_circuits} we compare the presented block encoding strategy to prior works. 

\section{Background} \label{sec:background}
To keep our discussion self-contained we only summarise the necessary mathematical details and notation required for our work. In this Section, we review the one-particle Green's function, the quantum singular-value transform algorithm and the single-impurity Anderson model that defines the physical system considered in our work.

\subsection{One-particle Green's Functions} \label{sec:G_functions}

The time-ordered single-particle Green's function (GF) at zero temperature in the frequency domain is defined in the Lehmann representation as \cite{lehmann1954eigenschaften, hjorth2017advanced}:
\begin{equation}
\label{eq:G_funct}
\begin{aligned}
    G_{ij}(z)  = G_{ij}^{(h)}(z) + G_{ij}^{(e)}(z),
\end{aligned}
\end{equation}
where:
\begin{subequations}
 \label{eq:GplusandGminus}
    \begin{equation}
    \label{eq:G_plus}
        \begin{aligned}
           G_{ij}^{(e)}(z) = \bra{\Psi_{0}} a_{i} \big( z - [H-E_{0}] \big)^{-1} a_{j}^{\dagger} \ket{\Psi_{0}},
        \end{aligned}
    \end{equation}
    \begin{equation}
    \label{eq:G_minus}
        \begin{aligned}
              G_{ij}^{(h)}(z) = \bra{\Psi_{0}} a_{j}^{\dagger} \big(z + [H-E_{0}] \big)^{-1}  a_{i}\ket{\Psi_{0}},
        \end{aligned}
    \end{equation}
\end{subequations}
For simplicity we have assumed $\ket{\Psi_{0}}$ to be non-degenerate, but this can be extended to degenerate ground-states at nonzero temperature. $G_{ij}^{(e)}(z)$ and $ G_{ij}^{(h)}(z)$ are called the advanced and retarded Green's function respectively, or the electron and hole excitation parts of the GF \cite{kosugi2020construction, tong2021fast}. Here $a_{i}$ and $ a_{i}^{\dagger}$ are fermionic creation and annihilation operators of an electron in the $i$-th spin orbital,  $\ket{\Psi_{0}}$ is the ground-state wavefunction, $E_{0}$ is the ground-state energy, $H$ is a  second quantized fermionic Hamiltonian and $z =\omega + i \delta$ is a complex frequency often interpreted as an energy shift \cite{tong2021fast}. The imaginary part of $z$, given by $\delta$, is small and required for convergence of the Fourier transform \cite{onida2002electronic}. Equation \ref{eq:G_funct} can be mapped to an equation involving qubit operators, by applying a fermionic-to-qubit transformation to the fermionic operators. For a given $z$, $G(z)\in \mathbb{C}^{N \times N}$ is an $N \times N$ matrix that is efficient to classically store, where $N$ is the number of spin orbitals (or qubits) describing the system. Equation \ref{eq:G_funct} shows how the $i$-th row and $j$-th column of $G(z)$ is calculated. Even though this matrix is efficient to store and manipulate classically, it should be noted that each entry requires solving an exponentially large problem. This is due to the size of the Hamiltonian scaling as $H \in \mathbb{C}^{2^{N} \times 2^{N}}$ or exponentially with the number of spin orbitals. Classically computing each entry in the Green's function quickly becomes intractable, as doing so requires inverting an exponentially large matrix. Such a Hilbert space is naturally expressed on a quantum computer with $N$ qubits. All that is  required is the ability to perform a matrix inverse on such a device. One way to do this is using the quantum singular-value transform algorithm. As will be discussed in the next section, this method provides a way to apply an (approximate) inverse of a block encoded operator onto a quantum state. %We note this approach does not give a user access to the singular-values or access to the inverted matrix. Algebrically 

% However, the quantum singular-value transform algorithm offers a way to calculate the matrix inverse on a quantum computer. 

A point to note when calculating the Green's function is that the ground-state $\ket{\Psi_{0}}$ must be known. In this work we assume it is known \textit{a priori} and can be efficiently prepared on a quantum device. However, in general the ground-state problem of a $l$-local Hamiltonian is QMA-complete for $l\geq 2$ (for $l=1$ the problem is in P) \cite{kempe2006complexity} and currently there are no known algorithms that can find a solution in polynomial time. How to find $\ket{\Psi_{0}}$ thus remains an open question and we do not consider this issue in the study presented here, as the toy system studied is classically tractable.

\subsection{Quantum Singular-Value Transform algorithm} \label{sec:QSVT_background}

A comprehensive review on quantum signal processing \cite{low2019hamiltonian} and the quantum singular-value transform \cite{gilyen2019quantum} can be found in \cite{martyn2021grand}. In this section we summarise the steps required to perform matrix inversion via QSVT. The algorithm can be broken down into four major steps:
\begin{enumerate}
    % \item Build a quantum circuit to prepre the ground-state ($\ket{\Psi_{0}}$).
    \item Construct a quantum circuit that block encodes a matrix.
    \item Generate the quantum signal-processing angles required to implement the desired function that will be applied to the singular values of the block encoded matrix. Here this will be an approximation of the inverse function: $f(x) \approx 1/x$.
    \item Construct the quantum circuit to implement the QSVT algorithm using the outputs of steps $1$ and $2$.
    \item Implement a Hadamard test to evaluate the real and complex parts of each entry in the Green's function.
\end{enumerate}
The following subsections review each of these steps, apart from the Hadamard test, which we did not implement in this work. A full analysis of step $4$ is given by Tong \textit{et al.} in \cite{tong2021fast}. %We assume the 1st step has been solved. This can be done through NISQ methods such as running VQE and also more fault tolerent procedures such as QPE. TODO maybe add imaginary time evolution.

\subsubsection{Linear Combination of Unitaries (block encoding)} \label{sec:block}

There are many different methods to block encode a matrix \cite{camps2022explicit, low2019hamiltonian}. In this paper, we focus on the linear combination of unitaries (LCU) approach, a technique to block encode any linear combinations of unitary operators \cite{childs2012hamiltonian, berry2015simulating}. Given such a matrix $A$:

\begin{equation}
 \label{eqn:A_def}
\begin{aligned}
	A &= \sum_{i=0}^{k-1} \alpha_{i} U_{i}, \text{ where} \\
 \| A\| &\leq \sum_{i=0}^{k-1} \bigg( |\alpha_{i}| \cdot \underbrace{\| U_{i}\|}_{=1} \bigg) = \sum_{i=0}^{k-1} |\alpha_{i}| = \|A \|_{1},
	\end{aligned}
\end{equation}
where, without loss of generality, we can assume $\alpha_{i}>0$ and $\alpha_{i} \in \mathbb{R}$ $\forall i$ by absorbing any complex phases and signs into the unitaries  $U_{j}$ \cite{childs2012hamiltonian, low2019hamiltonian, ralli2021implementation}.  Given a list of $\alpha_{i}$ and each $U_{i}$, which are assumed to be easy to implement as controlled operations on a quantum device, the block encoding can be constructed using the oracles \cite{ low2019hamiltonian, ralli2021implementation}:

\begin{equation}
 \label{eqn:prep_oracle}
\begin{aligned}
	\prep &= \sum_{i=0}^{k-1} \sqrt{\frac{\alpha_{i}}{ \| A\|_{1}}} \ket{i}\bra{0}_{p}+ \hdots \\
 &= \begin{bmatrix}
\sqrt{\big(\frac{\alpha_{0}}{\|A\|_{1}}\big)} & \cdot & \hdots \\ 
\sqrt{\big(\frac{\alpha_{1}}{\|A\|_{1}}\big)} & \cdot & \hdots  \\ 
\vdots &  \ddots & \hdots   \\ 
\sqrt{\big(\frac{\alpha_{k-1}}{\|A\|_{1}}\big)} & \cdot  &  \hdots
\end{bmatrix},
	\end{aligned}
\end{equation}
and
\begin{equation}
 \label{eqn:select_oracle}
\begin{aligned}
	U_{\select} &= \sum_{i=0}^{k-1} \Big( \ket{i}\bra{i}_{p} \otimes U_{i} \Big).
	\end{aligned}
\end{equation}
Here the subscript $s$ denotes the system register and $p$ the $prep$ (ancilla) register.  The number of prep qubits required will be $n_{p}=\lceil \log_{2}(k)\rceil$, where $k$ is the number of unitaries in the LCU.

The $\prep$ or ``Prepare'' oracle  is a unitary that prepares the state  $\ket{P} = \sum_{i=0}^{k-1} \sqrt{\frac{\alpha_{i}}{\|A\|_{1}} }\ket{i}$ from the all zero state on the $prep$ register - i.e. $\ket{\bar{0}} \mapsto \ket{P}$.  This is why in equation \ref{eqn:prep_oracle} only the first column of the $\prep$ unitary is defined. As discussed in \cite{ralli2021implementation}, the other columns can be take any value providing that $\prep$ remains unitary.  This means there is a lot of freedom in how to construct this operator. If one simply finds the quantum circuit that realises $\ket{\bar{0}}_{p} \mapsto \ket{P}_{p}$,  then the circuit's action on the other basis states are automatically accounted for and the whole of $\prep$ will be defined \cite{ralli2021implementation}. The only quantum circuit requirement is being able to generate any real quantum state from the all-zero state on the prep register. There are many different proposals on how to prepare arbitrary quantum states \cite{Long2001, Mottonen2005, Markov2006prep, Araujo2021}. Following the approaches given in both \cite{Markov2006prep, Araujo2021}, a real quantum state can be generated using multiplexed $R_{y}$ rotations with the number of single-qubit and CNOT gates both scaling as $\mathcal{O}(2^{n_{p}})$. As the number of $prep$ qubits scales logarithmically with the number of terms in $A$, $\mathcal{O}(\log_{2}|A|)$, the number of single-qubit and CNOT gates will scale linearly as $\mathcal{O}(|A|)$ for the $\prep$ part of the block-encoding circuit. 

The desired LCU block encoding is achieved by performing $\prep^{\dagger} U_{\select}\prep$ and post selecting on the all-zero state on the prep qubit register. We can check this via the following proof \cite{ low2019hamiltonian}:

\begin{figure*}[t]
\centering
 \scalebox{0.85}{\input{circuits/block_encoding.tikz}}
\caption{Circuit performing block encoding of $M = \sum_{i}^{|M|} c_{i}P_{i}$, where $\sum_{i}|c_{i}|=1$. Here the $\prep$ unitary (equation  \ref{eqn:prep_oracle}) prepares the following quantum state: $\ket{\overline{0}}_{prep} \mapsto \sum_{i}\sqrt{c_{i}}\ket{P}^{prep}$. The select operator is then performed (equation  \ref{eqn:select_oracle}) followed by $\prep^{\dagger}$. The overall circuit produces the following final state $\big(\mathcal{I}^{prep} \ket{\bar{0}}^{prep} \otimes M \ket{\psi}^{sys} \big)+ \sqrt{1- \| M \ket{\psi}^{sys} \|^{2}}\ket{\perp}$. Post selecting the all zero state on the ``preparation'' register ($\ket{\bar{0}}^{prep} $) results in $M$ being applied to the system state $\ket{\psi}_{sys}$ \cite{low2017optimal, ralli2021implementation}. Note the number of qubits needed by the system register is defined by the physical problem, $\ket{\psi}$, and the number of preparation qubits required is $\lceil \log_{2}(|M|) \rceil$. \label{fig:block_encode_circ} }
\end{figure*}

% The overall circuit produces the following final state $\big(\mathcal{I}^{prep} \ket{\bar{0}}^{prep} \otimes M \ket{\psi}^{sys} \big)+ \sqrt{ 1- \| M \ket{\psi}^{sys} \|^{2}} \ket{\perp}^{\substack{prep \\ sys}}$.  

\begin{equation}
\begin{aligned}
&\bigg[ \bra{\bar{0}}_{p} \otimes I_{s} \bigg] (\prep^{\dagger} \otimes I_{s} ) U_{\select} \big(   \prep \otimes I_{s}\big)\bigg[ \ket{\bar{0}}_{p} \otimes I_{s} \bigg] \\
&=\bigg[ \bra{P}_{p} \otimes I_{s} \bigg] U_{\select} \bigg[ \ket{P}_{p} \otimes I_{s} \bigg] \\
&=  \bigg[ \bra{P}_{p} \otimes I_{s} \bigg] U_{\select} \Bigg( \underbrace{\sum_{i=0}^{k-1} \sqrt{\frac{\alpha_{i}}{\|A\|_{1}} }\ket{i}_{p}}_{\ket{P}_{p} }\otimes I_{s}  \Bigg) \\
 &=  \bigg[ \bra{P}_{p} \otimes I_{s} \bigg] \Bigg( \underbrace{ \sum_{i=0}^{k-1} \sqrt{\frac{\alpha_{i}}{\|A\|_{1}} }\ket{i}_{p}\otimes U_{i}}_{\text{equation \ref{eqn:select_oracle}}}  \Bigg) \\
	&= \Bigg( \sum_{j=0}^{k-1} \sqrt{\frac{\alpha_{j}}{\|A\|_{1}} }\bra{j}\otimes I_{s}  \Bigg) \Bigg( \sum_{i=0}^{k-1} \sqrt{\frac{\alpha_{i}}{\|A\|_{1}} }\ket{i}_{p} \otimes U_{i}  \Bigg) \\
	&= \sum_{j=0}^{k-1}\sum_{i=0}^{k-1} \frac{\sqrt{\alpha_{i}\alpha_{j}}}{\|A\|_{1}} \bra{j} i \rangle_{p} \otimes U_{i} \\
	&= \frac{1}{\|A\|_{1}} \sum_{i=0}^{k-1} \alpha_{i} U_{i}.
	\end{aligned}
\end{equation}
This is implemented according to the circuit in Figure \ref{fig:block_encode_circ}. The probability of success for this block encoding is $(\|A\|_{1})^{-2}\bra{\psi}_{s} A^{\dagger} A \ket{\psi}_{s}$. As discussed in \cite{ralli2021implementation}, $A$ is not necessarily unitary and so $A^{\dagger}A$ may not equal $I$. The probability of success therefore depends on the system state $\ket{\psi}$ and the $1$-norm of the block-encoded matrix. Amplitude amplification  \cite{grover1998quantum, brassard2002quantum, yoder2014fixed} and oblivious amplitude amplification \cite{berry2014exponential, yan2022fixed} can then be used to increase the probability of success \cite{gilyen2019quantum}.

In the literature, it is common to see  $(\alpha, \kappa, \epsilon)$-block encodings. Here $\alpha$ is a normalisation factor of the block-encoded matrix, $\kappa$ is the number of extra ancillary qubits required to implement the block encoding and $\epsilon$ is the error of the block encoding. 

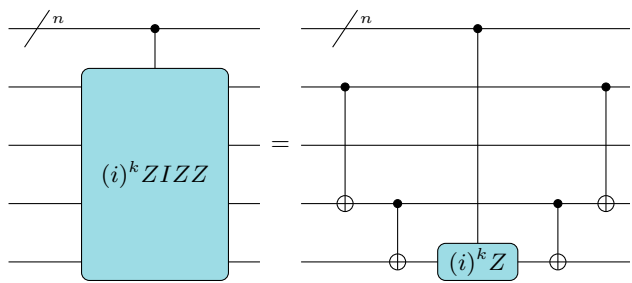
\begin{figure}[t]
\centering
\providecommand{\ket}[1]{\left |#1\right\rangle}
\providecommand{\bra}[1]{\left\langle #1|\right}
\definecolor{mygreen}{RGB}{34,139,33}
\definecolor{myblue}{RGB}{157,220,229}
\definecolor{myred}{RGB}{255,99,98}
\begin{tikzpicture}[scale=1.000000,x=1pt,y=1pt]
\filldraw[color=white] (0.000000, -11.000000) rectangle (232.666667, 99.000000);
% Drawing wires
% Line 12: q1 W
\draw[color=black] (0.000000,88.000000) -- (232.666667,88.000000);
% Line 13: q2 W
\draw[color=black] (0.000000,66.000000) -- (232.666667,66.000000);
% Line 14: q3 W
\draw[color=black] (0.000000,44.000000) -- (232.666667,44.000000);
% Line 15: q4 W
\draw[color=black] (0.000000,22.000000) -- (232.666667,22.000000);
% Line 16: q5 W
\draw[color=black] (0.000000,0.000000) -- (232.666667,0.000000);
% Done with wires; drawing gates
% Line 18: q1 / n
\draw (6.000000, 81.000000) -- (15.333333, 95.000000);
\draw (13.000000, 91.500000) node[right] {$\scriptstyle{n}$};
% Line 19: q2 q3 q4 q5 G:state width=55 $(i)^{k}ZIZZ$ q1
\draw[rounded corners=3pt] (54.833333,88.000000) -- (54.833333,0.000000);
\begin{scope}[rounded corners=3pt]
\begin{scope}
\draw[fill=myblue] (54.833333, 33.000000) +(-45.000000:38.890873pt and 56.568542pt) -- +(45.000000:38.890873pt and 56.568542pt) -- +(135.000000:38.890873pt and 56.568542pt) -- +(225.000000:38.890873pt and 56.568542pt) -- cycle;
\clip (54.833333, 33.000000) +(-45.000000:38.890873pt and 56.568542pt) -- +(45.000000:38.890873pt and 56.568542pt) -- +(135.000000:38.890873pt and 56.568542pt) -- +(225.000000:38.890873pt and 56.568542pt) -- cycle;
\draw (54.833333, 33.000000) node {$(i)^{k}ZIZZ$};
\end{scope}
\end{scope}
\filldraw (54.833333, 88.000000) circle(1.500000pt);
% Line 20: =
\draw[fill=white,color=white] (94.333333, -7.000000) rectangle (109.333333, 95.000000);
\draw (101.833333, 44.000000) node {$=$};
% Line 21: q1 / n
\draw (121.333333, 81.000000) -- (130.666667, 95.000000);
\draw (128.333333, 91.500000) node[right] {$\scriptstyle{n}$};
% Line 23: q4 C q2
\draw (126.000000,66.000000) -- (126.000000,22.000000);
\begin{scope}
\draw[fill=white] (126.000000, 22.000000) circle(3.000000pt);
\clip (126.000000, 22.000000) circle(3.000000pt);
\draw (123.000000, 22.000000) -- (129.000000, 22.000000);
\draw (126.000000, 19.000000) -- (126.000000, 25.000000);
\end{scope}
\filldraw (126.000000, 66.000000) circle(1.500000pt);
% Line 24: q5 C q4
\draw (145.666667,22.000000) -- (145.666667,0.000000);
\begin{scope}
\draw[fill=white] (145.666667, 0.000000) circle(3.000000pt);
\clip (145.666667, 0.000000) circle(3.000000pt);
\draw (142.666667, 0.000000) -- (148.666667, 0.000000);
\draw (145.666667, -3.000000) -- (145.666667, 3.000000);
\end{scope}
\filldraw (145.666667, 22.000000) circle(1.500000pt);
% Line 25: q5 G:state width=30 $(i)^{k}Z$ q1
\draw[rounded corners=3pt] (175.666667,88.000000) -- (175.666667,0.000000);
\begin{scope}[rounded corners=3pt]
\begin{scope}
\draw[fill=myblue] (175.666667, -0.000000) +(-45.000000:21.213203pt and 9.899495pt) -- +(45.000000:21.213203pt and 9.899495pt) -- +(135.000000:21.213203pt and 9.899495pt) -- +(225.000000:21.213203pt and 9.899495pt) -- cycle;
\clip (175.666667, -0.000000) +(-45.000000:21.213203pt and 9.899495pt) -- +(45.000000:21.213203pt and 9.899495pt) -- +(135.000000:21.213203pt and 9.899495pt) -- +(225.000000:21.213203pt and 9.899495pt) -- cycle;
\draw (175.666667, -0.000000) node {$(i)^{k}Z$};
\end{scope}
\end{scope}
\filldraw (175.666667, 88.000000) circle(1.500000pt);
% Line 26: q5 C q4
\draw (205.666667,22.000000) -- (205.666667,0.000000);
\begin{scope}
\draw[fill=white] (205.666667, 0.000000) circle(3.000000pt);
\clip (205.666667, 0.000000) circle(3.000000pt);
\draw (202.666667, 0.000000) -- (208.666667, 0.000000);
\draw (205.666667, -3.000000) -- (205.666667, 3.000000);
\end{scope}
\filldraw (205.666667, 22.000000) circle(1.500000pt);
% Line 27: q4 C q2
\draw (223.666667,66.000000) -- (223.666667,22.000000);
\begin{scope}
\draw[fill=white] (223.666667, 22.000000) circle(3.000000pt);
\clip (223.666667, 22.000000) circle(3.000000pt);
\draw (220.666667, 22.000000) -- (226.666667, 22.000000);
\draw (223.666667, 19.000000) -- (223.666667, 25.000000);
\end{scope}
\filldraw (223.666667, 66.000000) circle(1.500000pt);
% Done with gates; drawing ending labels
% Done with ending labels; drawing cut lines and comments
% Done with comments
\end{tikzpicture}
\caption{Circuit template to perform an $n$-control Pauli operator composed of single-qubit Pauli $Z$ and $I$ matrices \cite{ralli2021implementation}. Figure \ref{fig:imagZgate} provides the circuit construction for the multi-control $(i)^{k}Z$ gate. The single-qubit gates: $\{S, S^{\dagger}, H \}$ can be used to convert this circuit into a general $n$-control $(i)^{k}P$ operator via a change of basis.}
\label{fig:cntrl_P_gate}
\end{figure}

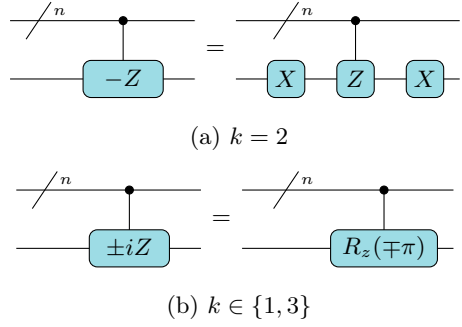
\begin{figure*}
     \centering
     \begin{subfigure}{0.9\textwidth}
         \centering
         \providecommand{\ket}[1]{\left |#1\right\rangle}
\providecommand{\bra}[1]{\left\langle #1|\right}
\definecolor{mygreen}{RGB}{34,139,33}
\definecolor{myblue}{RGB}{157,220,229}
\definecolor{myred}{RGB}{255,99,98}
\begin{tikzpicture}[scale=1.000000,x=1pt,y=1pt]
\filldraw[color=white] (0.000000, -11.000000) rectangle (168.333333, 33.000000);
% Drawing wires
% Line 12: q1 W
\draw[color=black] (0.000000,22.000000) -- (168.333333,22.000000);
% Line 13: q2 W
\draw[color=black] (0.000000,0.000000) -- (168.333333,0.000000);
% Done with wires; drawing gates
% Line 15: q1 / n
\draw (6.000000, 15.000000) -- (15.333333, 29.000000);
\draw (13.000000, 25.500000) node[right] {$\scriptstyle{n}$};
% Line 16: q2 G:state width=30 $-Z$ q1
\draw[rounded corners=3pt] (42.333333,22.000000) -- (42.333333,0.000000);
\begin{scope}[rounded corners=3pt]
\begin{scope}
\draw[fill=myblue] (42.333333, -0.000000) +(-45.000000:21.213203pt and 9.899495pt) -- +(45.000000:21.213203pt and 9.899495pt) -- +(135.000000:21.213203pt and 9.899495pt) -- +(225.000000:21.213203pt and 9.899495pt) -- cycle;
\clip (42.333333, -0.000000) +(-45.000000:21.213203pt and 9.899495pt) -- +(45.000000:21.213203pt and 9.899495pt) -- +(135.000000:21.213203pt and 9.899495pt) -- +(225.000000:21.213203pt and 9.899495pt) -- cycle;
\draw (42.333333, -0.000000) node {$-Z$};
\end{scope}
\end{scope}
\filldraw (42.333333, 22.000000) circle(1.500000pt);
% Line 17: =
\draw[fill=white,color=white] (69.333333, -7.000000) rectangle (84.333333, 29.000000);
\draw (76.833333, 11.000000) node {$=$};
% Line 18: q1 / n
\draw (98.666667, 15.000000) -- (108.000000, 29.000000);
\draw (105.666667, 25.500000) node[right] {$\scriptstyle{n}$};
% Line 19: q2 G:state $X$
\begin{scope}[rounded corners=3pt]
\begin{scope}
\draw[fill=myblue] (103.333333, -0.000000) +(-45.000000:9.899495pt and 9.899495pt) -- +(45.000000:9.899495pt and 9.899495pt) -- +(135.000000:9.899495pt and 9.899495pt) -- +(225.000000:9.899495pt and 9.899495pt) -- cycle;
\clip (103.333333, -0.000000) +(-45.000000:9.899495pt and 9.899495pt) -- +(45.000000:9.899495pt and 9.899495pt) -- +(135.000000:9.899495pt and 9.899495pt) -- +(225.000000:9.899495pt and 9.899495pt) -- cycle;
\draw (103.333333, -0.000000) node {$X$};
\end{scope}
\end{scope}
% Line 20: q2 G:state $Z$ q1
\draw[rounded corners=3pt] (129.333333,22.000000) -- (129.333333,0.000000);
\begin{scope}[rounded corners=3pt]
\begin{scope}
\draw[fill=myblue] (129.333333, -0.000000) +(-45.000000:9.899495pt and 9.899495pt) -- +(45.000000:9.899495pt and 9.899495pt) -- +(135.000000:9.899495pt and 9.899495pt) -- +(225.000000:9.899495pt and 9.899495pt) -- cycle;
\clip (129.333333, -0.000000) +(-45.000000:9.899495pt and 9.899495pt) -- +(45.000000:9.899495pt and 9.899495pt) -- +(135.000000:9.899495pt and 9.899495pt) -- +(225.000000:9.899495pt and 9.899495pt) -- cycle;
\draw (129.333333, -0.000000) node {$Z$};
\end{scope}
\end{scope}
\filldraw (129.333333, 22.000000) circle(1.500000pt);
% Line 21: q2 G:state $X$
\begin{scope}[rounded corners=3pt]
\begin{scope}
\draw[fill=myblue] (155.333333, -0.000000) +(-45.000000:9.899495pt and 9.899495pt) -- +(45.000000:9.899495pt and 9.899495pt) -- +(135.000000:9.899495pt and 9.899495pt) -- +(225.000000:9.899495pt and 9.899495pt) -- cycle;
\clip (155.333333, -0.000000) +(-45.000000:9.899495pt and 9.899495pt) -- +(45.000000:9.899495pt and 9.899495pt) -- +(135.000000:9.899495pt and 9.899495pt) -- +(225.000000:9.899495pt and 9.899495pt) -- cycle;
\draw (155.333333, -0.000000) node {$X$};
\end{scope}
\end{scope}
% Done with gates; drawing ending labels
% Done with ending labels; drawing cut lines and comments
% Done with comments
\end{tikzpicture}
         \caption{$k=2$}
         \label{fig:negZgate}
     \end{subfigure}
     \hfill
     \begin{subfigure}{0.9\textwidth}
         \centering
         \providecommand{\ket}[1]{\left |#1\right\rangle}
\providecommand{\bra}[1]{\left\langle #1|\right}
\definecolor{mygreen}{RGB}{34,139,33}
\definecolor{myblue}{RGB}{157,220,229}
\definecolor{myred}{RGB}{255,99,98}
\begin{tikzpicture}[scale=1.000000,x=1pt,y=1pt]
\filldraw[color=white] (0.000000, -11.000000) rectangle (163.666667, 33.000000);
% Drawing wires
% Line 12: q1 W
\draw[color=black] (0.000000,22.000000) -- (163.666667,22.000000);
% Line 13: q2 W
\draw[color=black] (0.000000,0.000000) -- (163.666667,0.000000);
% Done with wires; drawing gates
% Line 15: q1 / n
\draw (6.000000, 15.000000) -- (15.333333, 29.000000);
\draw (13.000000, 25.500000) node[right] {$\scriptstyle{n}$};
% Line 16: q2 G:state width=30 $\pm iZ$ q1
\draw[rounded corners=3pt] (42.333333,22.000000) -- (42.333333,0.000000);
\begin{scope}[rounded corners=3pt]
\begin{scope}
\draw[fill=myblue] (42.333333, -0.000000) +(-45.000000:21.213203pt and 9.899495pt) -- +(45.000000:21.213203pt and 9.899495pt) -- +(135.000000:21.213203pt and 9.899495pt) -- +(225.000000:21.213203pt and 9.899495pt) -- cycle;
\clip (42.333333, -0.000000) +(-45.000000:21.213203pt and 9.899495pt) -- +(45.000000:21.213203pt and 9.899495pt) -- +(135.000000:21.213203pt and 9.899495pt) -- +(225.000000:21.213203pt and 9.899495pt) -- cycle;
\draw (42.333333, -0.000000) node {$\pm iZ$};
\end{scope}
\end{scope}
\filldraw (42.333333, 22.000000) circle(1.500000pt);
% Line 17: =
\draw[fill=white,color=white] (69.333333, -7.000000) rectangle (84.333333, 29.000000);
\draw (76.833333, 11.000000) node {$=$};
% Line 18: q1 / n
\draw (96.333333, 15.000000) -- (105.666667, 29.000000);
\draw (103.333333, 25.500000) node[right] {$\scriptstyle{n}$};
% Line 19: q2 G:state width=40 $R_{z}(\mp \pi)$ q1
\draw[rounded corners=3pt] (137.666667,22.000000) -- (137.666667,0.000000);
\begin{scope}[rounded corners=3pt]
\begin{scope}
\draw[fill=myblue] (137.666667, -0.000000) +(-45.000000:28.284271pt and 9.899495pt) -- +(45.000000:28.284271pt and 9.899495pt) -- +(135.000000:28.284271pt and 9.899495pt) -- +(225.000000:28.284271pt and 9.899495pt) -- cycle;
\clip (137.666667, -0.000000) +(-45.000000:28.284271pt and 9.899495pt) -- +(45.000000:28.284271pt and 9.899495pt) -- +(135.000000:28.284271pt and 9.899495pt) -- +(225.000000:28.284271pt and 9.899495pt) -- cycle;
\draw (137.666667, -0.000000) node {$R_{z}(\mp \pi)$};
\end{scope}
\end{scope}
\filldraw (137.666667, 22.000000) circle(1.500000pt);
% Done with gates; drawing ending labels
% Done with ending labels; drawing cut lines and comments
% Done with comments
\end{tikzpicture}
         \caption{$k \in \{1,3\}$}
         \label{fig:imagZgate}
     \end{subfigure}

    \caption{Circuit construction of multicontrol single qubit $(i)^{k}Z$ gate, for $k \in \{0,1,2,3\}$. Circuit for $k=0$ is not explicitly pictured, as it is the trivial case of a multicontrol single-qubit $Z$ gate. This is the same as Figure \ref{fig:negZgate} with both $X$ gates removed.}
    \label{fig:phaseZgates}
\end{figure*}

In this work, the LCU is given as a linear combination of Pauli operators. The ``$\select$'' oracle (equation \ref{eqn:select_oracle}) applies a controlled version of each of these Pauli operators on the system register, controlled by the prep register. This requires performing multi-control Pauli operators with phases $\{ i,-i,1,-1 \}$. Following the work in \cite{ralli2021implementation}, this can achieved using the template given in Figure \ref{fig:cntrl_P_gate}. The relevant phases are then obtained via the following identities:

\begin{subequations}
 \label{eq:pauli_phases}
    \begin{equation}
    \label{eq:neg_sign}
	-Z =XZX,
    \end{equation}
\begin{equation}
    \label{eq:imag}
R_{z}(\mp \pi)= e^{\mp i \frac{\pi}{2}Z}  = \pm i Z.
\end{equation}
\end{subequations}
These can be implemented according to the circuit templates summarised in Figure \ref{fig:phaseZgates}. By performing a change of basis on certain qubits, using the single-qubit gates $\{S, S^{\dagger}, H \}$, the circuit proposed in Figure \ref{fig:cntrl_P_gate} can be used to generate any multicontrol Pauli operator with a $\pm 1, \pm i$ phase. We note the ordering of unitaries in   equation \ref{eqn:select_oracle} is arbitrary, but an optimal ordering can lead to significant circuit simplifications. We leave this as an open question, but note the work in \cite{ralli2021implementation}, \cite{hastings2014improving} and \cite{cowtan2019phase} can readily be applied to this problem.  

 % If this construction is decomposed into $CNOT$ and $U3$ gates, then $4 n^{2} - 4 n + 2$  $CNOT$ and $4 n^{2} - 2$ $U3$ gates are required.

To determine the overall circuit cost to implement  $U_{\select}$ (equation \ref{eqn:select_oracle} and Figure \ref{fig:block_encode_circ}), we need to determine the cost of implementing a multicontrol $Z$ gate and multicontrol $R_{z}$ gate. Following the proposal by da Silva and Park, any $n$-control single-qubit gate can be decomposed with $\mathcal{O}(n^{2})$ single qubit and CNOT gates with linear depth \cite{da2022linear}. For an $n$-control single qubit $Z$ gate with $n \leq 6$, the approach outlined in \cite{bullock2003smaller} and  \cite{shende2005synthesis} (theorem 8)  requires fewer two-qubit gates, where the number of single-qubit and CNOT gates required scales as $\mathcal{O}(2^{n})$ respectively. In general, using the work of da Silva and Park makes the cost of performing a multicontrol Pauli operator via the template in Figure  \ref{fig:cntrl_P_gate} scale as:
\begin{enumerate}
    \item $O(2n_{s})$ single-qubit gates, required to implement a change of basis.
    \item $O(2[n_{s}-1])$ CNOT gates, performing the ladder of CNOT gates on the system register.
    \item $O(n_{c}^{2})$ CNOT and single-qubit gates for the multicontrol $i^{k}Z$ gate.
\end{enumerate}
%$O(2n_{s})$ single qubit gates (required to implement a change of basis), $O(2[n_{s}-1])$ CNOT gates (representing the ladder of CNOT gates) and $O(n_{c}^{2})$ CNOT and single qubit gates for the multicontrol $i^{k}Z$ gate.
Here $n_{c}$ is the number of control qubits and $n_{s}$ is the number of `system' qubits the Pauli operator acts on. The single-qubit and CNOT gate cost per $n_{c}$-controlled Pauli operator scales linearly in system qubits and quadratically in control qubits as  $O(n_{c}^{2}+n_{s})$. The overall cost of implementing $U_{\select}$ via the circuits presented will depend on the number of $n_{c}$-controlled Pauli operators that are performed. Looking at Equation  \ref{eqn:select_oracle}, we see that $k$ operators are needed bringing the final cost to $O(k[n_{c}^{2}+n_{s}])$ single and two-qubit gates. As $k = |A|$ and $n_{c} = \lceil \log_{2}(|A|) \rceil$, we can write the final scaling  of single-qubit and CNOT gates as  $\mathcal{O}\Big( |A| ( \lceil \log_{2}(|A|) \rceil^{2} + n_{s} ) \Big)$. Table \ref{table:circuit_summary} provides a summary for the scaling of each part of the circuit.

 %\onecolumngrid
\begin{table*}[t]
\begin{adjustbox}{width=1\textwidth}
\begin{tabular}{llccl}
\hline
\hline
Citation              & Circuit / Gate                     & CNOT                                                         & Single                                                       & Comments                                                                                                                                                                                         \\
\hline
\hline
Shende \textit{et al.} \cite{Markov2006prep} & \begin{tabular}[c]{@{}l@{}} Circuit to prepare  any  \\ real amplitude $n$-qubit state \end{tabular} & $\mathcal{O}(2^{n})$                                         & $\mathcal{O}(2^{n})$                                         & \begin{tabular}[c]{@{}l@{}} Useful for $\prep$ part of LCU method, where $n=\lceil \log_{2}(|A|) \rceil$, \\ and thus $\mathcal{O}(2^{\log_{2}(|A|)}) =\mathcal{O}(|A|)$.  \end{tabular} \\

Silva and Park \cite{da2022linear}        & $n_{c}$-control $Z$                 & $\mathcal{O}(n_{c}^{2})$                                     & $\mathcal{O}(n_{c}^{2})$                                     & -                                                                                                                                                              \\
Bullock and Markov \cite{bullock2003smaller}    & $n_{c}$-control $Z$                 & $\mathcal{O}(2^{n_{c}})$                                      & $\mathcal{O}(2^{n_{c}})$                                          & For $n_{c}\leq 6$, requires fewer CNOT gates than Silva and Park approach. \\ \hline 
This work             & $n_{c}$-control $P$                 & $\mathcal{O}(n_{c}^{2}+ n_{s})$                              & $\mathcal{O}(n_{c}^{2} + n_{s})$                              & Circuit illustrated in Figure \ref{fig:cntrl_P_gate} up to single-qubit change of bases.                        \\ 
This work             & $\select$  &  $\mathcal{O}\Bigg( |A| \bigg( \lceil \log_{2}(|A|) \rceil^{2} + n_{s} \bigg) \Bigg)$ &$\mathcal{O}\Bigg( |A| \bigg( \lceil \log_{2}(|A|) \rceil^{2} + n_{s} \bigg) \Bigg)$ & \begin{tabular}[c]{@{}l@{}} \hline $U_{\select}$ circuit cost for a linear combination of $|A|$ Pauli \\  operators. Circuit template given in Figure \ref{fig:block_encode_circ}. \end{tabular}   \\                                                         \hline
\hline
\end{tabular}
\end{adjustbox}
\caption{Circuit scaling summary for different unitaries required to implement a block encoding of a linear combination of Pauli operators via the LCU method. Here $|A|$ denotes the number of Pauli operators in $A$ (Equation  \ref{eqn:A_def}), $n_{c}$ denotes the number of control qubits and $n_{s}$ denotes the number of system qubits.}
\label{table:circuit_summary}
\end{table*}
% \twocolumngrid
% 

The SIAM Hamiltonian considered in this paper is  constructed as a linear combination of Pauli operators  (equation \ref{eq:H_hubb}). Each individual Pauli operator $P_{i}$ is a unitary Hermitian operator that is easy to implement on a quantum computer as a controlled operation. However, rather than block encoding $H$, we  block encode the following complex shifted Hamiltonians:
\begin{subequations}
\label{eq:A_B}
    \begin{equation}
    \label{eq:A}
        \begin{aligned}
           B_{(e)} = \big( z - [H-E_{0}] \big)^{\dagger},
        \end{aligned}
    \end{equation}
    \begin{equation}
    \label{eq:B}
        \begin{aligned}
        C_{(h)} = \big( z + [H-E_{0}]\big)^{\dagger},
        \end{aligned}
    \end{equation}
\end{subequations}
The reason why the Hermitian conjugate is taken, in equations \ref{eq:A} and \ref{eq:B}, stems from how a matrix inverse can be obtained from a singular-value decomposition. If we write the  singular vector decomposition of an arbitrary matrix $A = U \Sigma V^{\dagger}$, then its inverse (if it exists) is $A^{-1} =  V \Sigma^{-1} U^{\dagger}$. Taking the Hermitian conjugate of $A$ yields $A^{\dagger} = V \Sigma  U^{\dagger}$ and to find the inverse of $A$ all that remains is to invert the singular values of $A^{\dagger}$.

The QSVT algorithm approximately inverts the singular values of a block encoded matrix. Performing this algorithm on the block encodings of $B_{(e)}$ and $C_{(h)}$ will therefore allow us to calculate the inverse parts of equations \ref{eq:G_plus} and \ref{eq:G_minus}. We calculate each inverse separately to reduce the circuit depth; however, it is possible to calculate the terms simultaneously using the method of adding different block encodings given in \cite{gilyen2019quantum, von2021quantum}. We decided not to implement this, as we wanted our approach to be more amenable to early fault-tolerant quantum computers.

As $z$ is a constant complex shift (has a real and imaginary component),  $B_{(e)}$ and $C_{(h)}$ are no longer Hermitian operators, whereas $H$ is. The Quantum Eigenvalue Transformation (QET) implements a polynomial transformation of a block-encoded Hermitian matrix when the polynomial of interest is represented by QSP \cite{low2019hamiltonian}. For a polynomial transformation of a general matrix the quantum singular-value transformation is used \cite{gilyen2019quantum}, hence why it is used in this work. We note that for a Hermitian matrix with a block-encoding input model, the quantum circuits of QET and QSVT can be the same \cite{dong2022ground}. Interestingly, it is possible to convert the non-Hermitian problem into a Hermitian one via matrix dilation, which requires an extra qubit - for further details  see Section 4.2 in \cite{chakraborty2018power}. This approach is unnecessary here, but the method of dilation is useful to know and we mention it in passing. This technique would be required if the quantum linear-system algorithm proposed by Harrow, Hassidim, and Lloyd (HHL) were to be used \cite{harrow2009quantum, cai2020quantum}.

Finally, we reiterate a comment made in \cite{martyn2021grand} on block encodings - much more work is necessary to find block encoding techniques specific to the relevant physical system. The LCU method is very general, but doesn't utilise any underlying structure of a problem and requires $\lceil \log_{2} (|A|) \rceil$ extra ancillary qubits. However, it is possible to make a $(\alpha,1,0)$ block encoding of any $n$-qubit matrix $A$ - i.e. only requiring a single ancilla qubit for a block encoding (see example 6.2 in \cite{lin2022lecture} and appendix D in \cite{lin2021real}). These schemes require a singular-value decomposition of $A$ and thus will not scale in a general setting. However, certain structures in particular physical problems may allow for more efficient encoding strategies. We leave this as an important open question.

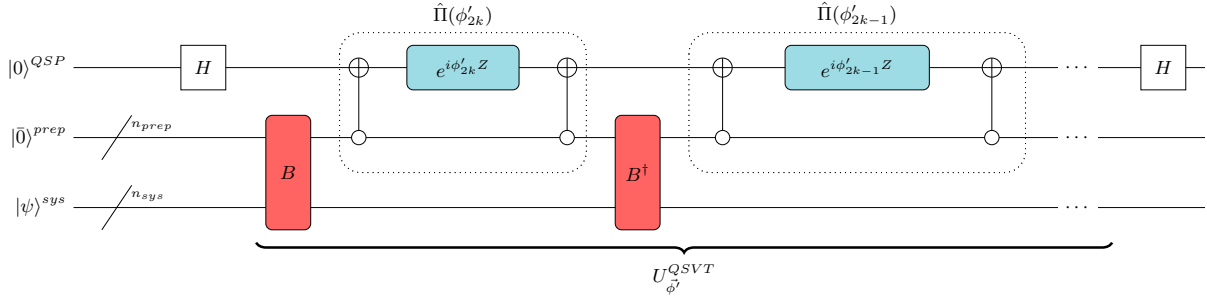
\begin{figure*}[t]
\centering
\scalebox{0.8}{%! \usetikzlibrary{decorations.pathreplacing,decorations.pathmorphing}
\providecommand{\ket}[1]{\left |#1\right\rangle}
\providecommand{\bra}[1]{\left\langle #1|\right}
\definecolor{mygreen}{RGB}{34,139,33}
\definecolor{myblue}{RGB}{157,220,229}
\definecolor{myred}{RGB}{255,99,98}
\begin{tikzpicture}[scale=1.500000,x=1pt,y=1pt]
\filldraw[color=white] (0.000000, -11.000000) rectangle (353.000000, 55.000000);
% Drawing wires
% Line 14: q0 W \ket{0}^{QSP}
\draw[color=black] (0.000000,44.000000) -- (353.000000,44.000000);
\draw[color=black] (0.000000,44.000000) node[left] {$\ket{0}^{QSP}$};
% Line 15: q1 W \ket{\bar{0}}^{prep}
\draw[color=black] (0.000000,22.000000) -- (353.000000,22.000000);
\draw[color=black] (0.000000,22.000000) node[left] {$\ket{\bar{0}}^{prep}$};
% Line 16: q2 W \ket{\psi}^{sys}
\draw[color=black] (0.000000,0.000000) -- (353.000000,0.000000);
\draw[color=black] (0.000000,0.000000) node[left] {$\ket{\psi}^{sys}$};
% Done with wires; drawing gates
% Line 18: q0 LABEL
% Line 19: q1 / n_{prep}
\draw (8.833333, 15.000000) -- (18.166667, 29.000000);
\draw (15.833333, 25.500000) node[right] {$\scriptstyle{n_{prep}}$};
% Line 20: q2 / n_{sys}
\draw (8.833333, -7.000000) -- (18.166667, 7.000000);
\draw (15.833333, 3.500000) node[right] {$\scriptstyle{n_{sys}}$};
% Line 23: q0 G $H$
\begin{scope}
\draw[fill=white] (40.500000, 44.000000) +(-45.000000:9.899495pt and 9.899495pt) -- +(45.000000:9.899495pt and 9.899495pt) -- +(135.000000:9.899495pt and 9.899495pt) -- +(225.000000:9.899495pt and 9.899495pt) -- cycle;
\clip (40.500000, 44.000000) +(-45.000000:9.899495pt and 9.899495pt) -- +(45.000000:9.899495pt and 9.899495pt) -- +(135.000000:9.899495pt and 9.899495pt) -- +(225.000000:9.899495pt and 9.899495pt) -- cycle;
\draw (40.500000, 44.000000) node {$H$};
\end{scope}
% Line 24: q1 LABEL
% Line 25: q2 LABEL
% Line 27: q1 q2 G:block $B$
\draw[rounded corners=3pt] (67.000000,22.000000) -- (67.000000,0.000000);
\begin{scope}[rounded corners=3pt]
\begin{scope}
\draw[fill=myred] (67.000000, 11.000000) +(-45.000000:9.899495pt and 25.455844pt) -- +(45.000000:9.899495pt and 25.455844pt) -- +(135.000000:9.899495pt and 25.455844pt) -- +(225.000000:9.899495pt and 25.455844pt) -- cycle;
\clip (67.000000, 11.000000) +(-45.000000:9.899495pt and 25.455844pt) -- +(45.000000:9.899495pt and 25.455844pt) -- +(135.000000:9.899495pt and 25.455844pt) -- +(225.000000:9.899495pt and 25.455844pt) -- cycle;
\draw (67.000000, 11.000000) node {$B$};
\end{scope}
\end{scope}
% Line 29: q0 C -q1
\draw (89.000000,44.000000) -- (89.000000,22.000000);
\begin{scope}
\draw[fill=white] (89.000000, 44.000000) circle(3.000000pt);
\clip (89.000000, 44.000000) circle(3.000000pt);
\draw (86.000000, 44.000000) -- (92.000000, 44.000000);
\draw (89.000000, 41.000000) -- (89.000000, 47.000000);
\end{scope}
\draw[fill=white] (89.000000, 22.000000) circle(2.250000pt);
% Line 30: q0 G:phase width=35  $e^{i\phi_{2k}'Z}$ % $\hat{\Pi}(\phi_{2k}')$
\draw (121.500000, 55.000000) node[text width=144pt,above,text centered] {$\hat{\Pi}(\phi_{2k}')$};
\begin{scope}[rounded corners=3pt]
\begin{scope}
\draw[fill=myblue] (121.500000, 44.000000) +(-45.000000:24.748737pt and 9.899495pt) -- +(45.000000:24.748737pt and 9.899495pt) -- +(135.000000:24.748737pt and 9.899495pt) -- +(225.000000:24.748737pt and 9.899495pt) -- cycle;
\clip (121.500000, 44.000000) +(-45.000000:24.748737pt and 9.899495pt) -- +(45.000000:24.748737pt and 9.899495pt) -- +(135.000000:24.748737pt and 9.899495pt) -- +(225.000000:24.748737pt and 9.899495pt) -- cycle;
\draw (121.500000, 44.000000) node {$e^{i\phi_{2k}'Z}$};
\end{scope}
\end{scope}
% Line 31: q0 C -q1
\draw (154.000000,44.000000) -- (154.000000,22.000000);
\begin{scope}
\draw[fill=white] (154.000000, 44.000000) circle(3.000000pt);
\clip (154.000000, 44.000000) circle(3.000000pt);
\draw (151.000000, 44.000000) -- (157.000000, 44.000000);
\draw (154.000000, 41.000000) -- (154.000000, 47.000000);
\end{scope}
\draw[fill=white] (154.000000, 22.000000) circle(2.250000pt);
% Line 33: q1 q2 G:block $B^{\dagger}$
\draw[rounded corners=3pt] (176.000000,22.000000) -- (176.000000,0.000000);
\begin{scope}[rounded corners=3pt]
\begin{scope}
\draw[fill=myred] (176.000000, 11.000000) +(-45.000000:9.899495pt and 25.455844pt) -- +(45.000000:9.899495pt and 25.455844pt) -- +(135.000000:9.899495pt and 25.455844pt) -- +(225.000000:9.899495pt and 25.455844pt) -- cycle;
\clip (176.000000, 11.000000) +(-45.000000:9.899495pt and 25.455844pt) -- +(45.000000:9.899495pt and 25.455844pt) -- +(135.000000:9.899495pt and 25.455844pt) -- +(225.000000:9.899495pt and 25.455844pt) -- cycle;
\draw (176.000000, 11.000000) node {$B^{\dagger}$};
\end{scope}
\end{scope}
% Line 35: q0 C -q1
\draw (202.500000,44.000000) -- (202.500000,22.000000);
\begin{scope}
\draw[fill=white] (202.500000, 44.000000) circle(3.000000pt);
\clip (202.500000, 44.000000) circle(3.000000pt);
\draw (199.500000, 44.000000) -- (205.500000, 44.000000);
\draw (202.500000, 41.000000) -- (202.500000, 47.000000);
\end{scope}
\draw[fill=white] (202.500000, 22.000000) circle(2.250000pt);
% Line 39: q2 LABEL
% Line 36: q0 G:phase width=45  $e^{i\phi_{2k-1}'Z}$  % $\hat{\Pi}(\phi_{2k-1}')$
\draw (244.500000, 55.000000) node[text width=144pt,above,text centered] {$\hat{\Pi}(\phi_{2k-1}')$};
\begin{scope}[rounded corners=3pt]
\begin{scope}
\draw[fill=myblue] (244.500000, 44.000000) +(-45.000000:31.819805pt and 9.899495pt) -- +(45.000000:31.819805pt and 9.899495pt) -- +(135.000000:31.819805pt and 9.899495pt) -- +(225.000000:31.819805pt and 9.899495pt) -- cycle;
\clip (244.500000, 44.000000) +(-45.000000:31.819805pt and 9.899495pt) -- +(45.000000:31.819805pt and 9.899495pt) -- +(135.000000:31.819805pt and 9.899495pt) -- +(225.000000:31.819805pt and 9.899495pt) -- cycle;
\draw (244.500000, 44.000000) node {$e^{i\phi_{2k-1}'Z}$};
\end{scope}
\end{scope}
% Line 40: q2 LABEL
% Line 37: q0 C -q1
\draw (286.500000,44.000000) -- (286.500000,22.000000);
\begin{scope}
\draw[fill=white] (286.500000, 44.000000) circle(3.000000pt);
\clip (286.500000, 44.000000) circle(3.000000pt);
\draw (283.500000, 44.000000) -- (289.500000, 44.000000);
\draw (286.500000, 41.000000) -- (286.500000, 47.000000);
\end{scope}
\draw[fill=white] (286.500000, 22.000000) circle(2.250000pt);
% Line 41: q2 LABEL
% Line 44: q0 LABEL ...
\draw[color=black] (313.500000, 44.000000) node [fill=white] {$\cdots$};
% Line 45: q1 LABEL ...
\draw[color=black] (313.500000, 22.000000) node [fill=white] {$\cdots$};
% Line 46: q2 LABEL ...
\draw[color=black] (313.500000, 0.000000) node [fill=white] {$\cdots$};
% Line 53: q0 G $H$
\begin{scope}
\draw[fill=white] (340.000000, 44.000000) +(-45.000000:9.899495pt and 9.899495pt) -- +(45.000000:9.899495pt and 9.899495pt) -- +(135.000000:9.899495pt and 9.899495pt) -- +(225.000000:9.899495pt and 9.899495pt) -- cycle;
\clip (340.000000, 44.000000) +(-45.000000:9.899495pt and 9.899495pt) -- +(45.000000:9.899495pt and 9.899495pt) -- +(135.000000:9.899495pt and 9.899495pt) -- +(225.000000:9.899495pt and 9.899495pt) -- cycle;
\draw (340.000000, 44.000000) node {$H$};
\end{scope}
% Done with gates; drawing ending labels
% Done with ending labels; drawing cut lines and comments
% Line 55: q0 q1 @ 3 5 color=black style=dotted,rounded_corners=10pt
\draw[draw opacity=1.000000,fill opacity=0.200000,color=black,dotted,rounded corners=10pt] (83.000000,55.000000) rectangle (160.000000,11.000000);
\draw[draw opacity=1.000000,fill opacity=0.200000,color=black,dotted,rounded corners=10pt] (83.000000,55.000000) rectangle (160.000000,11.000000);
% Line 58: q0 q1 @ 7 9 color=black style=dotted,rounded_corners=10pt
\draw[draw opacity=1.000000,fill opacity=0.200000,color=black,dotted,rounded corners=10pt] (192.000000,55.000000) rectangle (297.000000,11.000000);
\draw[draw opacity=1.000000,fill opacity=0.200000,color=black,dotted,rounded corners=10pt] (192.000000,55.000000) rectangle (297.000000,11.000000);
% Line 63: @ 2 10 %% $U_{\vec{\phi'}}^{QSVT}$
\draw[decorate,decoration={brace,mirror,amplitude = 4.666667pt},very thick] (57.000000,-11.000000) -- (324.000000,-11.000000);
\draw (190.500000, -15.666667) node[text width=144pt,below,text centered] {$U_{\vec{\phi'}}^{QSVT}$};
% Done with comments
\end{tikzpicture}}
\caption{Quantum circuit to implement the QSVT for an odd degree polynomial (equation \ref{eq:QSVT_eq}) \cite{gilyen2019quantum}. In this paper, the polynomial approximation of $f(x) \approx 1/x$ is odd degree. The zero controlled NOT gates have zero controls on all the ``prep'' qubits. As the polynomial approximation of the inverse function is real, we use the $\{ \ket{+},  \ket{-} \}$ signal basis on the QSP qubit \cite{martyn2021grand}. This is why the Hadamard gates are present in the circuit. Each block encoding unitary $B$ is constructed according to the approach outlined in Figure \ref{fig:block_encode_circ}.}
\label{fig:QSVT}
\end{figure*}

% For a general operator $O$, a block encoding encodes the matrix as:

% \begin{equation}
% \label{eq:block}
% \begin{aligned}
%     O_{block} &= \bra{0}_{a} \otimes I_{s} \big( G_{a}^{\dagger} \otimes I_{s} \big) U  \big( \otimes I_{s} \otimes G_{a} \big) I_{s} \otimes \ket{0}_{a} \\
%     &= \bra{G}_{a} \otimes I_{s}  U  I_{s} \otimes \ket{G}_{a} \\
%     &= mat \\
% \end{aligned}
% \end{equation}

% A full mathematical breakdown is provided in Appendix \ref{sec:block_encoding}, along with how to construct the quantum circuits required.

\subsubsection{Matrix inversion via quantum signal processing} \label{sec:QSP_inverison}

To perform the QSVT, one needs to be able to generate the quantum signal processing  angles \cite{low2016methodology, low2017optimal, low2019hamiltonian} to implement a function (or usually some approximation of a desired function). On a single qubit, QSP is usually defined as \cite{low2016methodology}:

\begin{equation}
\label{eq:QSP}
\begin{aligned}
    U_{\vec{\phi}}(a)  =e^{(+i \phi_{0} Z)} \prod_{k=1}^{d} W(a) e^{(+i \phi_{k} Z)} = \begin{bmatrix} \mathtt{P}(a) &  *\\ * & * \end{bmatrix},
\end{aligned}
\end{equation}
where $a \in [-1, 1]$, $W(a)= R_{x}[2 \cos^{-1}(a)]$ and $\mathtt{P}$ is a polynomial with degree at most the length of the sequence of QSP phases ($\leq d$). The constraints on what sort of polynomials can be implemented using this technique are covered in \cite{low2016methodology}. In equation \ref{eq:QSP}, once the polynomial to be implemented is fixed and the QSP angles are defined, all the $\{\phi_{k} | k=0,1,\hdots, d \} \in\vec{\phi}$ remain fixed. The only free variable remaining is $a$. It is therefore always possible to plot $\mathtt{P}(a)$ by simply calculating: $\bra{0} U_{\vec{\phi}}(a) \ket{0} = \mathtt{P}(a)$, where one scans over $-1 \leq a \leq 1$. 

We treat how the angles in $\vec{\phi}$ are calculated as a ``black-box'', further details are covered in \cite{haah2019product, chao2020finding, dong2021efficient, martyn2021grand}. Once the phases $\vec{\phi}$ have been calculated for a particular polynomial, they can be reused and  never have to be calculated again. The pyqsp \cite{pyqsp} and QSPPACK \cite{QSPPACK} open-source libraries allow users to generate different sequences of QSP angles for many different functions. An algorithm proposed by Haah in \cite{haah2019product} gives a rigorous analysis of how to find the angle sequence corresponding to a supplied polynomial that has a runtime scaling as $\mathcal{O}(d^{3}\text{polylog}(d/\epsilon))$, for a degree-$d$ polynomial. This returns a set of QSP angles for a uniform $\epsilon$-approximating polynomial over the interval $[-1,1]$. 

In this paper, we require an implementation of the inverse function. What is somewhat problematic about $1 / x$ is the discontinuity at $x=0$. Instead of approximating $1 / x$ over the full range, we approximate it over  $[-1,-\frac{1}{\kappa}]  \cup [\frac{1}{\kappa}, 1]$.  The existence of such an odd polynomial is guaranteed in Corollary 69 of \cite{gilyen2018quantum}. Importantly, the approximation of $1/x$ used in QSVT requires all the singular values  of the block-encoded matrix  $\{ \sigma \}$ to be $\sigma \geq 1/ \kappa \; \; \forall \sigma$,  otherwise they fall into the region where the polynomial approximation of $1/ x$ is ill defined. %Theorem 2 in \cite{gilyen2019quantum} allows such functions to be applied to a block encoded matrix.  

Extending the single-qubit QSP (equation \ref{eq:QSP}) to higher dimensions is discussed in \cite{gilyen2019quantum} (see theorem 2), where ideas from qubitization  \cite{low2019hamiltonian} and two-dimensional
invariant subspaces coming from Camille Jordan’s Lemma \cite{jordan1875essai} are used. Their results show how to apply certain polynomials to a block encoded matrix:

\begin{equation}
\label{eq:block_struc}
\begin{aligned}
B = 
 \begin{bmatrix} A =\sum_{j} \sigma_{j} \ket{w_{j}} \bra{v_{j}} &  *\\ * & * \end{bmatrix}.
\end{aligned}
\end{equation}
Here $A$ is written in its singular-value decomposition. The location of $A$ in $B$ is determined by certain projectors $\hat{\Pi}$ \cite{martyn2021grand}, in this work: $\ket{\overline{0}}\bra{\overline{0}}$. The QSVT circuit, for odd $d$,  can be built as \cite{gilyen2018quantum, gilyen2019quantum, martyn2021grand}:

\begin{equation}
\label{eq:QSVT_eq}
\begin{aligned}
U_{\vec{\phi'}}^{QSVT} &= 
 \hat{\Pi}(\phi_{0}') B \Bigg( \prod_{k=1}^{(d-1)/2}\hat{\Pi}(\phi_{2k-1}') B^{\dagger} \hat{\Pi}(\phi_{2k}') B \Bigg) \\
 &= \begin{bmatrix} \sum_{j} \mathtt{P}(\sigma_{j}) \ket{w_{j}} \bra{v_{j}} &  *\\ * & * \end{bmatrix},
\end{aligned}
\end{equation}
where:
\begin{equation}
\label{eq:angles}
\begin{aligned}
\phi_{l}' = \begin{cases}
       \phi_{0} + \phi_{d} +  \frac{(d-1)\pi}{2} , & \text{if}\ l = 0  \\
       \phi_{l} - \frac{\pi}{2}, & \text{if } l \in \{ 1,2,\hdots, d-1  \} 
    \end{cases}.
\end{aligned}
\end{equation}
Note the QSP phases $\vec{\phi} \in \mathbb{R}^{d+1}$ (equation \ref{eq:QSP}), have been modified to $\vec{\phi'} \in \mathbb{R}^{d
}$ for QSVT. This accounts for $W(a)$ not being a reflection operator, which is better suited to the qubitization formalism  \cite{low2019hamiltonian}.% See \cite{martyn2021grand} for further details.
% https://arxiv.org/pdf/2105.02859.pdf (eq 12 and 14)

In summary, equation \ref{eq:QSVT_eq} shows how a polynomial transform is applied to the singular values $\{ \sigma_{k}\} $ of $A$ (equation \ref{eq:block_struc}). We assumed $A$ to be a square matrix in our analysis, but this is not necessary \cite{martyn2021grand}. Figure \ref{fig:QSVT} summarises the QSVT circuit, where $\hat{\Pi}(\phi)$ is given be a multi zero-controlled $X$ gate targeted on the QSP qubit and controlled by the ``prep'' qubits (see Figure 1 in \cite{dong2021efficient}), followed by an $R_{z}$ rotation on the QSP qubit followed by another multi zero-controlled $X$ gate: $ \bigg( \big[ \ket{\bar{0}}\bra{\bar{0}}_{prep} \big] \otimes  X_{QSP}) + \big[I^{\otimes n_{prep}} - \ket{\bar{0}}\bra{\bar{0}}_{prep} \big] \otimes I_{QSP} \bigg)$.

\begin{figure*}[t]
    \centering
    \includegraphics[width=0.75\textwidth]{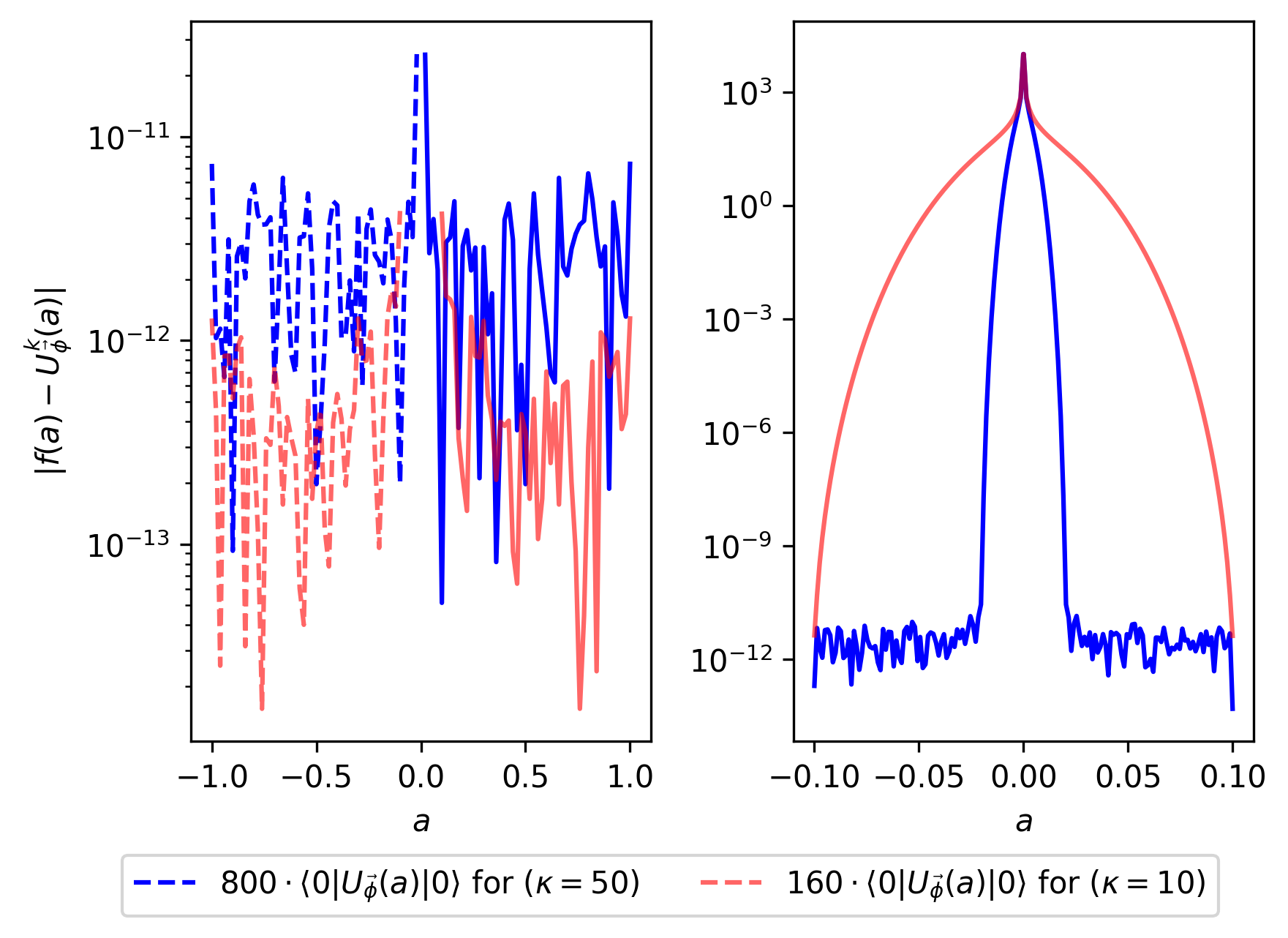}
    \caption{Absolute error of quantum signal processing (QSP) circuit approximating $f(a) =1/a$.  Each polynomial approximation is defined over the range $[-1,-\frac{1}{\kappa}]  \cup [\frac{1}{\kappa}, 1]$. The blue and red data represent different  approximations, $\kappa=10$ and $\kappa=50$, of the inverse function. These are $303$ and $1519$ degree polynomials respectively.
    The phases $\vec{\phi}$ to implement these functions via QSP are supplied in the Supporting Material. Note no data is calculated at $a=0$. The left figure is plotted over the domain $a\in \{[-1,-\frac{1}{\kappa}] \}  \cup \{[\frac{1}{\kappa}, 1]\}$,  the dashed data goes over the range $[ -1,-\frac{1}{\kappa} ]$ and the solid data over $[\frac{1}{\kappa}, 1]$. The constant multipliers given in the Figure legend account for the normalization factors required by each polynomial to ensure they lie between $\pm 1$.
        \label{fig:inv_x_qsp}}
\end{figure*}

% \caption{Single qubit quantum signal processing (QSP) circuit approximating $f(a) =1/a$. Each approximation is defined over the range $[-1,-\frac{1}{\kappa}]  \cup [\frac{1}{\kappa}, 1]$.  The top two plots take $a$ in the interval:  $ [-1,-\frac{1}{\kappa}]  \cup [\frac{1}{\kappa}, 1]$. The bottom two plots take $a$ in the interval: $ [-\frac{1}{\kappa}, 0)  \cup (0,\frac{1}{\kappa}]$. The degree of polynomial used to approximate $\kappa=50$ and $\kappa=10$ is $303$ and $1519$ respectively. The phases $\vec{\phi}$ to implement these functions via QSP are supplied in the Support Information.}

% page 45
% https://arxiv.org/pdf/2008.13295.pdf

\subsection{Single-impurity Anderson model} \label{sec:SIAM}
A common model used to describe strongly correlated electron systems in thermodynamic equilibrium is the Hubbard Hamiltonian. However, classical simulation of this model is severely limited by how many fermionic orbitals can be described, due to the exponential increase of the Hilbert space. Dynamical mean field theory (DMFT) was developed to solve this issue, where the physics of a many-body problem is captured via a single-impurity that is coupled self-consistently to a fermonic host (bath) \cite{kotliar2004strongly}. In the limit of a lattice with infinite dimensions, for the Hubbard model with infinite coordination number (nearest neighbours),  DMFT exactly maps the solution of the Hubbard model to that of the Anderson impurity model. This is because interacting electrons in the Hubbard model in the thermodynamic limit (infinite lattice sites) are modelled by a single-impurity site coupled to an electronic bath (infinite bath sites) that tunnel into the impurity site \cite{georges1996dynamical, kotliar2004strongly, steckmann2021simulating}. Crucially, DMFT is derived in the limit of infinite lattice coordination; however, for finite dimensions it can still provide good approximations and allow interesting phenomena to be explored \cite{caffarel94, georges96}. 

In this paper, we consider a two-site one-dimensional single-impurity Anderson model defined by the Hamiltonian \cite{kreula2016few}:

% https://cqwbkpro.s3.eu-west-2.amazonaws.com/wp-content/uploads/2021/07/27144258/191010_CQ_Dynamical-Mean-Field-Theory-Algorithm-and-Experiment-on-Quantum-Computers.pdf
% eq19
\begin{equation}
\label{eq:H_hubb}
\begin{aligned}
    H = &\frac{U}{4} Z_{1}Z_{3} + \bigg( \frac{\mu}{2} -  \frac{U}{4} \bigg) \big( Z_{1} + Z_{3} \big) - \frac{\epsilon_{2}}{2} \big( Z_{2} + Z_{4} \big) \\
    & + \frac{V}{2} \big( X_{1}X_{2} +Y_{1}Y_{2} + X_{3}X_{4} + Y_{3}Y_{4}\big).
\end{aligned}
\end{equation}
Details on this Hamiltonian are provided in \cite{potthoff2001two, kreula2016few, rungger2019dynamical}.  Note $H$ is written under the Jordan-Wigner transformation, which allowed the fermionic operators to be mapped to spin operators acting on qubits \cite{Jordan1928}. Qubit index $1$ ($3$) represents the impurity spin-up (spin-down) site and index $2$ ($4$) represents the spin up (spin down) bath site. Here, $U$ is the onsite Coulomb repulsion, $\mu$ is the chemical potential that controls the electron filling in the grand canonical ensemble\footnote{ A grand canonical ensemble is a generalization of the canonical ensemble (that represents the possible states of a mechanical system in thermal equilibrium with a heat bath at a fixed temperature), where the restriction to a definite number of particles is removed. An example of this is in chemistry, where the number of each molecular species is not conserved but the number of atoms is. For example: $4A + 2B \rightarrow A_{4}B_{2}$, where there are six molecules (particles) on the left and only one on the right, but always six atoms.}, $\epsilon_{2}$ describes the on-site energy of the non-interacting bath site $2$, and $V$ is the interaction of this bath site with the impurity. Interestingly, $H$ is equivalent for spin-up and spin-down electrons and so the self-energy of the impurity only needs to be calculated for one spin site \cite{rungger2019dynamical}.
% % pg 9
% % https://scholar.harvard.edu/files/schwartz/files/7-ensembles.pdf
To solve equation \ref{eq:H_hubb} via DMFT, i.e. find the parameters of the effective model, one needs to consider the Green's function of the lattice problem $G_{lat}(z)$ and impurity $G_{imp}(z)$. For infinite bath sites $G_{lat}(z)=G_{imp}(z)$. In practice, only a finite number of bath sites can be used and so the difference between $G_{lat}(z)$ and $G_{imp}(z)$ is minimised. In this work we consider $2$-site DMFT under the particle-hole (ph) symmetric case, where $\mu = \frac{U}{2}$ and $\epsilon_{2}=0$ \cite{rungger2019dynamical}. The only impurity parameter is therefore $V$. For a fixed $U$ and given threshold $\zeta$, the following steps are taken \cite{kreula2016few}:

\begin{enumerate}
    \item For a fixed $U$, guess an initial on-site energy $V$, thus determining $H$ (equation \ref{eq:H_hubb}).
    \item Calculate the Green's function of the Hamiltonian $G_{ij}(z)$ (equation \ref{eq:G_funct}). \begin{enumerate}
    \item In this work, each element of $G_{ij}(z)$ is determined by the quantum singular-value transform.
  \end{enumerate}
    \item From $G_{ij}(z)$ define $G_{imp}(z)$ \begin{enumerate}
    \item This is achieved by selecting the elements of $G_{ij}(z)$ that correspond to the impurity site.
  \end{enumerate}
    \item Calculate the quasi-particle weight $z_{qp}=(1-\frac{Im[\Sigma_{imp} (i \delta)]}{\delta})^{-1}$.\begin{enumerate}
    \item The self energy can be obtained as: $\Sigma_{imp}(z) = G_{imp}^{0}(z)^{-1} - G_{imp}(z)^{-1}$, where the noninteracting Green's function is defined as $G_{imp}^{0}(z) = (z -\epsilon_{\alpha} + \mu - \frac{|V|^{2}}{z})^{-1}$ \cite{kreula2016few, rungger2019dynamical}.
    \item Due to particle hole symmetry, $Im[\Sigma_{imp} (i \delta)]$ is a single number due to spin-up and -down self-energies being the same for the impurity site.
  \end{enumerate}
    \item Set $V_{new} = \sqrt{z_{qp}}$
    \item If $|V_{new} - V|\leq \zeta$ then the bath parameter (and so DMFT) has converged. Otherwise, set $V=V_{new}$ and repeat from step $2$.
\end{enumerate}
For the 2-site model considered here, there is an analytic form for $V$ \cite{rungger2019dynamical, potthoff2001two}:

\begin{equation}
\label{eq:V_true}
    V = \begin{cases}
  \sqrt{1-\big(\frac{U}{6}\big)^{2}}, & \text{if}\ U<6 \\
  0, & \text{if}\ U\geq 6
\end{cases}.
\end{equation}
In the work presented, rather than optimising for $V$ at different fixed $U$, we use equation \ref{eq:V_true} to determine the optimal $V$ before calculating the Green's function. The goal is to investigate calculation of the Green's function via QSVT, not performing DMFT self-consistent optimisation loops.

\section{Method} \label{sec:method}
We numerically investigated the performance of calculating the Green's fucntion for the two-site Anderson model via the QSVT algorithm. To build the qubit Hamiltonian,  Quantinuum's InQuanto package was utilized \cite{inquanto, inquanto_prod}. The circuits required to perform QSVT were then constructed using PyTket \cite{sivarajah2020t}. Importantly, we only built the QSVT circuit to perform matrix inversion.  First, we generated the QSP phase angles in the open-source python library QSPPACK \cite{QSPPACK}. The phases $\vec{\phi}$ obtained (for the different polynomial approximations of the inverse function -
$k=10$ and $k=50$) are supplied in the Supporting Material. Next, for each $z$, we built two quantum circuits that performed $T \approx \big( z - [H-E_{0}] \big)^{-1}$ and $W \approx \big(z + [H-E_{0}] \big)^{-1}$ via the quantum singular value transform algorithm - see Figure \ref{fig:QSVT}. This required the block-encoding circuits for $(\|B_{(e)}\|_{1} ,3,0)$ and $(\|C_{(h)}\|_{1} ,3,0)$, except for the SIAM Hamiltonian defined for $U=8$ and $V=0$ where a $(\|B_{(e)}\|_{1} ,1,0)$ and $(\|C_{(h)}\|_{1} ,1,0)$ was used, due to certain Pauli operators having a coefficient of zero. Each block encoding was constructed according to the template in Figure \ref{fig:block_encode_circ}. We note here, that QSP only approximates the true inverse function via a polynomial, hence the approximately equal use. In all instances, the complex part of $z= \omega + i\delta$ was fixed to be $\delta = 0.1$. This was chosen to ensure all the singular values of each block encoded matrix were above $0.02$.

After each quantum circuit was built, a noise-free classical simulation was performed giving the unitary of the whole QSVT cirucit for each $z$ value. We post-select into the correct block of the unitary (see equation \ref{eq:QSVT_eq}) to obtain the transformed matrix. For each pair of quantum circuits, we denote these post-selected matrices $T$ and $W$.  We then classically determined $G_{ij}(z)$   by evaluating $\bra{\Psi_{0}} a_{i} T a_{j}^{\dagger} \ket{\Psi_{0}}$ and $\bra{\Psi_{0}} a_{j}^{\dagger} W  a_{i}\ket{\Psi_{0}}$ (equation \ref{eq:G_plus} and \ref{eq:G_minus}) for all $i,j$, where $i$ and $j$ run over all qubit indices using the standard linear algebra python libraries \cite{Numpy, SciPy}. The ground state $\ket{\Psi_{0}}$ used in each calculation was obtained by diagonalizing $H$ on a classical computer for particular $(U,V)$ parameterizations. For each QSVT simulation, we also calculated the exact classical solution, where the Green's function was calculated via matrix inversion performed on classical hardware.

\section{Results and Discussion} \label{sec:discussion}
\begin{figure}[t]
    \centering
    \includegraphics[width=0.5\textwidth]{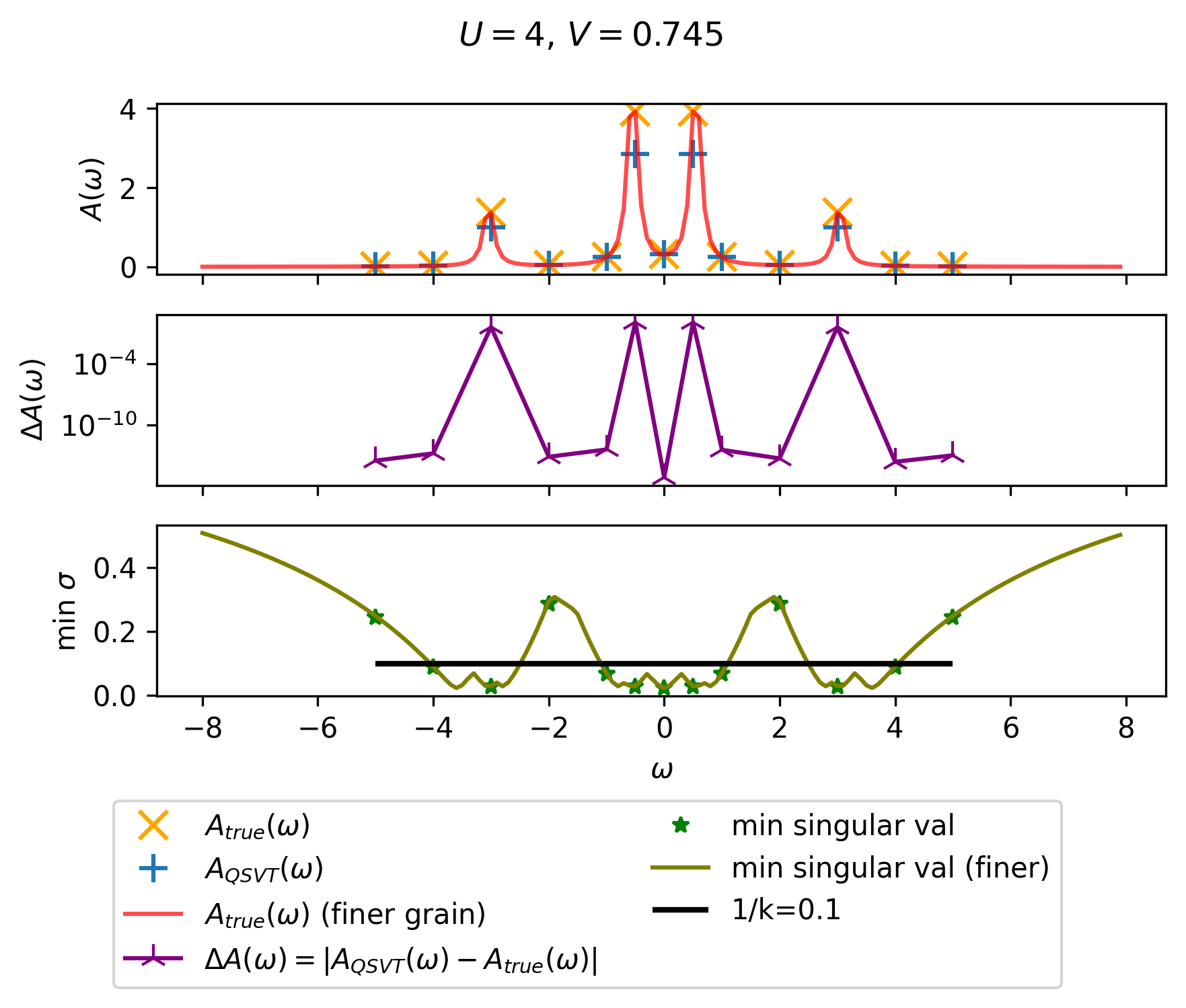}
    \caption{ (top) Spectral function  (equation \ref{eq:spectral_F}) of single-particle Anderson model for $U=4$ and $V=0.745$. The red line and orange points shows the spectral function being calculated via exact diagonalization. The blue points show the spectral function calculated via the quantum singular value transform  with $k=10$. (middle) Plot shows absolute error of spectral function $\Delta A(\omega) = |A_{QSVT}(\omega) - A_{true}(\omega) |$. (bottom) Plot shows the minimum singular value of the matrix to undergo inversion via the QSVT. Any singular below the black line is outside the region of where the approximation of $1/x$ is well defined.}
        \label{fig:Green_U4_K10}
\end{figure}

The QSVT applies a function, defined by the classically pre-computed $\vec{\phi}$ angles, to the singular values  of a (block encoded) matrix. At no point does a user have access to the singular values; it can be shown that the algorithm just applies a function to the singular values: aka $M = U \Sigma V^{\dagger} \mapsto U f(\Sigma) V^{\dagger}$. However, not knowing what the singular values are has consequences when implementing matrix inversion via QSVT. As discussed in Section \ref{sec:QSP_inverison}, the inverse function is not defined at $x=0$ and so is approximated over the domain $[-1,-\frac{1}{\kappa}]  \cup [\frac{1}{\kappa}, 1]$. If any singular value of the matrix to invert falls outside of this range it will not be transformed properly. This leads to a dilemma, where one needs to know the singular values to determine an appropriate $k$; however, knowing the singular values is the same as solving the inversion problem.  This issue can be resolved in two ways.

The first strategy makes the approximation of $1/x$ arbitrarily small, by using a very large value of $k$. This is somewhat similar to how conventional computers perform mathematical operations to machine precision. However, the degree of polynomial approximating the inverse function scales as \cite{martyn2021grand} :

% https://arxiv.org/pdf/2105.02859.pdf eq 81
 \begin{equation}
\begin{aligned}
    d = \mathcal{O}(k \log(k/\epsilon)).
\end{aligned}
\end{equation}
What this means is as a better polynomial approximation is used (higher $k$ value) the greater the degree of the resulting polynomial will be. As the circuit depth of QSVT scales as $\mathcal{O}(d)$ repeats of the block encoded circuit \cite{martyn2021grand}, using an arbitrarily large $k$ could unnecessarily increase the circuit depth of a given problem. 

The second approach to determine a valid $k$, is to estimate the magnitude of the lowest singular value of a matrix $M\in \mathbb{C}^{n \times n}$ \cite{hong1992lower, piazza2002upper,huang2008estimation, zou2010estimation, zou2012lower}. In  \cite{zou2012lower}, it is shown that:
 
 \begin{equation}
\label{eq:min_sig_value}
\begin{aligned}
    \sigma_{min} \geq |det(M)| \cdot \bigg(\frac{n-1}{\| M \|_{F}^{2} - l^{2}}\bigg)^{(n-1)/2} = \sigma_{min}^{\text{approx}},
\end{aligned}
\end{equation}
which provides a lower bound on the magnitude of the smallest singular value $\sigma_{min}$ of a non-singular $n \times n$ complex matrix $M$. Here, $l= |det(M)| \cdot \big(\frac{n-1}{\| M \|_{F}^{2}}\big)^{(n-1)/2}$ and $\| M \|_{F}$ is the Frobenius norm. Evaluating equation \ref{eq:min_sig_value}, allows $k$ to be determined as: $k \geq 1/\sigma_{min}^{\text{approx}}$. However, this approach requires the determinant of the matrix to be found, which can be costly. A further approximation could be used to estimate $|det(M)|$, such as using the methods in \cite{bai1996some, ipsen2011determinant}.

% We note the preconditioning approach by Tong \textit{et al.} can increase the can also be utilised to transform the problem to have a larger $\sigma_{min}$. 

% \onecolumngrid

\begin{figure}[b]
    \centering
    \includegraphics[width=0.5\textwidth]{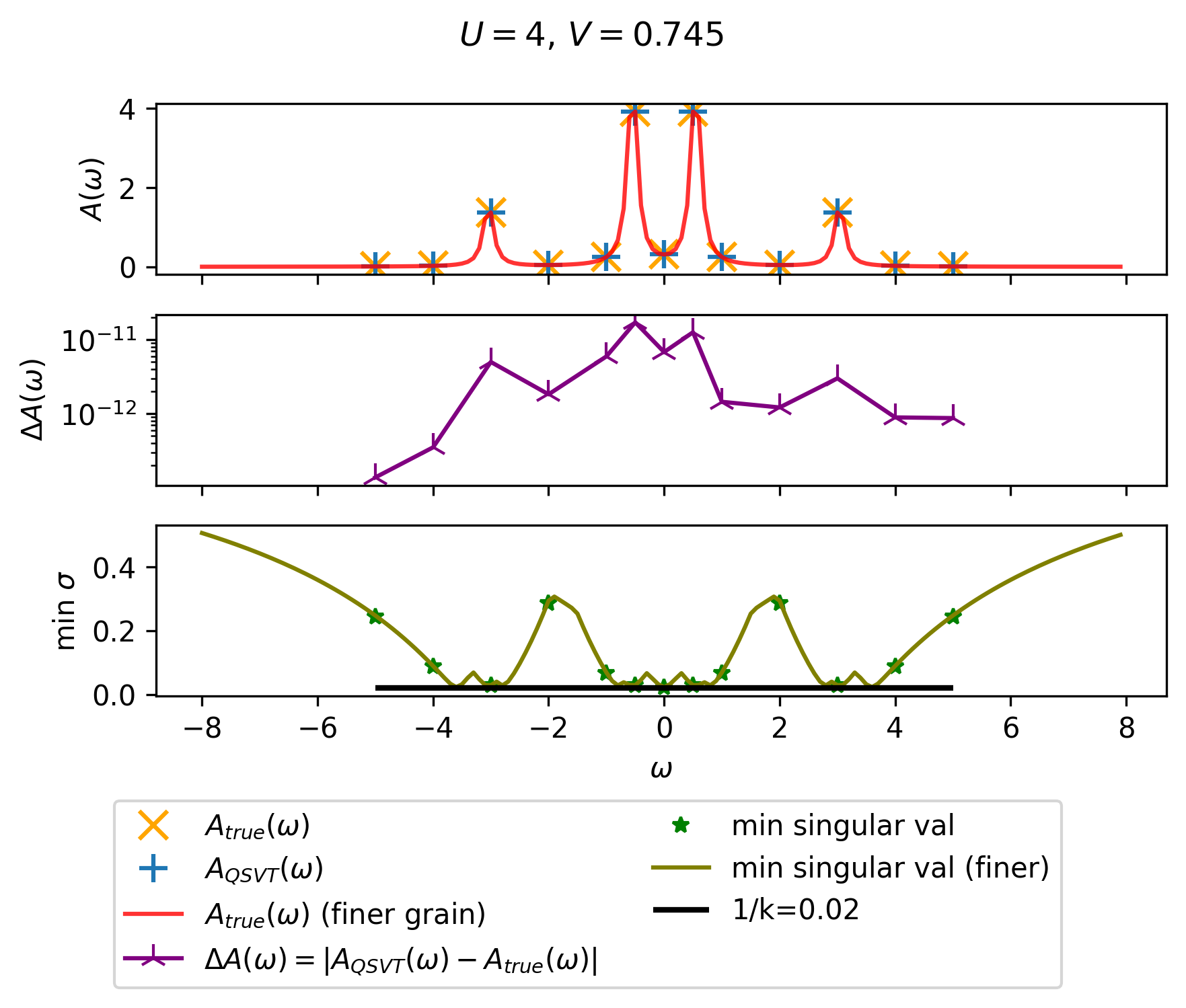}
    \caption{(top) Spectral function  (equation \ref{eq:spectral_F}) of single particle Anderson model for $U=4$ and $V=0.745$. The red line and orange points shows the spectral function function being calculated via exact diagonalization. The blue points show the spectral function calculated via the quantum singular value transform  with $k=50$. (middle) Plot shows absolute error of spectral function $\Delta A(\omega) = |A_{QSVT}(\omega) - A_{true}(\omega) |$. (bottom) Plot shows the minimum singular value of the matrix to undergo inversion via the QSVT. Any singular below the black line is outside the region of where the approximation of $1/x$ is well defined.}
        \label{fig:Green_U4_K50}
\end{figure}
 
The two-site SIAM considered in this work is defined on four qubits and so classically performing a singular value decomposition (SVD) of a $(16 \times 16)$ matrix was possible. Therefore, in order to find appropriate $k$, rather than using equation \ref{eq:min_sig_value}, we used the true $\sigma_{min}$. We found all the singular values were above $0.02$ and so used a $k=50$ approximation. This represented a scenario where all the singular values would be inverted properly via the QSVT.  We also simulated a $k=10$ polynomial approximation, where some of the singular values lay below $0.1$ and so wouldn't be inverted properly. The goal was to see what effect this would have. The $k=10$ and $k=50$ polynomial approximations of $1/x$  were represented by $303$ and $1519$ degree  polynomials respectively. Figure \ref{fig:inv_x_qsp} illustrates the  errors of these polynomial approximations compared to the true inverse function.   

For each $k$, we calculated the single particle Green's function at different $z$ for different $U,V$ parameters via QSVT. The outputs were then used to plot the spectral function, which is defined as:

\begin{equation}
\label{eq:spectral_F}
    % A(\omega) = \abs{\frac{1}{\pi} Im\big[G(\omega) \big]}
\begin{aligned}
    A(\omega) &= -\frac{1}{\pi} Tr\Bigg( Im\big[G_{H}(\omega) \big] \Bigg).
\end{aligned}
\end{equation}
The results for $U=4$ and $V=0.745$ are given in Figures \ref{fig:Green_U4_K10} and \ref{fig:Green_U4_K50}. The results for the other $U,V$ regimes are given in Appendix \ref{sec:A_omega_plots}. As this problem is defined over $4$ qubits, exact diagonalization solutions were possible to compute classically. In Figures \ref{fig:Green_U4_K10}, \ref{fig:Green_U4_K50} and those in Appendix \ref{sec:A_omega_plots}, we provide the lowest singular value of the matrix to undergo QSVT matrix inversion. As expected, we observe that whenever the singular value lies below $1/k$, the error in the spectral function becomes large. This can be seen for $k=10$, where in Figure  \ref{fig:Green_U4_K10} errors due to $\sigma$ being below $0.1$ can sometimes differ from the real answer by $\Delta A(\omega) \approx 1$. Whereas, in Figure \ref{fig:Green_U4_K50} all singular values lie above $0.02$ and the error in $A(\omega)$ remains at $\Delta A(\omega) \approx 10^{-12}$.

\begin{figure}[t]
% https://journals-aps-org.libproxy.ucl.ac.uk/prb/pdf/10.1103/PhysRevB.64.165114 
% eq 3
%% interacting DOS given below eq 25!
% AND
% http://www.physics.rutgers.edu/~kotliar/papers/anderson.pdf
% EQ 1.15
% AND
% https://epjquantumtechnology.springeropen.com/articles/10.1140/epjqt/s40507-016-0049-1
% EQ 7
% and START of section 5!
    \centering
    \includegraphics[width=0.8\columnwidth]{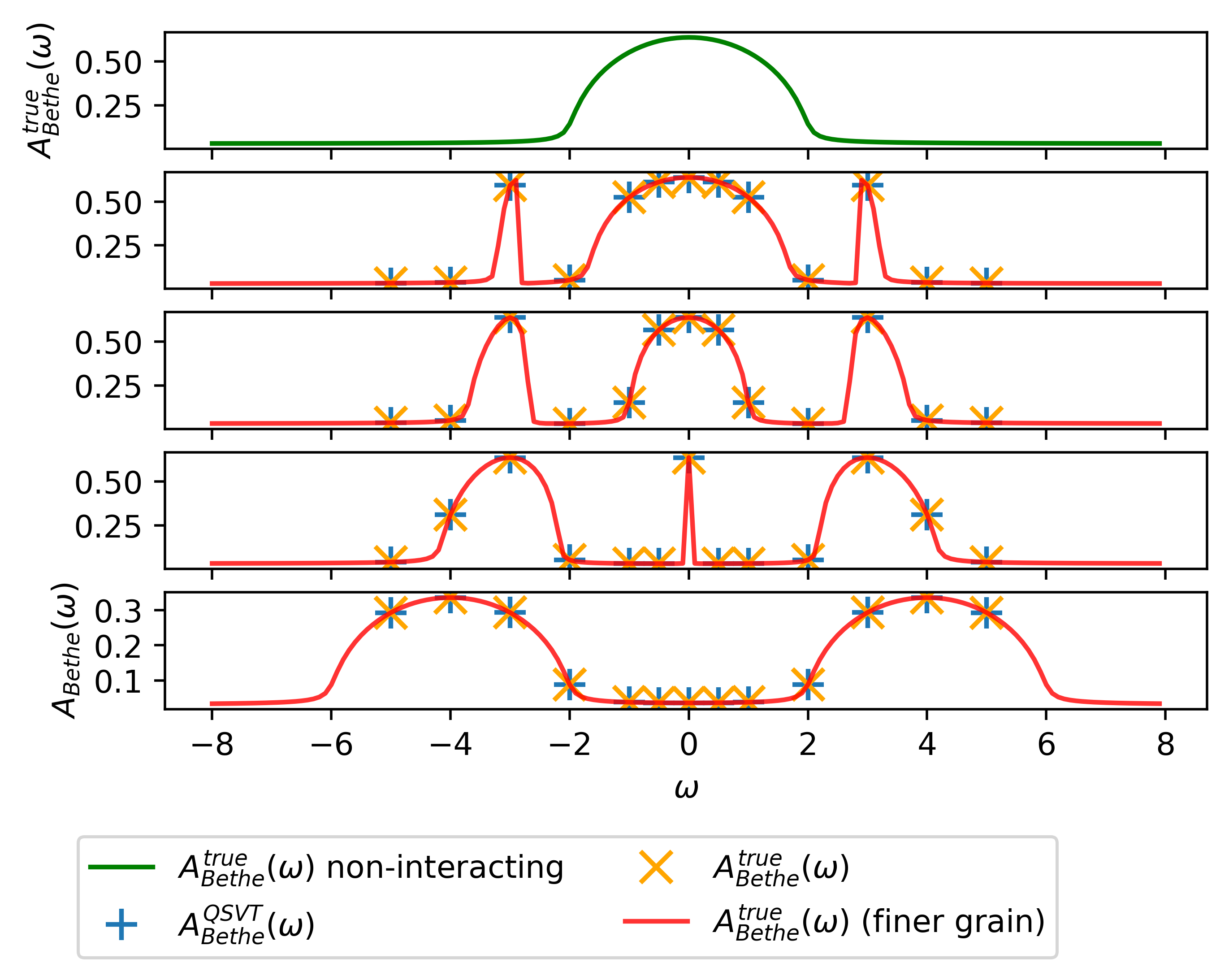}
    \caption{Metal to insulator Mott transition of the two-site Anderson impurity model. Plots show the density of states on the Bethe lattice. The first plot gives the non-interacting system, followed by $(U,V)$ combinations of $(2,0.943), (4,0.745), (5.99, 0.058), (8, 0)$. Note the free density of states on the Bethe lattice with infinite coordination is  $\rho_{0}(x) = \frac{1}{2\pi t^{2}} \sqrt{4t^{2}- x^{2}}$ (in this work $t=1$) and the interacting density of states is $\rho(\omega)=\rho_{0}(\omega + \mu - \Sigma_{imp}[\omega])$ \cite{potthoff2001two, kreula2016few, rungger2019dynamical}. The non-interacting Green's function is defined as $G_{imp}^{0}(\omega) = (\omega + i \delta -\epsilon_{\alpha} + \mu - \frac{|V|^{2}}{\omega+i \delta})^{-1}$ (in this work $\epsilon_{\alpha}=0$ and $\mu = \frac{U}{2}$) \cite{kreula2016few, rungger2019dynamical}. For the final plot $(8, 0)$, particle hole symmetry was broken and the spin-up and -down parts of $\Sigma_{imp}[\omega])$ were treated separately and combined.}
        \label{fig:mott_k50}
\end{figure}

%% https://physics.stackexchange.com/questions/29542/what-do-the-poles-of-a-green-function-mean-physically

Looking at the Green's function definition in equation \ref{eq:G_funct}, we see that when the real part of $z$ is equal to $\pm \lambda_{i}\mp E_{0}$ (where $\lambda_{i}$ is an eigenvalue of $H$), then the real part of the denominator in either the advanced or retard Green's function vanishes leading to a so called ``pole''. For a non-interacting system, where the eigenfunctions are represented by single-configuration states, the amplitudes and Lehmann energies $\omega$ are equal to the eigenfunctions and eigenvalues of the corresponding one-electron Hamiltonian \cite{onida2002electronic}.  The spectral function consists of a set of peaks at those eigenvalues and each peak is associated with a particle \cite{onida2002electronic, aryasetiawan1998gw}.  When interactions are considered, the eigenfunctions are no longer single-configuration states, instead they are in general normalised linear combinations of them. There will now be more non-vanishing contributions to the spectral function,  by merging these contributions they will form a structure which can be thought of as deriving from peaks when the interactions are turned off \cite{onida2002electronic}. This allows one to work in a particle-like picture; however, each peak is now associated with a ``quasiparticle'' \cite{onida2002electronic, aryasetiawan1998gw}. The spreading of the peak contains information about many-body correlation effects in the interacting system  \cite{onida2002electronic}. The spectral function is usually peaked at each energy $E_{i} = \epsilon_{i} + Re \Delta \Sigma_{i}(E_{i})$, with a lifetime given by $1/ Im \Sigma_{i}(E_{i})$ where $\Sigma$ is the self-energy operator. Further details on this are discussed in \cite{aryasetiawan1998gw}.

From the calculated Green's functions, we then plot the Mott phase transition by plotting the density of states on the Bethe lattice \cite{eckstein2005hopping} for different $U,V$ regimes at different $z$. The $k=50$ result is summarised in Figure  \ref{fig:mott_k50}. Qualitatively, our results are similar to related work -  see Figure 3 in \cite{zhang1993mott} and Figure 4 in \cite{sriluckshmy2021fully}. We provide error details and the $k=10$ result in Appendix \ref{sec:mott_appendix}. From Figure \ref{fig:mott_k50} we see that the QSVT result matches well with the true (exact diagonalization) results. The error of the QSVT approach, for the $k=50$ polynomial approximation, for each data point is around the $10^{-12}$ level. Wheres, we obtain much larger error for the $k=10$ approximations ranging from the $10^{-12}$ to $10^{-2}$ level. This is due to errors occurring when the singular values are below $0.1$, which are then not inverted properly leading to errors in the calculated single particle Green's function and thus the spectral function. 

%% SEE:
% 1. https://arxiv.org/pdf/cond-mat/9712013.pdf
% 2. Onida, Reining, and Rubio: Density-functional vs many-body

\section{Improved LCU Circuits} \label{sec:LCU_circuits}

In Section  \ref{sec:block}, we reviewed the LCU technique \cite{childs2012hamiltonian, low2017optimal} and then proposed an efficient way to construct the ``$\select$'' operator when defined as a linear combination of Pauli operators.  This strategy came from a combination of ideas presented in \cite{ralli2021implementation, bullock2003smaller, da2022linear} - see Table \ref{table:circuit_summary} which summaries the different costs. The overall cost to implement a LCU block encoding via this approach has a circuit cost scaling as  $O\big( \underbrace{|A|}_{\prep}  + \underbrace{|A|  \log_{2}(|A|)^{2}+ |A|n_{s}}_{\select} + \underbrace{|A|}_{\prep^{\dagger}}\big)$ CNOT and single-qubit gates.  Here $n_{s}$ is the number of system qubits. Our work is distinct from the approach in \cite{low2018trading}, which is a more fault tolerant approach to the problem, \cite{wan2021exponentially} that focuses on second-quantised fermionic Hamiltonians, and \cite{camps2022fable}, which uses ideas of multiplexors to reduce the gate-complexity. The FABLE approach is a different block encoding technique to the LCU method, where a defined matrix is decomposed into a block encoding. This makes a direct comparison hard. The FABLE method also requires $n+1$ ancillary qubits, rather than $\lceil \log_{2}(|A|) \rceil$ for a LCU, and has a single- and two-qubit gate cost scaling as $\mathcal{O}(2^{2n})$. We re-iterate that these block encoding methods differ in how the matrix to block encoding is defined and thus their expected applications are different. However, for problems where the Hamiltonian is defined as a linear combination of unitaries, the circuit implementation of LCU presented in this work will result in quantum circuits with exponentially fewer gates. The intuition comes from the LCU method requiring exponentially fewer ancillary qubits to perform the block encoding. For example, in \cite{camps2022fable} they apply the FABLE approach to different Hubbard Hamiltonians. In this scenario our approach will require exponentially fewer single- and two-qubit gates.

\section{Conclusions} \label{sec:conclusion}
The work presented here shows how the quantum singular transform can be used to calculate the Green's function in the Lehmann representation, given a sufficient polynomial approximation of the inverse function. In our noise-free simulations, we observe large errors when performing matrix inversion via QSVT, if the singular values of the matrix to invert fall outside the well-defined domain of the polynomial approximation of the inverse function. Our results indicate that care must be taken to ensure this doesn't happen. Equation \ref{eq:min_sig_value} offers a route to determining a proper approximation level; however, as discussed in the main text, this may require further approximations to evaluate. 

We show the metal insulator phase transition of the two-site Anderson model calculated for the Green's functions obtained from noise-free QSVT simulations (Figure \ref{fig:mott_k50}). For $k=50$ we find the absolute error from the true answer remains at the $10^{-12}$ level. For the $k=10$ result, where some of the singular values of the problem lie outside the domain where the polynomial approximation is correctly defined, we obtain much larger errors.

Finally, we described a new circuit strategy to implement the $\select$ oracle in the LCU technique. This uses ideas presented by Ralli \textit{et al.} in \cite{ralli2021implementation} and da Silva \textit{et al.} in \cite{da2022linear}. The overall circuit cost to perform a block encoding of any matrix supplied as a linear combination of Pauli operators scales as $O\big( |A|(\lceil \log_{2}(|A|) \rceil^{2}+n_{s}) \big)$ CNOT and single-qubit gates. Here $n_{s}$ is the number of system qubits and $|A|$ is the number of Pauli operators in the matrix defined in the LCU.

An interesting avenue for future work is to investigate whether ``phase gadgets'' can further improve the circuit cost of the $\select$ operator \cite{cowtan2019phase}. In detail, can ZX-calculus maximize gate cancellations between multicontrol Pauli gates in $U_{\select}$ implemented, according to the template outlined in Figure \ref{fig:cntrl_P_gate}?

% Future work will look into  building this LCU construction into the open-source PyTket compiler, so it can easily be utilised by the quantum computing community. 

\section*{Acknowledgements} \label{sec:acknowledgements}
The authors would like to thank Hans Hon Sang Chan, Silas Dilkes and Yao Tang for helpful discussions along with Yuta Kikuchi and Ifan Williams for feedback on the manuscript.

\bibliographystyle{apsrev4-1.bst}
\bibliography{references.bib}

\begin{thebibliography}{85}%
\makeatletter
\providecommand \@ifxundefined [1]{%
 \@ifx{#1\undefined}
}%
\providecommand \@ifnum [1]{%
 \ifnum #1\expandafter \@firstoftwo
 \else \expandafter \@secondoftwo
 \fi
}%
\providecommand \@ifx [1]{%
 \ifx #1\expandafter \@firstoftwo
 \else \expandafter \@secondoftwo
 \fi
}%
\providecommand \natexlab [1]{#1}%
\providecommand \enquote  [1]{``#1''}%
\providecommand \bibnamefont  [1]{#1}%
\providecommand \bibfnamefont [1]{#1}%
\providecommand \citenamefont [1]{#1}%
\providecommand \href@noop [0]{\@secondoftwo}%
\providecommand \href [0]{\begingroup \@sanitize@url \@href}%
\providecommand \@href[1]{\@@startlink{#1}\@@href}%
\providecommand \@@href[1]{\endgroup#1\@@endlink}%
\providecommand \@sanitize@url [0]{\catcode `\\12\catcode `\$12\catcode
  `\&12\catcode `\#12\catcode `\^12\catcode `\_12\catcode `\%12\relax}%
\providecommand \@@startlink[1]{}%
\providecommand \@@endlink[0]{}%
\providecommand \url  [0]{\begingroup\@sanitize@url \@url }%
\providecommand \@url [1]{\endgroup\@href {#1}{\urlprefix }}%
\providecommand \urlprefix  [0]{URL }%
\providecommand \Eprint [0]{\href }%
\providecommand \doibase [0]{http://dx.doi.org/}%
\providecommand \selectlanguage [0]{\@gobble}%
\providecommand \bibinfo  [0]{\@secondoftwo}%
\providecommand \bibfield  [0]{\@secondoftwo}%
\providecommand \translation [1]{[#1]}%
\providecommand \BibitemOpen [0]{}%
\providecommand \bibitemStop [0]{}%
\providecommand \bibitemNoStop [0]{.\EOS\space}%
\providecommand \EOS [0]{\spacefactor3000\relax}%
\providecommand \BibitemShut  [1]{\csname bibitem#1\endcsname}%
\let\auto@bib@innerbib\@empty
%</preamble>
\bibitem [{\citenamefont {Green}(1889)}]{green1889essay}%
  \BibitemOpen
  \bibfield  {author} {\bibinfo {author} {\bibfnamefont {G.}~\bibnamefont
  {Green}},\ }\href@noop {} {\emph {\bibinfo {title} {An essay on the
  application of mathematical analysis to the theories of electricity and
  magnetism}}},\ Vol.~\bibinfo {volume} {3}\ (\bibinfo  {publisher} {author},\
  \bibinfo {year} {1889})\BibitemShut {NoStop}%
\bibitem [{\citenamefont {Duffy}(2015)}]{duffy2015green}%
  \BibitemOpen
  \bibfield  {author} {\bibinfo {author} {\bibfnamefont {D.~G.}\ \bibnamefont
  {Duffy}},\ }\href@noop {} {\emph {\bibinfo {title} {Green's functions with
  applications}}}\ (\bibinfo  {publisher} {Chapman and Hall/CRC},\ \bibinfo
  {year} {2015})\BibitemShut {NoStop}%
\bibitem [{\citenamefont {Challis}\ and\ \citenamefont
  {Sheard}(2003)}]{challis2003green}%
  \BibitemOpen
  \bibfield  {author} {\bibinfo {author} {\bibfnamefont {L.}~\bibnamefont
  {Challis}}\ and\ \bibinfo {author} {\bibfnamefont {F.}~\bibnamefont
  {Sheard}},\ }\href@noop {} {\bibfield  {journal} {\bibinfo  {journal}
  {Physics Today}\ }\textbf {\bibinfo {volume} {56}},\ \bibinfo {pages} {41}
  (\bibinfo {year} {2003})}\BibitemShut {NoStop}%
\bibitem [{Nob()}]{Nobel1965}%
  \BibitemOpen
  \href@noop {} {\enquote {\bibinfo {title} {The nobel prize in physics
  1965},}\ }\bibinfo {howpublished}
  {\url{https://www.nobelprize.org/prizes/physics/1965/summary/}},\ \bibinfo
  {note} {accessed: 2022-08-15}\BibitemShut {NoStop}%
\bibitem [{\citenamefont {Schwinger}(1993)}]{schwinger1993greening}%
  \BibitemOpen
  \bibfield  {author} {\bibinfo {author} {\bibfnamefont {J.}~\bibnamefont
  {Schwinger}},\ }\href@noop {} {\bibfield  {journal} {\bibinfo  {journal}
  {arXiv preprint hep-ph/9310283}\ } (\bibinfo {year} {1993})}\BibitemShut
  {NoStop}%
\bibitem [{\citenamefont {Szabo}\ and\ \citenamefont
  {Ostlund}(2012)}]{szabo2012modern}%
  \BibitemOpen
  \bibfield  {author} {\bibinfo {author} {\bibfnamefont {A.}~\bibnamefont
  {Szabo}}\ and\ \bibinfo {author} {\bibfnamefont {N.~S.}\ \bibnamefont
  {Ostlund}},\ }\href@noop {} {\emph {\bibinfo {title} {Modern quantum
  chemistry: introduction to advanced electronic structure theory}}}\ (\bibinfo
   {publisher} {Courier Corporation},\ \bibinfo {year} {2012})\BibitemShut
  {NoStop}%
\bibitem [{\citenamefont {Onida}\ \emph {et~al.}(2002)\citenamefont {Onida},
  \citenamefont {Reining},\ and\ \citenamefont {Rubio}}]{onida2002electronic}%
  \BibitemOpen
  \bibfield  {author} {\bibinfo {author} {\bibfnamefont {G.}~\bibnamefont
  {Onida}}, \bibinfo {author} {\bibfnamefont {L.}~\bibnamefont {Reining}}, \
  and\ \bibinfo {author} {\bibfnamefont {A.}~\bibnamefont {Rubio}},\
  }\href@noop {} {\bibfield  {journal} {\bibinfo  {journal} {Reviews of modern
  physics}\ }\textbf {\bibinfo {volume} {74}},\ \bibinfo {pages} {601}
  (\bibinfo {year} {2002})}\BibitemShut {NoStop}%
\bibitem [{\citenamefont {March}(1999)}]{farid1999electron}%
  \BibitemOpen
  \bibfield  {author} {\bibinfo {author} {\bibfnamefont {N.~H.}\ \bibnamefont
  {March}},\ }\href {\doibase 10.1142/p174} {\emph {\bibinfo {title} {Electron
  Correlation in the Solid State}}}\ (\bibinfo  {publisher} {Imperial College
  Press and Distributed by World Scientific Publishing Co},\ \bibinfo {year}
  {1999})\BibitemShut {NoStop}%
\bibitem [{\citenamefont {Rungger}\ \emph {et~al.}(2019)\citenamefont
  {Rungger}, \citenamefont {Fitzpatrick}, \citenamefont {Chen}, \citenamefont
  {Alderete}, \citenamefont {Apel}, \citenamefont {Cowtan}, \citenamefont
  {Patterson}, \citenamefont {Ramo}, \citenamefont {Zhu}, \citenamefont
  {Nguyen} \emph {et~al.}}]{rungger2019dynamical}%
  \BibitemOpen
  \bibfield  {author} {\bibinfo {author} {\bibfnamefont {I.}~\bibnamefont
  {Rungger}}, \bibinfo {author} {\bibfnamefont {N.}~\bibnamefont
  {Fitzpatrick}}, \bibinfo {author} {\bibfnamefont {H.}~\bibnamefont {Chen}},
  \bibinfo {author} {\bibfnamefont {C.}~\bibnamefont {Alderete}}, \bibinfo
  {author} {\bibfnamefont {H.}~\bibnamefont {Apel}}, \bibinfo {author}
  {\bibfnamefont {A.}~\bibnamefont {Cowtan}}, \bibinfo {author} {\bibfnamefont
  {A.}~\bibnamefont {Patterson}}, \bibinfo {author} {\bibfnamefont {D.~M.}\
  \bibnamefont {Ramo}}, \bibinfo {author} {\bibfnamefont {Y.}~\bibnamefont
  {Zhu}}, \bibinfo {author} {\bibfnamefont {N.~H.}\ \bibnamefont {Nguyen}},
  \emph {et~al.},\ }\href@noop {} {\bibfield  {journal} {\bibinfo  {journal}
  {arXiv preprint arXiv:1910.04735}\ } (\bibinfo {year} {2019})}\BibitemShut
  {NoStop}%
\bibitem [{\citenamefont {Cai}\ \emph {et~al.}(2020)\citenamefont {Cai},
  \citenamefont {Fang}, \citenamefont {Fan},\ and\ \citenamefont
  {Li}}]{cai2020quantum}%
  \BibitemOpen
  \bibfield  {author} {\bibinfo {author} {\bibfnamefont {X.}~\bibnamefont
  {Cai}}, \bibinfo {author} {\bibfnamefont {W.-H.}\ \bibnamefont {Fang}},
  \bibinfo {author} {\bibfnamefont {H.}~\bibnamefont {Fan}}, \ and\ \bibinfo
  {author} {\bibfnamefont {Z.}~\bibnamefont {Li}},\ }\href@noop {} {\bibfield
  {journal} {\bibinfo  {journal} {Physical Review Research}\ }\textbf {\bibinfo
  {volume} {2}},\ \bibinfo {pages} {033324} (\bibinfo {year}
  {2020})}\BibitemShut {NoStop}%
\bibitem [{\citenamefont {Endo}\ \emph {et~al.}(2020)\citenamefont {Endo},
  \citenamefont {Kurata},\ and\ \citenamefont
  {Nakagawa}}]{endo2020calculation}%
  \BibitemOpen
  \bibfield  {author} {\bibinfo {author} {\bibfnamefont {S.}~\bibnamefont
  {Endo}}, \bibinfo {author} {\bibfnamefont {I.}~\bibnamefont {Kurata}}, \ and\
  \bibinfo {author} {\bibfnamefont {Y.~O.}\ \bibnamefont {Nakagawa}},\
  }\href@noop {} {\bibfield  {journal} {\bibinfo  {journal} {Physical Review
  Research}\ }\textbf {\bibinfo {volume} {2}},\ \bibinfo {pages} {033281}
  (\bibinfo {year} {2020})}\BibitemShut {NoStop}%
\bibitem [{\citenamefont {Chen}\ \emph {et~al.}(2021)\citenamefont {Chen},
  \citenamefont {Nusspickel}, \citenamefont {Tilly}, \citenamefont {Booth}
  \emph {et~al.}}]{chen2021variational}%
  \BibitemOpen
  \bibfield  {author} {\bibinfo {author} {\bibfnamefont {H.}~\bibnamefont
  {Chen}}, \bibinfo {author} {\bibfnamefont {M.}~\bibnamefont {Nusspickel}},
  \bibinfo {author} {\bibfnamefont {J.}~\bibnamefont {Tilly}}, \bibinfo
  {author} {\bibfnamefont {G.~H.}\ \bibnamefont {Booth}},  \emph {et~al.},\
  }\href@noop {} {\bibfield  {journal} {\bibinfo  {journal} {Physical Review
  A}\ }\textbf {\bibinfo {volume} {104}},\ \bibinfo {pages} {032405} (\bibinfo
  {year} {2021})}\BibitemShut {NoStop}%
\bibitem [{\citenamefont {Jamet}\ \emph {et~al.}(2021)\citenamefont {Jamet},
  \citenamefont {Agarwal}, \citenamefont {Lupo}, \citenamefont {Browne},
  \citenamefont {Weber},\ and\ \citenamefont {Rungger}}]{jamet2021krylov}%
  \BibitemOpen
  \bibfield  {author} {\bibinfo {author} {\bibfnamefont {F.}~\bibnamefont
  {Jamet}}, \bibinfo {author} {\bibfnamefont {A.}~\bibnamefont {Agarwal}},
  \bibinfo {author} {\bibfnamefont {C.}~\bibnamefont {Lupo}}, \bibinfo {author}
  {\bibfnamefont {D.~E.}\ \bibnamefont {Browne}}, \bibinfo {author}
  {\bibfnamefont {C.}~\bibnamefont {Weber}}, \ and\ \bibinfo {author}
  {\bibfnamefont {I.}~\bibnamefont {Rungger}},\ }\href@noop {} {\bibfield
  {journal} {\bibinfo  {journal} {arXiv preprint arXiv:2105.13298}\ } (\bibinfo
  {year} {2021})}\BibitemShut {NoStop}%
\bibitem [{\citenamefont {Sakurai}\ \emph {et~al.}(2022)\citenamefont
  {Sakurai}, \citenamefont {Mizukami},\ and\ \citenamefont
  {Shinaoka}}]{sakurai2022hybrid}%
  \BibitemOpen
  \bibfield  {author} {\bibinfo {author} {\bibfnamefont {R.}~\bibnamefont
  {Sakurai}}, \bibinfo {author} {\bibfnamefont {W.}~\bibnamefont {Mizukami}}, \
  and\ \bibinfo {author} {\bibfnamefont {H.}~\bibnamefont {Shinaoka}},\
  }\href@noop {} {\bibfield  {journal} {\bibinfo  {journal} {Physical Review
  Research}\ }\textbf {\bibinfo {volume} {4}},\ \bibinfo {pages} {023219}
  (\bibinfo {year} {2022})}\BibitemShut {NoStop}%
\bibitem [{\citenamefont {Zhu}\ \emph {et~al.}(2022)\citenamefont {Zhu},
  \citenamefont {Nakagawa}, \citenamefont {Zhang}, \citenamefont {Li},\ and\
  \citenamefont {Guo}}]{zhu2022calculating}%
  \BibitemOpen
  \bibfield  {author} {\bibinfo {author} {\bibfnamefont {J.}~\bibnamefont
  {Zhu}}, \bibinfo {author} {\bibfnamefont {Y.~O.}\ \bibnamefont {Nakagawa}},
  \bibinfo {author} {\bibfnamefont {Y.-S.}\ \bibnamefont {Zhang}}, \bibinfo
  {author} {\bibfnamefont {C.-F.}\ \bibnamefont {Li}}, \ and\ \bibinfo {author}
  {\bibfnamefont {G.-C.}\ \bibnamefont {Guo}},\ }\href@noop {} {\bibfield
  {journal} {\bibinfo  {journal} {New Journal of Physics}\ }\textbf {\bibinfo
  {volume} {24}},\ \bibinfo {pages} {043030} (\bibinfo {year}
  {2022})}\BibitemShut {NoStop}%
\bibitem [{\citenamefont {Steckmann}\ \emph {et~al.}(2021)\citenamefont
  {Steckmann}, \citenamefont {Keen}, \citenamefont {Kemper}, \citenamefont
  {Dumitrescu},\ and\ \citenamefont {Wang}}]{steckmann2021simulating}%
  \BibitemOpen
  \bibfield  {author} {\bibinfo {author} {\bibfnamefont {T.}~\bibnamefont
  {Steckmann}}, \bibinfo {author} {\bibfnamefont {T.}~\bibnamefont {Keen}},
  \bibinfo {author} {\bibfnamefont {A.~F.}\ \bibnamefont {Kemper}}, \bibinfo
  {author} {\bibfnamefont {E.~F.}\ \bibnamefont {Dumitrescu}}, \ and\ \bibinfo
  {author} {\bibfnamefont {Y.}~\bibnamefont {Wang}},\ }\href@noop {} {\bibfield
   {journal} {\bibinfo  {journal} {arXiv preprint arXiv:2112.05688}\ }
  (\bibinfo {year} {2021})}\BibitemShut {NoStop}%
\bibitem [{\citenamefont {Bauer}\ \emph {et~al.}(2016)\citenamefont {Bauer},
  \citenamefont {Wecker}, \citenamefont {Millis}, \citenamefont {Hastings},\
  and\ \citenamefont {Troyer}}]{bauer2016hybrid}%
  \BibitemOpen
  \bibfield  {author} {\bibinfo {author} {\bibfnamefont {B.}~\bibnamefont
  {Bauer}}, \bibinfo {author} {\bibfnamefont {D.}~\bibnamefont {Wecker}},
  \bibinfo {author} {\bibfnamefont {A.~J.}\ \bibnamefont {Millis}}, \bibinfo
  {author} {\bibfnamefont {M.~B.}\ \bibnamefont {Hastings}}, \ and\ \bibinfo
  {author} {\bibfnamefont {M.}~\bibnamefont {Troyer}},\ }\href@noop {}
  {\bibfield  {journal} {\bibinfo  {journal} {Physical Review X}\ }\textbf
  {\bibinfo {volume} {6}},\ \bibinfo {pages} {031045} (\bibinfo {year}
  {2016})}\BibitemShut {NoStop}%
\bibitem [{\citenamefont {Kosugi}\ and\ \citenamefont
  {Matsushita}(2020)}]{kosugi2020construction}%
  \BibitemOpen
  \bibfield  {author} {\bibinfo {author} {\bibfnamefont {T.}~\bibnamefont
  {Kosugi}}\ and\ \bibinfo {author} {\bibfnamefont {Y.-i.}\ \bibnamefont
  {Matsushita}},\ }\href@noop {} {\bibfield  {journal} {\bibinfo  {journal}
  {Physical Review A}\ }\textbf {\bibinfo {volume} {101}},\ \bibinfo {pages}
  {012330} (\bibinfo {year} {2020})}\BibitemShut {NoStop}%
\bibitem [{\citenamefont {Tong}\ \emph {et~al.}(2021)\citenamefont {Tong},
  \citenamefont {An}, \citenamefont {Wiebe},\ and\ \citenamefont
  {Lin}}]{tong2021fast}%
  \BibitemOpen
  \bibfield  {author} {\bibinfo {author} {\bibfnamefont {Y.}~\bibnamefont
  {Tong}}, \bibinfo {author} {\bibfnamefont {D.}~\bibnamefont {An}}, \bibinfo
  {author} {\bibfnamefont {N.}~\bibnamefont {Wiebe}}, \ and\ \bibinfo {author}
  {\bibfnamefont {L.}~\bibnamefont {Lin}},\ }\href@noop {} {\bibfield
  {journal} {\bibinfo  {journal} {Physical Review A}\ }\textbf {\bibinfo
  {volume} {104}},\ \bibinfo {pages} {032422} (\bibinfo {year}
  {2021})}\BibitemShut {NoStop}%
\bibitem [{\citenamefont {Gily{\'e}n}\ \emph {et~al.}(2019)\citenamefont
  {Gily{\'e}n}, \citenamefont {Su}, \citenamefont {Low},\ and\ \citenamefont
  {Wiebe}}]{gilyen2019quantum}%
  \BibitemOpen
  \bibfield  {author} {\bibinfo {author} {\bibfnamefont {A.}~\bibnamefont
  {Gily{\'e}n}}, \bibinfo {author} {\bibfnamefont {Y.}~\bibnamefont {Su}},
  \bibinfo {author} {\bibfnamefont {G.~H.}\ \bibnamefont {Low}}, \ and\
  \bibinfo {author} {\bibfnamefont {N.}~\bibnamefont {Wiebe}},\ }in\ \href@noop
  {} {\emph {\bibinfo {booktitle} {Proceedings of the 51st Annual ACM SIGACT
  Symposium on Theory of Computing}}}\ (\bibinfo {year} {2019})\ pp.\ \bibinfo
  {pages} {193--204}\BibitemShut {NoStop}%
\bibitem [{\citenamefont {Anderson}(1961)}]{anderson1961localized}%
  \BibitemOpen
  \bibfield  {author} {\bibinfo {author} {\bibfnamefont {P.~W.}\ \bibnamefont
  {Anderson}},\ }\href@noop {} {\bibfield  {journal} {\bibinfo  {journal}
  {Physical Review}\ }\textbf {\bibinfo {volume} {124}},\ \bibinfo {pages} {41}
  (\bibinfo {year} {1961})}\BibitemShut {NoStop}%
\bibitem [{\citenamefont {Potthoff}(2001)}]{potthoff2001two}%
  \BibitemOpen
  \bibfield  {author} {\bibinfo {author} {\bibfnamefont {M.}~\bibnamefont
  {Potthoff}},\ }\href@noop {} {\bibfield  {journal} {\bibinfo  {journal}
  {Physical Review B}\ }\textbf {\bibinfo {volume} {64}},\ \bibinfo {pages}
  {165114} (\bibinfo {year} {2001})}\BibitemShut {NoStop}%
\bibitem [{\citenamefont {Lehmann}(1954)}]{lehmann1954eigenschaften}%
  \BibitemOpen
  \bibfield  {author} {\bibinfo {author} {\bibfnamefont {H.}~\bibnamefont
  {Lehmann}},\ }\href@noop {} {\bibfield  {journal} {\bibinfo  {journal} {Il
  Nuovo Cimento (1943-1954)}\ }\textbf {\bibinfo {volume} {11}},\ \bibinfo
  {pages} {342} (\bibinfo {year} {1954})}\BibitemShut {NoStop}%
\bibitem [{\citenamefont {Hjorth-Jensen}\ \emph {et~al.}(2017)\citenamefont
  {Hjorth-Jensen}, \citenamefont {Lombardo},\ and\ \citenamefont
  {Van~Kolck}}]{hjorth2017advanced}%
  \BibitemOpen
  \bibfield  {author} {\bibinfo {author} {\bibfnamefont {M.}~\bibnamefont
  {Hjorth-Jensen}}, \bibinfo {author} {\bibfnamefont {M.~P.}\ \bibnamefont
  {Lombardo}}, \ and\ \bibinfo {author} {\bibfnamefont {U.}~\bibnamefont
  {Van~Kolck}},\ }\href@noop {} {\bibfield  {journal} {\bibinfo  {journal}
  {Springer Lecture Notes in Physics}\ }\textbf {\bibinfo {volume} {936}}
  (\bibinfo {year} {2017})}\BibitemShut {NoStop}%
\bibitem [{\citenamefont {Kempe}\ \emph {et~al.}(2006)\citenamefont {Kempe},
  \citenamefont {Kitaev},\ and\ \citenamefont {Regev}}]{kempe2006complexity}%
  \BibitemOpen
  \bibfield  {author} {\bibinfo {author} {\bibfnamefont {J.}~\bibnamefont
  {Kempe}}, \bibinfo {author} {\bibfnamefont {A.}~\bibnamefont {Kitaev}}, \
  and\ \bibinfo {author} {\bibfnamefont {O.}~\bibnamefont {Regev}},\
  }\href@noop {} {\bibfield  {journal} {\bibinfo  {journal} {Siam journal on
  computing}\ }\textbf {\bibinfo {volume} {35}},\ \bibinfo {pages} {1070}
  (\bibinfo {year} {2006})}\BibitemShut {NoStop}%
\bibitem [{\citenamefont {Low}\ and\ \citenamefont
  {Chuang}(2019)}]{low2019hamiltonian}%
  \BibitemOpen
  \bibfield  {author} {\bibinfo {author} {\bibfnamefont {G.~H.}\ \bibnamefont
  {Low}}\ and\ \bibinfo {author} {\bibfnamefont {I.~L.}\ \bibnamefont
  {Chuang}},\ }\href@noop {} {\bibfield  {journal} {\bibinfo  {journal}
  {Quantum}\ }\textbf {\bibinfo {volume} {3}},\ \bibinfo {pages} {163}
  (\bibinfo {year} {2019})}\BibitemShut {NoStop}%
\bibitem [{\citenamefont {Martyn}\ \emph {et~al.}(2021)\citenamefont {Martyn},
  \citenamefont {Rossi}, \citenamefont {Tan},\ and\ \citenamefont
  {Chuang}}]{martyn2021grand}%
  \BibitemOpen
  \bibfield  {author} {\bibinfo {author} {\bibfnamefont {J.~M.}\ \bibnamefont
  {Martyn}}, \bibinfo {author} {\bibfnamefont {Z.~M.}\ \bibnamefont {Rossi}},
  \bibinfo {author} {\bibfnamefont {A.~K.}\ \bibnamefont {Tan}}, \ and\
  \bibinfo {author} {\bibfnamefont {I.~L.}\ \bibnamefont {Chuang}},\
  }\href@noop {} {\bibfield  {journal} {\bibinfo  {journal} {PRX Quantum}\
  }\textbf {\bibinfo {volume} {2}},\ \bibinfo {pages} {040203} (\bibinfo {year}
  {2021})}\BibitemShut {NoStop}%
\bibitem [{\citenamefont {Camps}\ \emph {et~al.}(2022)\citenamefont {Camps},
  \citenamefont {Lin}, \citenamefont {Van~Beeumen},\ and\ \citenamefont
  {Yang}}]{camps2022explicit}%
  \BibitemOpen
  \bibfield  {author} {\bibinfo {author} {\bibfnamefont {D.}~\bibnamefont
  {Camps}}, \bibinfo {author} {\bibfnamefont {L.}~\bibnamefont {Lin}}, \bibinfo
  {author} {\bibfnamefont {R.}~\bibnamefont {Van~Beeumen}}, \ and\ \bibinfo
  {author} {\bibfnamefont {C.}~\bibnamefont {Yang}},\ }\href@noop {} {\bibfield
   {journal} {\bibinfo  {journal} {arXiv preprint arXiv:2203.10236}\ }
  (\bibinfo {year} {2022})}\BibitemShut {NoStop}%
\bibitem [{\citenamefont {Childs}\ and\ \citenamefont
  {Wiebe}(2012)}]{childs2012hamiltonian}%
  \BibitemOpen
  \bibfield  {author} {\bibinfo {author} {\bibfnamefont {A.~M.}\ \bibnamefont
  {Childs}}\ and\ \bibinfo {author} {\bibfnamefont {N.}~\bibnamefont {Wiebe}},\
  }\href@noop {} {\bibfield  {journal} {\bibinfo  {journal} {arXiv preprint
  arXiv:1202.5822}\ } (\bibinfo {year} {2012})}\BibitemShut {NoStop}%
\bibitem [{\citenamefont {Berry}\ \emph {et~al.}(2015)\citenamefont {Berry},
  \citenamefont {Childs}, \citenamefont {Cleve}, \citenamefont {Kothari},\ and\
  \citenamefont {Somma}}]{berry2015simulating}%
  \BibitemOpen
  \bibfield  {author} {\bibinfo {author} {\bibfnamefont {D.~W.}\ \bibnamefont
  {Berry}}, \bibinfo {author} {\bibfnamefont {A.~M.}\ \bibnamefont {Childs}},
  \bibinfo {author} {\bibfnamefont {R.}~\bibnamefont {Cleve}}, \bibinfo
  {author} {\bibfnamefont {R.}~\bibnamefont {Kothari}}, \ and\ \bibinfo
  {author} {\bibfnamefont {R.~D.}\ \bibnamefont {Somma}},\ }\href@noop {}
  {\bibfield  {journal} {\bibinfo  {journal} {Physical Review Letters}\
  }\textbf {\bibinfo {volume} {114}},\ \bibinfo {pages} {090502} (\bibinfo
  {year} {2015})}\BibitemShut {NoStop}%
\bibitem [{\citenamefont {Ralli}\ \emph {et~al.}(2021)\citenamefont {Ralli},
  \citenamefont {Love}, \citenamefont {Tranter},\ and\ \citenamefont
  {Coveney}}]{ralli2021implementation}%
  \BibitemOpen
  \bibfield  {author} {\bibinfo {author} {\bibfnamefont {A.}~\bibnamefont
  {Ralli}}, \bibinfo {author} {\bibfnamefont {P.~J.}\ \bibnamefont {Love}},
  \bibinfo {author} {\bibfnamefont {A.}~\bibnamefont {Tranter}}, \ and\
  \bibinfo {author} {\bibfnamefont {P.~V.}\ \bibnamefont {Coveney}},\
  }\href@noop {} {\bibfield  {journal} {\bibinfo  {journal} {Physical Review
  Research}\ }\textbf {\bibinfo {volume} {3}},\ \bibinfo {pages} {033195}
  (\bibinfo {year} {2021})}\BibitemShut {NoStop}%
\bibitem [{\citenamefont {Long}\ and\ \citenamefont {Sun}(2001)}]{Long2001}%
  \BibitemOpen
  \bibfield  {author} {\bibinfo {author} {\bibfnamefont {G.-L.}\ \bibnamefont
  {Long}}\ and\ \bibinfo {author} {\bibfnamefont {Y.}~\bibnamefont {Sun}},\
  }\href {\doibase 10.1103/PhysRevA.64.014303} {\bibfield  {journal} {\bibinfo
  {journal} {Physical Review A}\ }\textbf {\bibinfo {volume} {64}},\ \bibinfo
  {pages} {014303} (\bibinfo {year} {2001})},\ \Eprint
  {http://arxiv.org/abs/0104030} {arXiv:0104030 [quant-ph]} \BibitemShut
  {NoStop}%
\bibitem [{\citenamefont {Mottonen}\ \emph {et~al.}(2004)\citenamefont
  {Mottonen}, \citenamefont {Vartiainen}, \citenamefont {Bergholm},\ and\
  \citenamefont {Salomaa}}]{Mottonen2005}%
  \BibitemOpen
  \bibfield  {author} {\bibinfo {author} {\bibfnamefont {M.}~\bibnamefont
  {Mottonen}}, \bibinfo {author} {\bibfnamefont {J.~J.}\ \bibnamefont
  {Vartiainen}}, \bibinfo {author} {\bibfnamefont {V.}~\bibnamefont
  {Bergholm}}, \ and\ \bibinfo {author} {\bibfnamefont {M.~M.}\ \bibnamefont
  {Salomaa}},\ }\href {http://arxiv.org/abs/quant-ph/0407010} {\bibfield
  {journal} {\bibinfo  {journal} {Quantum Information and Computation}\
  }\textbf {\bibinfo {volume} {5}},\ \bibinfo {pages} {467} (\bibinfo {year}
  {2004})},\ \Eprint {http://arxiv.org/abs/0407010} {arXiv:0407010 [quant-ph]}
  \BibitemShut {NoStop}%
\bibitem [{\citenamefont {Shende}\ \emph {et~al.}(2006)\citenamefont {Shende},
  \citenamefont {Bullock},\ and\ \citenamefont {Markov}}]{Markov2006prep}%
  \BibitemOpen
  \bibfield  {author} {\bibinfo {author} {\bibfnamefont {V.}~\bibnamefont
  {Shende}}, \bibinfo {author} {\bibfnamefont {S.}~\bibnamefont {Bullock}}, \
  and\ \bibinfo {author} {\bibfnamefont {I.}~\bibnamefont {Markov}},\ }\href
  {\doibase 10.1109/TCAD.2005.855930} {\bibfield  {journal} {\bibinfo
  {journal} {IEEE Transactions on Computer-Aided Design of Integrated Circuits
  and Systems}\ }\textbf {\bibinfo {volume} {25}},\ \bibinfo {pages} {1000}
  (\bibinfo {year} {2006})}\BibitemShut {NoStop}%
\bibitem [{\citenamefont {Araujo}\ \emph {et~al.}(2021)\citenamefont {Araujo},
  \citenamefont {Park}, \citenamefont {Petruccione},\ and\ \citenamefont
  {da~Silva}}]{Araujo2021}%
  \BibitemOpen
  \bibfield  {author} {\bibinfo {author} {\bibfnamefont {I.~F.}\ \bibnamefont
  {Araujo}}, \bibinfo {author} {\bibfnamefont {D.~K.}\ \bibnamefont {Park}},
  \bibinfo {author} {\bibfnamefont {F.}~\bibnamefont {Petruccione}}, \ and\
  \bibinfo {author} {\bibfnamefont {A.~J.}\ \bibnamefont {da~Silva}},\ }\href
  {\doibase 10.1038/s41598-021-85474-1} {\bibfield  {journal} {\bibinfo
  {journal} {Scientific Reports}\ }\textbf {\bibinfo {volume} {11}},\ \bibinfo
  {pages} {6329} (\bibinfo {year} {2021})},\ \Eprint
  {http://arxiv.org/abs/2008.01511} {arXiv:2008.01511} \BibitemShut {NoStop}%
\bibitem [{\citenamefont {Low}\ and\ \citenamefont
  {Chuang}(2017)}]{low2017optimal}%
  \BibitemOpen
  \bibfield  {author} {\bibinfo {author} {\bibfnamefont {G.~H.}\ \bibnamefont
  {Low}}\ and\ \bibinfo {author} {\bibfnamefont {I.~L.}\ \bibnamefont
  {Chuang}},\ }\href@noop {} {\bibfield  {journal} {\bibinfo  {journal}
  {Physical Review Letters}\ }\textbf {\bibinfo {volume} {118}},\ \bibinfo
  {pages} {010501} (\bibinfo {year} {2017})}\BibitemShut {NoStop}%
\bibitem [{\citenamefont {Grover}(1998)}]{grover1998quantum}%
  \BibitemOpen
  \bibfield  {author} {\bibinfo {author} {\bibfnamefont {L.~K.}\ \bibnamefont
  {Grover}},\ }\href@noop {} {\bibfield  {journal} {\bibinfo  {journal}
  {Physical Review Letters}\ }\textbf {\bibinfo {volume} {80}},\ \bibinfo
  {pages} {4329} (\bibinfo {year} {1998})}\BibitemShut {NoStop}%
\bibitem [{\citenamefont {Brassard}\ \emph {et~al.}(2002)\citenamefont
  {Brassard}, \citenamefont {Hoyer}, \citenamefont {Mosca},\ and\ \citenamefont
  {Tapp}}]{brassard2002quantum}%
  \BibitemOpen
  \bibfield  {author} {\bibinfo {author} {\bibfnamefont {G.}~\bibnamefont
  {Brassard}}, \bibinfo {author} {\bibfnamefont {P.}~\bibnamefont {Hoyer}},
  \bibinfo {author} {\bibfnamefont {M.}~\bibnamefont {Mosca}}, \ and\ \bibinfo
  {author} {\bibfnamefont {A.}~\bibnamefont {Tapp}},\ }\href@noop {} {\bibfield
   {journal} {\bibinfo  {journal} {Contemporary Mathematics}\ }\textbf
  {\bibinfo {volume} {305}},\ \bibinfo {pages} {53} (\bibinfo {year}
  {2002})}\BibitemShut {NoStop}%
\bibitem [{\citenamefont {Yoder}\ \emph {et~al.}(2014)\citenamefont {Yoder},
  \citenamefont {Low},\ and\ \citenamefont {Chuang}}]{yoder2014fixed}%
  \BibitemOpen
  \bibfield  {author} {\bibinfo {author} {\bibfnamefont {T.~J.}\ \bibnamefont
  {Yoder}}, \bibinfo {author} {\bibfnamefont {G.~H.}\ \bibnamefont {Low}}, \
  and\ \bibinfo {author} {\bibfnamefont {I.~L.}\ \bibnamefont {Chuang}},\
  }\href@noop {} {\bibfield  {journal} {\bibinfo  {journal} {Physical Review
  Letters}\ }\textbf {\bibinfo {volume} {113}},\ \bibinfo {pages} {210501}
  (\bibinfo {year} {2014})}\BibitemShut {NoStop}%
\bibitem [{\citenamefont {Berry}\ \emph {et~al.}(2014)\citenamefont {Berry},
  \citenamefont {Childs}, \citenamefont {Cleve}, \citenamefont {Kothari},\ and\
  \citenamefont {Somma}}]{berry2014exponential}%
  \BibitemOpen
  \bibfield  {author} {\bibinfo {author} {\bibfnamefont {D.~W.}\ \bibnamefont
  {Berry}}, \bibinfo {author} {\bibfnamefont {A.~M.}\ \bibnamefont {Childs}},
  \bibinfo {author} {\bibfnamefont {R.}~\bibnamefont {Cleve}}, \bibinfo
  {author} {\bibfnamefont {R.}~\bibnamefont {Kothari}}, \ and\ \bibinfo
  {author} {\bibfnamefont {R.~D.}\ \bibnamefont {Somma}},\ }in\ \href@noop {}
  {\emph {\bibinfo {booktitle} {Proceedings of the forty-sixth annual ACM
  symposium on Theory of computing}}}\ (\bibinfo {year} {2014})\ pp.\ \bibinfo
  {pages} {283--292}\BibitemShut {NoStop}%
\bibitem [{\citenamefont {Yan}\ \emph {et~al.}(2022)\citenamefont {Yan},
  \citenamefont {Wei}, \citenamefont {Jiang}, \citenamefont {Wang},
  \citenamefont {Duan}, \citenamefont {Ma},\ and\ \citenamefont
  {Long}}]{yan2022fixed}%
  \BibitemOpen
  \bibfield  {author} {\bibinfo {author} {\bibfnamefont {B.}~\bibnamefont
  {Yan}}, \bibinfo {author} {\bibfnamefont {S.}~\bibnamefont {Wei}}, \bibinfo
  {author} {\bibfnamefont {H.}~\bibnamefont {Jiang}}, \bibinfo {author}
  {\bibfnamefont {H.}~\bibnamefont {Wang}}, \bibinfo {author} {\bibfnamefont
  {Q.}~\bibnamefont {Duan}}, \bibinfo {author} {\bibfnamefont {Z.}~\bibnamefont
  {Ma}}, \ and\ \bibinfo {author} {\bibfnamefont {G.-L.}\ \bibnamefont
  {Long}},\ }\href@noop {} {\bibfield  {journal} {\bibinfo  {journal}
  {Scientific Reports}\ }\textbf {\bibinfo {volume} {12}},\ \bibinfo {pages}
  {1} (\bibinfo {year} {2022})}\BibitemShut {NoStop}%
\bibitem [{\citenamefont {Hastings}\ \emph {et~al.}(2014)\citenamefont
  {Hastings}, \citenamefont {Wecker}, \citenamefont {Bauer},\ and\
  \citenamefont {Troyer}}]{hastings2014improving}%
  \BibitemOpen
  \bibfield  {author} {\bibinfo {author} {\bibfnamefont {M.~B.}\ \bibnamefont
  {Hastings}}, \bibinfo {author} {\bibfnamefont {D.}~\bibnamefont {Wecker}},
  \bibinfo {author} {\bibfnamefont {B.}~\bibnamefont {Bauer}}, \ and\ \bibinfo
  {author} {\bibfnamefont {M.}~\bibnamefont {Troyer}},\ }\href@noop {}
  {\bibfield  {journal} {\bibinfo  {journal} {arXiv preprint arXiv:1403.1539}\
  } (\bibinfo {year} {2014})}\BibitemShut {NoStop}%
\bibitem [{\citenamefont {Cowtan}\ \emph {et~al.}(2019)\citenamefont {Cowtan},
  \citenamefont {Dilkes}, \citenamefont {Duncan}, \citenamefont {Simmons},\
  and\ \citenamefont {Sivarajah}}]{cowtan2019phase}%
  \BibitemOpen
  \bibfield  {author} {\bibinfo {author} {\bibfnamefont {A.}~\bibnamefont
  {Cowtan}}, \bibinfo {author} {\bibfnamefont {S.}~\bibnamefont {Dilkes}},
  \bibinfo {author} {\bibfnamefont {R.}~\bibnamefont {Duncan}}, \bibinfo
  {author} {\bibfnamefont {W.}~\bibnamefont {Simmons}}, \ and\ \bibinfo
  {author} {\bibfnamefont {S.}~\bibnamefont {Sivarajah}},\ }\href@noop {}
  {\bibfield  {journal} {\bibinfo  {journal} {arXiv preprint arXiv:1906.01734}\
  } (\bibinfo {year} {2019})}\BibitemShut {NoStop}%
\bibitem [{\citenamefont {da~Silva}\ and\ \citenamefont
  {Park}(2022)}]{da2022linear}%
  \BibitemOpen
  \bibfield  {author} {\bibinfo {author} {\bibfnamefont {A.~J.}\ \bibnamefont
  {da~Silva}}\ and\ \bibinfo {author} {\bibfnamefont {D.~K.}\ \bibnamefont
  {Park}},\ }\href@noop {} {\bibfield  {journal} {\bibinfo  {journal} {arXiv
  preprint arXiv:2203.11882}\ } (\bibinfo {year} {2022})}\BibitemShut {NoStop}%
\bibitem [{\citenamefont {Bullock}\ and\ \citenamefont
  {Markov}(2003)}]{bullock2003smaller}%
  \BibitemOpen
  \bibfield  {author} {\bibinfo {author} {\bibfnamefont {S.~S.}\ \bibnamefont
  {Bullock}}\ and\ \bibinfo {author} {\bibfnamefont {I.~L.}\ \bibnamefont
  {Markov}},\ }\href@noop {} {\bibfield  {journal} {\bibinfo  {journal} {arXiv
  preprint quant-ph/0303039}\ } (\bibinfo {year} {2003})}\BibitemShut {NoStop}%
\bibitem [{\citenamefont {Shende}\ \emph {et~al.}(2005)\citenamefont {Shende},
  \citenamefont {Bullock},\ and\ \citenamefont {Markov}}]{shende2005synthesis}%
  \BibitemOpen
  \bibfield  {author} {\bibinfo {author} {\bibfnamefont {V.~V.}\ \bibnamefont
  {Shende}}, \bibinfo {author} {\bibfnamefont {S.~S.}\ \bibnamefont {Bullock}},
  \ and\ \bibinfo {author} {\bibfnamefont {I.~L.}\ \bibnamefont {Markov}},\
  }in\ \href@noop {} {\emph {\bibinfo {booktitle} {Proceedings of the 2005 Asia
  and South Pacific Design Automation Conference}}}\ (\bibinfo {year} {2005})\
  pp.\ \bibinfo {pages} {272--275}\BibitemShut {NoStop}%
\bibitem [{\citenamefont {von Burg}\ \emph {et~al.}(2021)\citenamefont {von
  Burg}, \citenamefont {Low}, \citenamefont {H{\"a}ner}, \citenamefont
  {Steiger}, \citenamefont {Reiher}, \citenamefont {Roetteler},\ and\
  \citenamefont {Troyer}}]{von2021quantum}%
  \BibitemOpen
  \bibfield  {author} {\bibinfo {author} {\bibfnamefont {V.}~\bibnamefont {von
  Burg}}, \bibinfo {author} {\bibfnamefont {G.~H.}\ \bibnamefont {Low}},
  \bibinfo {author} {\bibfnamefont {T.}~\bibnamefont {H{\"a}ner}}, \bibinfo
  {author} {\bibfnamefont {D.~S.}\ \bibnamefont {Steiger}}, \bibinfo {author}
  {\bibfnamefont {M.}~\bibnamefont {Reiher}}, \bibinfo {author} {\bibfnamefont
  {M.}~\bibnamefont {Roetteler}}, \ and\ \bibinfo {author} {\bibfnamefont
  {M.}~\bibnamefont {Troyer}},\ }\href@noop {} {\bibfield  {journal} {\bibinfo
  {journal} {Physical Review Research}\ }\textbf {\bibinfo {volume} {3}},\
  \bibinfo {pages} {033055} (\bibinfo {year} {2021})}\BibitemShut {NoStop}%
\bibitem [{\citenamefont {Dong}\ \emph
  {et~al.}(2022{\natexlab{a}})\citenamefont {Dong}, \citenamefont {Lin},\ and\
  \citenamefont {Tong}}]{dong2022ground}%
  \BibitemOpen
  \bibfield  {author} {\bibinfo {author} {\bibfnamefont {Y.}~\bibnamefont
  {Dong}}, \bibinfo {author} {\bibfnamefont {L.}~\bibnamefont {Lin}}, \ and\
  \bibinfo {author} {\bibfnamefont {Y.}~\bibnamefont {Tong}},\ }\href@noop {}
  {\bibfield  {journal} {\bibinfo  {journal} {PRX Quantum}\ }\textbf {\bibinfo
  {volume} {3}},\ \bibinfo {pages} {040305} (\bibinfo {year}
  {2022}{\natexlab{a}})}\BibitemShut {NoStop}%
\bibitem [{\citenamefont {Chakraborty}\ \emph {et~al.}(2018)\citenamefont
  {Chakraborty}, \citenamefont {Gily{\'e}n},\ and\ \citenamefont
  {Jeffery}}]{chakraborty2018power}%
  \BibitemOpen
  \bibfield  {author} {\bibinfo {author} {\bibfnamefont {S.}~\bibnamefont
  {Chakraborty}}, \bibinfo {author} {\bibfnamefont {A.}~\bibnamefont
  {Gily{\'e}n}}, \ and\ \bibinfo {author} {\bibfnamefont {S.}~\bibnamefont
  {Jeffery}},\ }\href@noop {} {\bibfield  {journal} {\bibinfo  {journal} {arXiv
  preprint arXiv:1804.01973}\ } (\bibinfo {year} {2018})}\BibitemShut {NoStop}%
\bibitem [{\citenamefont {Harrow}\ \emph {et~al.}(2009)\citenamefont {Harrow},
  \citenamefont {Hassidim},\ and\ \citenamefont {Lloyd}}]{harrow2009quantum}%
  \BibitemOpen
  \bibfield  {author} {\bibinfo {author} {\bibfnamefont {A.~W.}\ \bibnamefont
  {Harrow}}, \bibinfo {author} {\bibfnamefont {A.}~\bibnamefont {Hassidim}}, \
  and\ \bibinfo {author} {\bibfnamefont {S.}~\bibnamefont {Lloyd}},\
  }\href@noop {} {\bibfield  {journal} {\bibinfo  {journal} {Physical Review
  Letters}\ }\textbf {\bibinfo {volume} {103}},\ \bibinfo {pages} {150502}
  (\bibinfo {year} {2009})}\BibitemShut {NoStop}%
\bibitem [{\citenamefont {Lin}(2022)}]{lin2022lecture}%
  \BibitemOpen
  \bibfield  {author} {\bibinfo {author} {\bibfnamefont {L.}~\bibnamefont
  {Lin}},\ }\href@noop {} {\bibfield  {journal} {\bibinfo  {journal} {arXiv
  preprint arXiv:2201.08309}\ } (\bibinfo {year} {2022})}\BibitemShut {NoStop}%
\bibitem [{\citenamefont {Lin}\ \emph {et~al.}(2021)\citenamefont {Lin},
  \citenamefont {Dilip}, \citenamefont {Green}, \citenamefont {Smith},\ and\
  \citenamefont {Pollmann}}]{lin2021real}%
  \BibitemOpen
  \bibfield  {author} {\bibinfo {author} {\bibfnamefont {S.-H.}\ \bibnamefont
  {Lin}}, \bibinfo {author} {\bibfnamefont {R.}~\bibnamefont {Dilip}}, \bibinfo
  {author} {\bibfnamefont {A.~G.}\ \bibnamefont {Green}}, \bibinfo {author}
  {\bibfnamefont {A.}~\bibnamefont {Smith}}, \ and\ \bibinfo {author}
  {\bibfnamefont {F.}~\bibnamefont {Pollmann}},\ }\href@noop {} {\bibfield
  {journal} {\bibinfo  {journal} {PRX Quantum}\ }\textbf {\bibinfo {volume}
  {2}},\ \bibinfo {pages} {010342} (\bibinfo {year} {2021})}\BibitemShut
  {NoStop}%
\bibitem [{\citenamefont {Low}\ \emph {et~al.}(2016)\citenamefont {Low},
  \citenamefont {Yoder},\ and\ \citenamefont {Chuang}}]{low2016methodology}%
  \BibitemOpen
  \bibfield  {author} {\bibinfo {author} {\bibfnamefont {G.~H.}\ \bibnamefont
  {Low}}, \bibinfo {author} {\bibfnamefont {T.~J.}\ \bibnamefont {Yoder}}, \
  and\ \bibinfo {author} {\bibfnamefont {I.~L.}\ \bibnamefont {Chuang}},\
  }\href@noop {} {\bibfield  {journal} {\bibinfo  {journal} {Physical Review
  X}\ }\textbf {\bibinfo {volume} {6}},\ \bibinfo {pages} {041067} (\bibinfo
  {year} {2016})}\BibitemShut {NoStop}%
\bibitem [{\citenamefont {Haah}(2019)}]{haah2019product}%
  \BibitemOpen
  \bibfield  {author} {\bibinfo {author} {\bibfnamefont {J.}~\bibnamefont
  {Haah}},\ }\href@noop {} {\bibfield  {journal} {\bibinfo  {journal}
  {Quantum}\ }\textbf {\bibinfo {volume} {3}},\ \bibinfo {pages} {190}
  (\bibinfo {year} {2019})}\BibitemShut {NoStop}%
\bibitem [{\citenamefont {Chao}\ \emph {et~al.}(2020)\citenamefont {Chao},
  \citenamefont {Ding}, \citenamefont {Gilyen}, \citenamefont {Huang},\ and\
  \citenamefont {Szegedy}}]{chao2020finding}%
  \BibitemOpen
  \bibfield  {author} {\bibinfo {author} {\bibfnamefont {R.}~\bibnamefont
  {Chao}}, \bibinfo {author} {\bibfnamefont {D.}~\bibnamefont {Ding}}, \bibinfo
  {author} {\bibfnamefont {A.}~\bibnamefont {Gilyen}}, \bibinfo {author}
  {\bibfnamefont {C.}~\bibnamefont {Huang}}, \ and\ \bibinfo {author}
  {\bibfnamefont {M.}~\bibnamefont {Szegedy}},\ }\href@noop {} {\bibfield
  {journal} {\bibinfo  {journal} {arXiv preprint arXiv:2003.02831}\ } (\bibinfo
  {year} {2020})}\BibitemShut {NoStop}%
\bibitem [{\citenamefont {Dong}\ \emph {et~al.}(2021)\citenamefont {Dong},
  \citenamefont {Meng}, \citenamefont {Whaley},\ and\ \citenamefont
  {Lin}}]{dong2021efficient}%
  \BibitemOpen
  \bibfield  {author} {\bibinfo {author} {\bibfnamefont {Y.}~\bibnamefont
  {Dong}}, \bibinfo {author} {\bibfnamefont {X.}~\bibnamefont {Meng}}, \bibinfo
  {author} {\bibfnamefont {K.~B.}\ \bibnamefont {Whaley}}, \ and\ \bibinfo
  {author} {\bibfnamefont {L.}~\bibnamefont {Lin}},\ }\href@noop {} {\bibfield
  {journal} {\bibinfo  {journal} {Physical Review A}\ }\textbf {\bibinfo
  {volume} {103}},\ \bibinfo {pages} {042419} (\bibinfo {year}
  {2021})}\BibitemShut {NoStop}%
\bibitem [{\citenamefont {Martyn}\ \emph {et~al.}(2022)\citenamefont {Martyn},
  \citenamefont {Tan}, \citenamefont {Huang},\ and\ \citenamefont
  {chuang}}]{pyqsp}%
  \BibitemOpen
  \bibfield  {author} {\bibinfo {author} {\bibfnamefont {J.}~\bibnamefont
  {Martyn}}, \bibinfo {author} {\bibfnamefont {A.}~\bibnamefont {Tan}},
  \bibinfo {author} {\bibfnamefont {C.}~\bibnamefont {Huang}}, \ and\ \bibinfo
  {author} {\bibfnamefont {I.}~\bibnamefont {chuang}},\ }\href@noop {}
  {\enquote {\bibinfo {title} {pyqsp},}\ }\bibinfo {howpublished}
  {\url{https://github.com/ichuang/pyqsp}} (\bibinfo {year} {2022})\BibitemShut
  {NoStop}%
\bibitem [{\citenamefont {Dong}\ \emph
  {et~al.}(2022{\natexlab{b}})\citenamefont {Dong}, \citenamefont {Meng},
  \citenamefont {Wang},\ and\ \citenamefont {Lin}}]{QSPPACK}%
  \BibitemOpen
  \bibfield  {author} {\bibinfo {author} {\bibfnamefont {Y.}~\bibnamefont
  {Dong}}, \bibinfo {author} {\bibfnamefont {X.}~\bibnamefont {Meng}}, \bibinfo
  {author} {\bibfnamefont {J.}~\bibnamefont {Wang}}, \ and\ \bibinfo {author}
  {\bibfnamefont {l.}~\bibnamefont {Lin}},\ }\href@noop {} {\enquote {\bibinfo
  {title} {{QSPPACK}},}\ }\bibinfo {howpublished}
  {\url{https://github.com/qsppack/QSPPACK}} (\bibinfo {year}
  {2022}{\natexlab{b}})\BibitemShut {NoStop}%
\bibitem [{\citenamefont {Gily{\'e}n}\ \emph {et~al.}(2018)\citenamefont
  {Gily{\'e}n}, \citenamefont {Su}, \citenamefont {Low},\ and\ \citenamefont
  {Wiebe}}]{gilyen2018quantum}%
  \BibitemOpen
  \bibfield  {author} {\bibinfo {author} {\bibfnamefont {A.}~\bibnamefont
  {Gily{\'e}n}}, \bibinfo {author} {\bibfnamefont {Y.}~\bibnamefont {Su}},
  \bibinfo {author} {\bibfnamefont {G.~H.}\ \bibnamefont {Low}}, \ and\
  \bibinfo {author} {\bibfnamefont {N.}~\bibnamefont {Wiebe}},\ }\href@noop {}
  {\bibfield  {journal} {\bibinfo  {journal} {arXiv preprint arXiv:1806.01838}\
  } (\bibinfo {year} {2018})}\BibitemShut {NoStop}%
\bibitem [{\citenamefont {Jordan}(1875)}]{jordan1875essai}%
  \BibitemOpen
  \bibfield  {author} {\bibinfo {author} {\bibfnamefont {C.}~\bibnamefont
  {Jordan}},\ }\href@noop {} {\bibfield  {journal} {\bibinfo  {journal}
  {Bulletin de la Soci{\'e}t{\'e} math{\'e}matique de France}\ }\textbf
  {\bibinfo {volume} {3}},\ \bibinfo {pages} {103} (\bibinfo {year}
  {1875})}\BibitemShut {NoStop}%
\bibitem [{\citenamefont {Kotliar}\ and\ \citenamefont
  {Vollhardt}(2004)}]{kotliar2004strongly}%
  \BibitemOpen
  \bibfield  {author} {\bibinfo {author} {\bibfnamefont {G.}~\bibnamefont
  {Kotliar}}\ and\ \bibinfo {author} {\bibfnamefont {D.}~\bibnamefont
  {Vollhardt}},\ }\href@noop {} {\bibfield  {journal} {\bibinfo  {journal}
  {Physics today}\ }\textbf {\bibinfo {volume} {57}},\ \bibinfo {pages} {53}
  (\bibinfo {year} {2004})}\BibitemShut {NoStop}%
\bibitem [{\citenamefont {Georges}\ \emph
  {et~al.}(1996{\natexlab{a}})\citenamefont {Georges}, \citenamefont {Kotliar},
  \citenamefont {Krauth},\ and\ \citenamefont
  {Rozenberg}}]{georges1996dynamical}%
  \BibitemOpen
  \bibfield  {author} {\bibinfo {author} {\bibfnamefont {A.}~\bibnamefont
  {Georges}}, \bibinfo {author} {\bibfnamefont {G.}~\bibnamefont {Kotliar}},
  \bibinfo {author} {\bibfnamefont {W.}~\bibnamefont {Krauth}}, \ and\ \bibinfo
  {author} {\bibfnamefont {M.~J.}\ \bibnamefont {Rozenberg}},\ }\href@noop {}
  {\bibfield  {journal} {\bibinfo  {journal} {Reviews of Modern Physics}\
  }\textbf {\bibinfo {volume} {68}},\ \bibinfo {pages} {13} (\bibinfo {year}
  {1996}{\natexlab{a}})}\BibitemShut {NoStop}%
\bibitem [{\citenamefont {Caffarel}\ and\ \citenamefont
  {Krauth}(1994)}]{caffarel94}%
  \BibitemOpen
  \bibfield  {author} {\bibinfo {author} {\bibfnamefont {M.}~\bibnamefont
  {Caffarel}}\ and\ \bibinfo {author} {\bibfnamefont {W.}~\bibnamefont
  {Krauth}},\ }\href@noop {} {\bibfield  {journal} {\bibinfo  {journal} {Phys.
  Rev. Lett.}\ }\textbf {\bibinfo {volume} {72}},\ \bibinfo {pages} {1545}
  (\bibinfo {year} {1994})}\BibitemShut {NoStop}%
\bibitem [{\citenamefont {Georges}\ \emph
  {et~al.}(1996{\natexlab{b}})\citenamefont {Georges}, \citenamefont {Kotliar},
  \citenamefont {Krauth},\ and\ \citenamefont {Rozenberg}}]{georges96}%
  \BibitemOpen
  \bibfield  {author} {\bibinfo {author} {\bibfnamefont {A.}~\bibnamefont
  {Georges}}, \bibinfo {author} {\bibfnamefont {G.}~\bibnamefont {Kotliar}},
  \bibinfo {author} {\bibfnamefont {W.}~\bibnamefont {Krauth}}, \ and\ \bibinfo
  {author} {\bibfnamefont {M.~J.}\ \bibnamefont {Rozenberg}},\ }\href@noop {}
  {\bibfield  {journal} {\bibinfo  {journal} {Rev. Mod. Phys.}\ }\textbf
  {\bibinfo {volume} {68}},\ \bibinfo {pages} {13} (\bibinfo {year}
  {1996}{\natexlab{b}})}\BibitemShut {NoStop}%
\bibitem [{\citenamefont {Kreula}\ \emph {et~al.}(2016)\citenamefont {Kreula},
  \citenamefont {Garc{\'\i}a-{\'A}lvarez}, \citenamefont {Lamata},
  \citenamefont {Clark}, \citenamefont {Solano},\ and\ \citenamefont
  {Jaksch}}]{kreula2016few}%
  \BibitemOpen
  \bibfield  {author} {\bibinfo {author} {\bibfnamefont {J.~M.}\ \bibnamefont
  {Kreula}}, \bibinfo {author} {\bibfnamefont {L.}~\bibnamefont
  {Garc{\'\i}a-{\'A}lvarez}}, \bibinfo {author} {\bibfnamefont
  {L.}~\bibnamefont {Lamata}}, \bibinfo {author} {\bibfnamefont {S.~R.}\
  \bibnamefont {Clark}}, \bibinfo {author} {\bibfnamefont {E.}~\bibnamefont
  {Solano}}, \ and\ \bibinfo {author} {\bibfnamefont {D.}~\bibnamefont
  {Jaksch}},\ }\href@noop {} {\bibfield  {journal} {\bibinfo  {journal} {EPJ
  Quantum Technology}\ }\textbf {\bibinfo {volume} {3}},\ \bibinfo {pages} {1}
  (\bibinfo {year} {2016})}\BibitemShut {NoStop}%
\bibitem [{\citenamefont {Jordan}\ and\ \citenamefont
  {Wigner}(1928)}]{Jordan1928}%
  \BibitemOpen
  \bibfield  {author} {\bibinfo {author} {\bibfnamefont {P.}~\bibnamefont
  {Jordan}}\ and\ \bibinfo {author} {\bibfnamefont {E.}~\bibnamefont
  {Wigner}},\ }\href {\doibase 10.1007/BF01331938} {\bibfield  {journal}
  {\bibinfo  {journal} {Zeitschrift fur Physik}\ }\textbf {\bibinfo {volume}
  {47}},\ \bibinfo {pages} {631} (\bibinfo {year} {1928})}\BibitemShut
  {NoStop}%
\bibitem [{\citenamefont {Tranter}\ \emph {et~al.}(2022)\citenamefont
  {Tranter}, \citenamefont {Paola}, \citenamefont {Ramo}, \citenamefont {David
  Zsolt~Manrique}, \citenamefont {Greene-Diniz}, \citenamefont {Christopoulou},
  \citenamefont {Polyak}, \citenamefont {Irfan~Khan}, \citenamefont {Kirsopp},
  \citenamefont {Yamamoto}, \citenamefont {Tudorovskaya}, \citenamefont
  {Krompiec},\ and\ \citenamefont {Fitzpatrick}}]{inquanto}%
  \BibitemOpen
  \bibfield  {author} {\bibinfo {author} {\bibfnamefont {A.}~\bibnamefont
  {Tranter}}, \bibinfo {author} {\bibfnamefont {C.~D.}\ \bibnamefont {Paola}},
  \bibinfo {author} {\bibfnamefont {D.~M.}\ \bibnamefont {Ramo}}, \bibinfo
  {author} {\bibfnamefont {D.~G.}\ \bibnamefont {David Zsolt~Manrique}},
  \bibinfo {author} {\bibfnamefont {G.}~\bibnamefont {Greene-Diniz}}, \bibinfo
  {author} {\bibfnamefont {G.}~\bibnamefont {Christopoulou}}, \bibinfo {author}
  {\bibfnamefont {I.}~\bibnamefont {Polyak}}, \bibinfo {author} {\bibfnamefont
  {J.~P.}\ \bibnamefont {Irfan~Khan}}, \bibinfo {author} {\bibfnamefont
  {J.}~\bibnamefont {Kirsopp}}, \bibinfo {author} {\bibfnamefont
  {K.}~\bibnamefont {Yamamoto}}, \bibinfo {author} {\bibfnamefont
  {M.}~\bibnamefont {Tudorovskaya}}, \bibinfo {author} {\bibfnamefont
  {M.}~\bibnamefont {Krompiec}}, \ and\ \bibinfo {author} {\bibfnamefont
  {N.}~\bibnamefont {Fitzpatrick}},\ }\href@noop {} {\enquote {\bibinfo {title}
  {Introduction to the inquanto computational chemistry platform for quantum
  computers},}\ }\bibinfo {howpublished}
  {\url{https://medium.com/cambridge-quantum-computing/4fced08d66cc}} (\bibinfo
  {year} {2022})\BibitemShut {NoStop}%
\bibitem [{\citenamefont {Quantinuum}(2022)}]{inquanto_prod}%
  \BibitemOpen
  \bibfield  {author} {\bibinfo {author} {\bibnamefont {Quantinuum}},\
  }\href@noop {} {\enquote {\bibinfo {title} {Inquanto},}\ }\bibinfo
  {howpublished} {\url{https://www.quantinuum.com/products/inquanto}} (\bibinfo
  {year} {2022})\BibitemShut {NoStop}%
\bibitem [{\citenamefont {Sivarajah}\ \emph {et~al.}(2020)\citenamefont
  {Sivarajah}, \citenamefont {Dilkes}, \citenamefont {Cowtan}, \citenamefont
  {Simmons}, \citenamefont {Edgington},\ and\ \citenamefont
  {Duncan}}]{sivarajah2020t}%
  \BibitemOpen
  \bibfield  {author} {\bibinfo {author} {\bibfnamefont {S.}~\bibnamefont
  {Sivarajah}}, \bibinfo {author} {\bibfnamefont {S.}~\bibnamefont {Dilkes}},
  \bibinfo {author} {\bibfnamefont {A.}~\bibnamefont {Cowtan}}, \bibinfo
  {author} {\bibfnamefont {W.}~\bibnamefont {Simmons}}, \bibinfo {author}
  {\bibfnamefont {A.}~\bibnamefont {Edgington}}, \ and\ \bibinfo {author}
  {\bibfnamefont {R.}~\bibnamefont {Duncan}},\ }\href@noop {} {\bibfield
  {journal} {\bibinfo  {journal} {Quantum Science and Technology}\ }\textbf
  {\bibinfo {volume} {6}},\ \bibinfo {pages} {014003} (\bibinfo {year}
  {2020})}\BibitemShut {NoStop}%
\bibitem [{\citenamefont {Harris}\ \emph {et~al.}(2020)\citenamefont {Harris},
  \citenamefont {Millman}, \citenamefont {van~der Walt}, \citenamefont
  {Gommers}, \citenamefont {Virtanen}, \citenamefont {Cournapeau},
  \citenamefont {Wieser}, \citenamefont {Taylor}, \citenamefont {Berg},
  \citenamefont {Smith}, \citenamefont {Kern}, \citenamefont {Picus},
  \citenamefont {Hoyer}, \citenamefont {van Kerkwijk}, \citenamefont {Brett},
  \citenamefont {Haldane}, \citenamefont {del R{\'{i}}o}, \citenamefont
  {Wiebe}, \citenamefont {Peterson}, \citenamefont {G{\'{e}}rard-Marchant},
  \citenamefont {Sheppard}, \citenamefont {Reddy}, \citenamefont {Weckesser},
  \citenamefont {Abbasi}, \citenamefont {Gohlke},\ and\ \citenamefont
  {Oliphant}}]{Numpy}%
  \BibitemOpen
  \bibfield  {author} {\bibinfo {author} {\bibfnamefont {C.~R.}\ \bibnamefont
  {Harris}}, \bibinfo {author} {\bibfnamefont {K.~J.}\ \bibnamefont {Millman}},
  \bibinfo {author} {\bibfnamefont {S.~J.}\ \bibnamefont {van~der Walt}},
  \bibinfo {author} {\bibfnamefont {R.}~\bibnamefont {Gommers}}, \bibinfo
  {author} {\bibfnamefont {P.}~\bibnamefont {Virtanen}}, \bibinfo {author}
  {\bibfnamefont {D.}~\bibnamefont {Cournapeau}}, \bibinfo {author}
  {\bibfnamefont {E.}~\bibnamefont {Wieser}}, \bibinfo {author} {\bibfnamefont
  {J.}~\bibnamefont {Taylor}}, \bibinfo {author} {\bibfnamefont
  {S.}~\bibnamefont {Berg}}, \bibinfo {author} {\bibfnamefont {N.~J.}\
  \bibnamefont {Smith}}, \bibinfo {author} {\bibfnamefont {R.}~\bibnamefont
  {Kern}}, \bibinfo {author} {\bibfnamefont {M.}~\bibnamefont {Picus}},
  \bibinfo {author} {\bibfnamefont {S.}~\bibnamefont {Hoyer}}, \bibinfo
  {author} {\bibfnamefont {M.~H.}\ \bibnamefont {van Kerkwijk}}, \bibinfo
  {author} {\bibfnamefont {M.}~\bibnamefont {Brett}}, \bibinfo {author}
  {\bibfnamefont {A.}~\bibnamefont {Haldane}}, \bibinfo {author} {\bibfnamefont
  {J.~F.}\ \bibnamefont {del R{\'{i}}o}}, \bibinfo {author} {\bibfnamefont
  {M.}~\bibnamefont {Wiebe}}, \bibinfo {author} {\bibfnamefont
  {P.}~\bibnamefont {Peterson}}, \bibinfo {author} {\bibfnamefont
  {P.}~\bibnamefont {G{\'{e}}rard-Marchant}}, \bibinfo {author} {\bibfnamefont
  {K.}~\bibnamefont {Sheppard}}, \bibinfo {author} {\bibfnamefont
  {T.}~\bibnamefont {Reddy}}, \bibinfo {author} {\bibfnamefont
  {W.}~\bibnamefont {Weckesser}}, \bibinfo {author} {\bibfnamefont
  {H.}~\bibnamefont {Abbasi}}, \bibinfo {author} {\bibfnamefont
  {C.}~\bibnamefont {Gohlke}}, \ and\ \bibinfo {author} {\bibfnamefont {T.~E.}\
  \bibnamefont {Oliphant}},\ }\href {\doibase 10.1038/s41586-020-2649-2}
  {\bibfield  {journal} {\bibinfo  {journal} {Nature}\ }\textbf {\bibinfo
  {volume} {585}},\ \bibinfo {pages} {357} (\bibinfo {year}
  {2020})}\BibitemShut {NoStop}%
\bibitem [{\citenamefont {Virtanen}\ \emph {et~al.}(2020)\citenamefont
  {Virtanen}, \citenamefont {Gommers}, \citenamefont {Oliphant}, \citenamefont
  {Haberland}, \citenamefont {Reddy}, \citenamefont {Cournapeau}, \citenamefont
  {Burovski}, \citenamefont {Peterson}, \citenamefont {Weckesser},
  \citenamefont {Bright}, \citenamefont {{van der Walt}}, \citenamefont
  {Brett}, \citenamefont {Wilson}, \citenamefont {Millman}, \citenamefont
  {Mayorov}, \citenamefont {Nelson}, \citenamefont {Jones}, \citenamefont
  {Kern}, \citenamefont {Larson}, \citenamefont {Carey}, \citenamefont {Polat},
  \citenamefont {Feng}, \citenamefont {Moore}, \citenamefont {{VanderPlas}},
  \citenamefont {Laxalde}, \citenamefont {Perktold}, \citenamefont {Cimrman},
  \citenamefont {Henriksen}, \citenamefont {Quintero}, \citenamefont {Harris},
  \citenamefont {Archibald}, \citenamefont {Ribeiro}, \citenamefont
  {Pedregosa}, \citenamefont {{van Mulbregt}},\ and\ \citenamefont {{SciPy 1.0
  Contributors}}}]{SciPy}%
  \BibitemOpen
  \bibfield  {author} {\bibinfo {author} {\bibfnamefont {P.}~\bibnamefont
  {Virtanen}}, \bibinfo {author} {\bibfnamefont {R.}~\bibnamefont {Gommers}},
  \bibinfo {author} {\bibfnamefont {T.~E.}\ \bibnamefont {Oliphant}}, \bibinfo
  {author} {\bibfnamefont {M.}~\bibnamefont {Haberland}}, \bibinfo {author}
  {\bibfnamefont {T.}~\bibnamefont {Reddy}}, \bibinfo {author} {\bibfnamefont
  {D.}~\bibnamefont {Cournapeau}}, \bibinfo {author} {\bibfnamefont
  {E.}~\bibnamefont {Burovski}}, \bibinfo {author} {\bibfnamefont
  {P.}~\bibnamefont {Peterson}}, \bibinfo {author} {\bibfnamefont
  {W.}~\bibnamefont {Weckesser}}, \bibinfo {author} {\bibfnamefont
  {J.}~\bibnamefont {Bright}}, \bibinfo {author} {\bibfnamefont {S.~J.}\
  \bibnamefont {{van der Walt}}}, \bibinfo {author} {\bibfnamefont
  {M.}~\bibnamefont {Brett}}, \bibinfo {author} {\bibfnamefont
  {J.}~\bibnamefont {Wilson}}, \bibinfo {author} {\bibfnamefont {K.~J.}\
  \bibnamefont {Millman}}, \bibinfo {author} {\bibfnamefont {N.}~\bibnamefont
  {Mayorov}}, \bibinfo {author} {\bibfnamefont {A.~R.~J.}\ \bibnamefont
  {Nelson}}, \bibinfo {author} {\bibfnamefont {E.}~\bibnamefont {Jones}},
  \bibinfo {author} {\bibfnamefont {R.}~\bibnamefont {Kern}}, \bibinfo {author}
  {\bibfnamefont {E.}~\bibnamefont {Larson}}, \bibinfo {author} {\bibfnamefont
  {C.~J.}\ \bibnamefont {Carey}}, \bibinfo {author} {\bibfnamefont
  {{\.I}.}~\bibnamefont {Polat}}, \bibinfo {author} {\bibfnamefont
  {Y.}~\bibnamefont {Feng}}, \bibinfo {author} {\bibfnamefont {E.~W.}\
  \bibnamefont {Moore}}, \bibinfo {author} {\bibfnamefont {J.}~\bibnamefont
  {{VanderPlas}}}, \bibinfo {author} {\bibfnamefont {D.}~\bibnamefont
  {Laxalde}}, \bibinfo {author} {\bibfnamefont {J.}~\bibnamefont {Perktold}},
  \bibinfo {author} {\bibfnamefont {R.}~\bibnamefont {Cimrman}}, \bibinfo
  {author} {\bibfnamefont {I.}~\bibnamefont {Henriksen}}, \bibinfo {author}
  {\bibfnamefont {E.~A.}\ \bibnamefont {Quintero}}, \bibinfo {author}
  {\bibfnamefont {C.~R.}\ \bibnamefont {Harris}}, \bibinfo {author}
  {\bibfnamefont {A.~M.}\ \bibnamefont {Archibald}}, \bibinfo {author}
  {\bibfnamefont {A.~H.}\ \bibnamefont {Ribeiro}}, \bibinfo {author}
  {\bibfnamefont {F.}~\bibnamefont {Pedregosa}}, \bibinfo {author}
  {\bibfnamefont {P.}~\bibnamefont {{van Mulbregt}}}, \ and\ \bibinfo {author}
  {\bibnamefont {{SciPy 1.0 Contributors}}},\ }\href {\doibase
  10.1038/s41592-019-0686-2} {\bibfield  {journal} {\bibinfo  {journal} {Nature
  Methods}\ }\textbf {\bibinfo {volume} {17}},\ \bibinfo {pages} {261}
  (\bibinfo {year} {2020})}\BibitemShut {NoStop}%
\bibitem [{\citenamefont {Hong}\ and\ \citenamefont
  {Pan}(1992)}]{hong1992lower}%
  \BibitemOpen
  \bibfield  {author} {\bibinfo {author} {\bibfnamefont {Y.}~\bibnamefont
  {Hong}}\ and\ \bibinfo {author} {\bibfnamefont {C.-T.}\ \bibnamefont {Pan}},\
  }\href@noop {} {\bibfield  {journal} {\bibinfo  {journal} {Linear Algebra and
  its Applications}\ }\textbf {\bibinfo {volume} {172}},\ \bibinfo {pages} {27}
  (\bibinfo {year} {1992})}\BibitemShut {NoStop}%
\bibitem [{\citenamefont {Piazza}\ and\ \citenamefont
  {Politi}(2002)}]{piazza2002upper}%
  \BibitemOpen
  \bibfield  {author} {\bibinfo {author} {\bibfnamefont {G.}~\bibnamefont
  {Piazza}}\ and\ \bibinfo {author} {\bibfnamefont {T.}~\bibnamefont
  {Politi}},\ }\href@noop {} {\bibfield  {journal} {\bibinfo  {journal}
  {Journal of Computational and Applied Mathematics}\ }\textbf {\bibinfo
  {volume} {143}},\ \bibinfo {pages} {141} (\bibinfo {year}
  {2002})}\BibitemShut {NoStop}%
\bibitem [{\citenamefont {Huang}(2008)}]{huang2008estimation}%
  \BibitemOpen
  \bibfield  {author} {\bibinfo {author} {\bibfnamefont {T.-Z.}\ \bibnamefont
  {Huang}},\ }\href@noop {} {\bibfield  {journal} {\bibinfo  {journal}
  {Computers \& Mathematics with Applications}\ }\textbf {\bibinfo {volume}
  {55}},\ \bibinfo {pages} {1075} (\bibinfo {year} {2008})}\BibitemShut
  {NoStop}%
\bibitem [{\citenamefont {Zou}\ and\ \citenamefont
  {Jiang}(2010)}]{zou2010estimation}%
  \BibitemOpen
  \bibfield  {author} {\bibinfo {author} {\bibfnamefont {L.}~\bibnamefont
  {Zou}}\ and\ \bibinfo {author} {\bibfnamefont {Y.}~\bibnamefont {Jiang}},\
  }\href@noop {} {\bibfield  {journal} {\bibinfo  {journal} {Linear algebra and
  its applications}\ }\textbf {\bibinfo {volume} {433}},\ \bibinfo {pages}
  {1203} (\bibinfo {year} {2010})}\BibitemShut {NoStop}%
\bibitem [{\citenamefont {Zou}(2012)}]{zou2012lower}%
  \BibitemOpen
  \bibfield  {author} {\bibinfo {author} {\bibfnamefont {L.}~\bibnamefont
  {Zou}},\ }\href@noop {} {\bibfield  {journal} {\bibinfo  {journal} {J. Math.
  Inequal}\ }\textbf {\bibinfo {volume} {6}},\ \bibinfo {pages} {625} (\bibinfo
  {year} {2012})}\BibitemShut {NoStop}%
\bibitem [{\citenamefont {Bai}\ \emph {et~al.}(1996)\citenamefont {Bai},
  \citenamefont {Fahey},\ and\ \citenamefont {Golub}}]{bai1996some}%
  \BibitemOpen
  \bibfield  {author} {\bibinfo {author} {\bibfnamefont {Z.}~\bibnamefont
  {Bai}}, \bibinfo {author} {\bibfnamefont {G.}~\bibnamefont {Fahey}}, \ and\
  \bibinfo {author} {\bibfnamefont {G.}~\bibnamefont {Golub}},\ }\href@noop {}
  {\bibfield  {journal} {\bibinfo  {journal} {Journal of Computational and
  Applied Mathematics}\ }\textbf {\bibinfo {volume} {74}},\ \bibinfo {pages}
  {71} (\bibinfo {year} {1996})}\BibitemShut {NoStop}%
\bibitem [{\citenamefont {Ipsen}\ and\ \citenamefont
  {Lee}(2011)}]{ipsen2011determinant}%
  \BibitemOpen
  \bibfield  {author} {\bibinfo {author} {\bibfnamefont {I.~C.}\ \bibnamefont
  {Ipsen}}\ and\ \bibinfo {author} {\bibfnamefont {D.~J.}\ \bibnamefont
  {Lee}},\ }\href@noop {} {\bibfield  {journal} {\bibinfo  {journal} {arXiv
  preprint arXiv:1105.0437}\ } (\bibinfo {year} {2011})}\BibitemShut {NoStop}%
\bibitem [{\citenamefont {Aryasetiawan}\ and\ \citenamefont
  {Gunnarsson}(1998)}]{aryasetiawan1998gw}%
  \BibitemOpen
  \bibfield  {author} {\bibinfo {author} {\bibfnamefont {F.}~\bibnamefont
  {Aryasetiawan}}\ and\ \bibinfo {author} {\bibfnamefont {O.}~\bibnamefont
  {Gunnarsson}},\ }\href@noop {} {\bibfield  {journal} {\bibinfo  {journal}
  {Reports on Progress in Physics}\ }\textbf {\bibinfo {volume} {61}},\
  \bibinfo {pages} {237} (\bibinfo {year} {1998})}\BibitemShut {NoStop}%
\bibitem [{\citenamefont {Eckstein}\ \emph {et~al.}(2005)\citenamefont
  {Eckstein}, \citenamefont {Kollar}, \citenamefont {Byczuk},\ and\
  \citenamefont {Vollhardt}}]{eckstein2005hopping}%
  \BibitemOpen
  \bibfield  {author} {\bibinfo {author} {\bibfnamefont {M.}~\bibnamefont
  {Eckstein}}, \bibinfo {author} {\bibfnamefont {M.}~\bibnamefont {Kollar}},
  \bibinfo {author} {\bibfnamefont {K.}~\bibnamefont {Byczuk}}, \ and\ \bibinfo
  {author} {\bibfnamefont {D.}~\bibnamefont {Vollhardt}},\ }\href@noop {}
  {\bibfield  {journal} {\bibinfo  {journal} {Physical Review B}\ }\textbf
  {\bibinfo {volume} {71}},\ \bibinfo {pages} {235119} (\bibinfo {year}
  {2005})}\BibitemShut {NoStop}%
\bibitem [{\citenamefont {Zhang}\ \emph {et~al.}(1993)\citenamefont {Zhang},
  \citenamefont {Rozenberg},\ and\ \citenamefont {Kotliar}}]{zhang1993mott}%
  \BibitemOpen
  \bibfield  {author} {\bibinfo {author} {\bibfnamefont {X.}~\bibnamefont
  {Zhang}}, \bibinfo {author} {\bibfnamefont {M.}~\bibnamefont {Rozenberg}}, \
  and\ \bibinfo {author} {\bibfnamefont {G.}~\bibnamefont {Kotliar}},\
  }\href@noop {} {\bibfield  {journal} {\bibinfo  {journal} {Physical Review
  Letters}\ }\textbf {\bibinfo {volume} {70}},\ \bibinfo {pages} {1666}
  (\bibinfo {year} {1993})}\BibitemShut {NoStop}%
\bibitem [{\citenamefont {Sriluckshmy}\ \emph {et~al.}(2021)\citenamefont
  {Sriluckshmy}, \citenamefont {Nusspickel}, \citenamefont {Fertitta},\ and\
  \citenamefont {Booth}}]{sriluckshmy2021fully}%
  \BibitemOpen
  \bibfield  {author} {\bibinfo {author} {\bibfnamefont {P.}~\bibnamefont
  {Sriluckshmy}}, \bibinfo {author} {\bibfnamefont {M.}~\bibnamefont
  {Nusspickel}}, \bibinfo {author} {\bibfnamefont {E.}~\bibnamefont
  {Fertitta}}, \ and\ \bibinfo {author} {\bibfnamefont {G.~H.}\ \bibnamefont
  {Booth}},\ }\href@noop {} {\bibfield  {journal} {\bibinfo  {journal}
  {Physical Review B}\ }\textbf {\bibinfo {volume} {103}},\ \bibinfo {pages}
  {085131} (\bibinfo {year} {2021})}\BibitemShut {NoStop}%
\bibitem [{\citenamefont {Low}\ \emph {et~al.}(2018)\citenamefont {Low},
  \citenamefont {Kliuchnikov},\ and\ \citenamefont
  {Schaeffer}}]{low2018trading}%
  \BibitemOpen
  \bibfield  {author} {\bibinfo {author} {\bibfnamefont {G.~H.}\ \bibnamefont
  {Low}}, \bibinfo {author} {\bibfnamefont {V.}~\bibnamefont {Kliuchnikov}}, \
  and\ \bibinfo {author} {\bibfnamefont {L.}~\bibnamefont {Schaeffer}},\
  }\href@noop {} {\bibfield  {journal} {\bibinfo  {journal} {arXiv preprint
  arXiv:1812.00954}\ } (\bibinfo {year} {2018})}\BibitemShut {NoStop}%
\bibitem [{\citenamefont {Wan}(2021)}]{wan2021exponentially}%
  \BibitemOpen
  \bibfield  {author} {\bibinfo {author} {\bibfnamefont {K.}~\bibnamefont
  {Wan}},\ }\href@noop {} {\bibfield  {journal} {\bibinfo  {journal} {Quantum}\
  }\textbf {\bibinfo {volume} {5}},\ \bibinfo {pages} {380} (\bibinfo {year}
  {2021})}\BibitemShut {NoStop}%
\bibitem [{\citenamefont {Camps}\ and\ \citenamefont
  {Van~Beeumen}(2022)}]{camps2022fable}%
  \BibitemOpen
  \bibfield  {author} {\bibinfo {author} {\bibfnamefont {D.}~\bibnamefont
  {Camps}}\ and\ \bibinfo {author} {\bibfnamefont {R.}~\bibnamefont
  {Van~Beeumen}},\ }\href@noop {} {\bibfield  {journal} {\bibinfo  {journal}
  {arXiv preprint arXiv:2205.00081}\ } (\bibinfo {year} {2022})}\BibitemShut
  {NoStop}%
\end{thebibliography}%

\onecolumngrid
\appendix
\addcontentsline{toc}{section}{Appendices}
\clearpage
\section*{Appendices}
\section{Spectral function plots} \label{sec:A_omega_plots}

\subsection{$k=10$ results}
\begin{figure}[h!]
    \centering
    \includegraphics[width=0.75\textwidth]{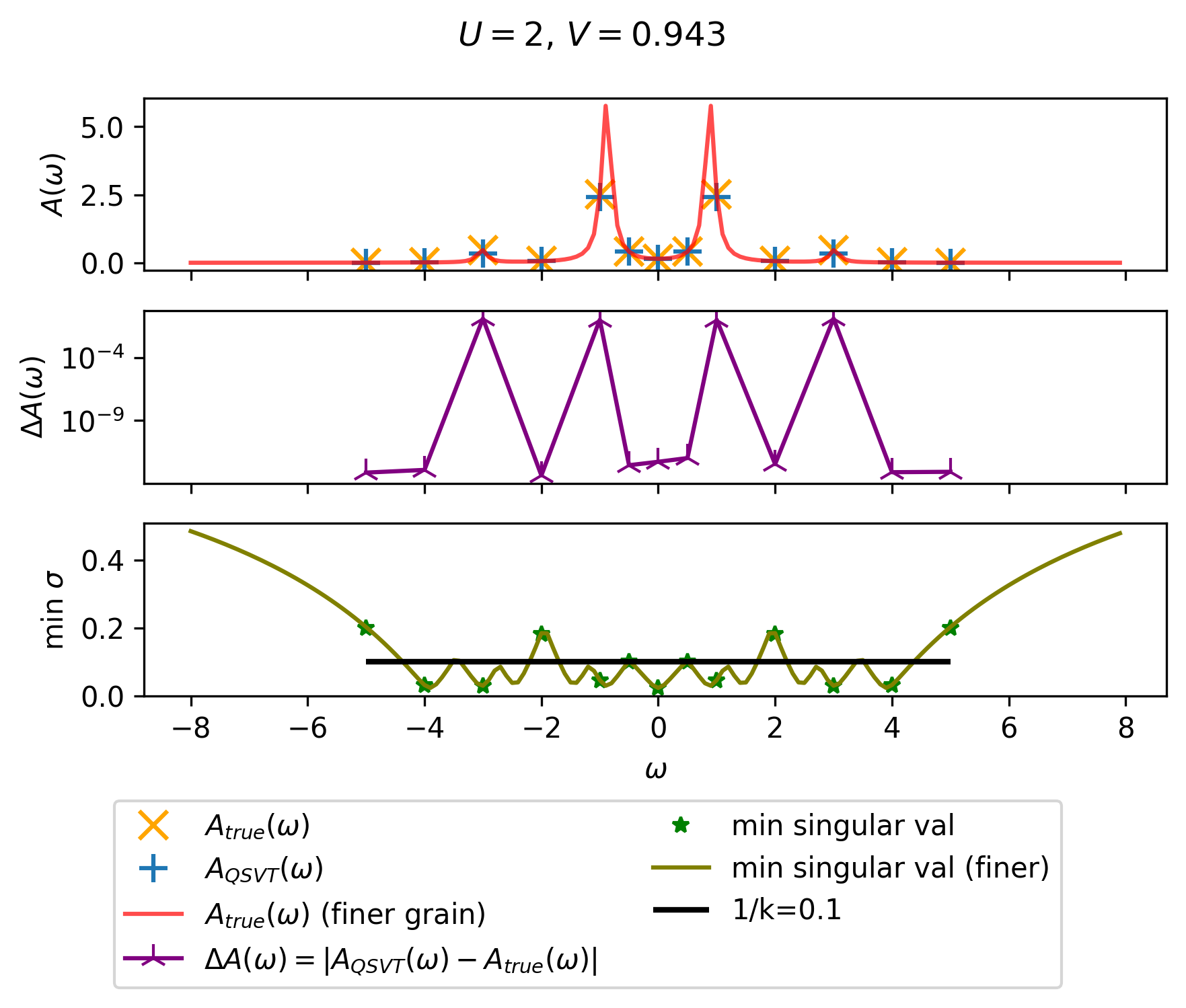}
    \caption{(top) Plot of spectral function  (equation \ref{eq:spectral_F}) of single particle Anderson model for $U=2$ and $V=0.943$. The red line and orange points shows the spectral function function being calculated via exact diagonalization. The blue points show the spectral function calculated via the quantum singular value transform with $k=10$. (middle) Plot shows absolute error of spectral function $\Delta A(\omega) = |A_{QSVT}(\omega) - A_{true}(\omega) |$. (bottom) Plot shows the minimum singular value of the matrix to undergo inversion via the QSVT. Any singular below the black line is outside the region of where the approximation of $1/x$ is well defined.}
        \label{fig:Green_U2_K10}
\end{figure}

\begin{figure}[h!]
    \centering
    \includegraphics[width=0.75\textwidth]{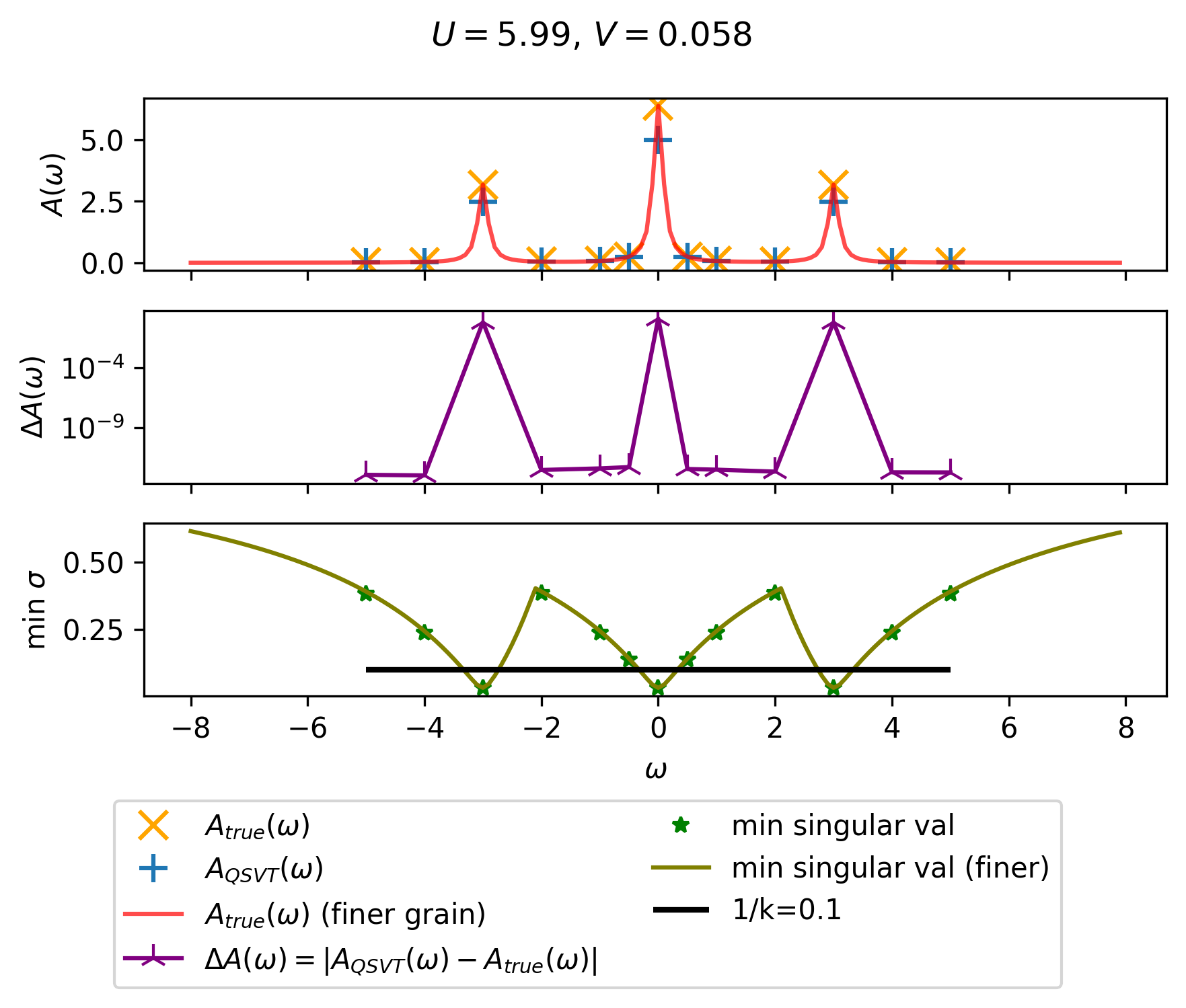}
    \caption{(top) Plot of spectral function  (equation \ref{eq:spectral_F}) of single-particle Anderson model for $U=5.99$ and $V=0.058$. The red line and orange points shows the spectral function function being calculated via exact diagonalization. The blue points show the spectral function calculated via the quantum singular value transform with $k=10$. (middle) Plot shows absolute error of spectral function $\Delta A(\omega) = |A_{QSVT}(\omega) - A_{true}(\omega) |$. (bottom) Plot shows the minimum singular value of the matrix to undergo inversion via the QSVT. Any singular below the black line is outside the region of where the approximation of $1/x$ is well defined.}
        \label{fig:Green_U5_K10}
\end{figure}

\begin{figure}[t]
    \centering
    \includegraphics[width=0.75\textwidth]{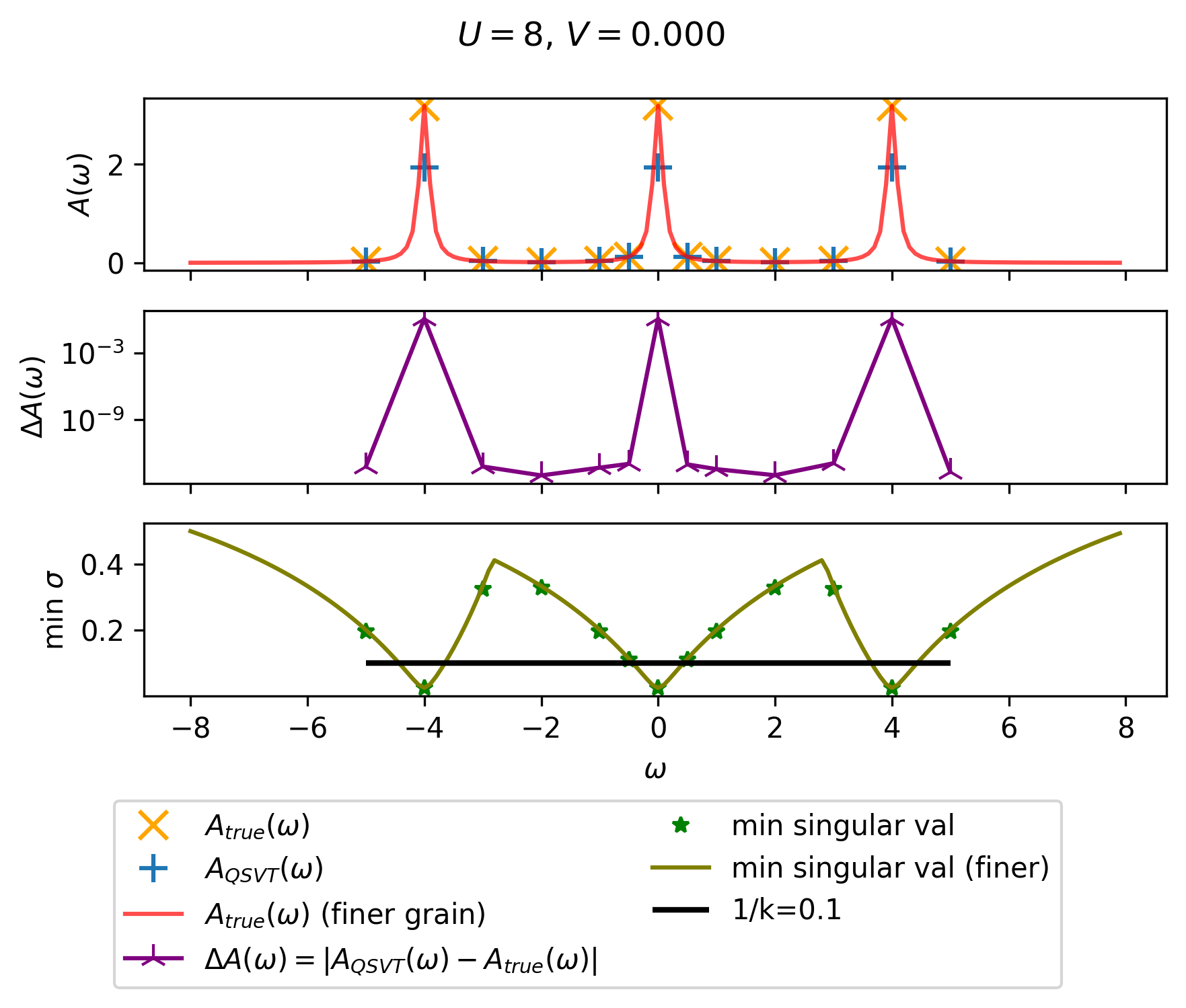}
    \caption{(top) Plot of spectral function  (equation \ref{eq:spectral_F}) of single-particle Anderson model for $U=8$ and $V=0$. The red line and orange points shows the spectral function function being calculated via exact diagonalization. The blue points show the spectral function calculated via the quantum singular value transform with $k=10$. (middle) Plot shows absolute error of spectral function $\Delta A(\omega) = |A_{QSVT}(\omega) - A_{true}(\omega) |$. (bottom) Plot shows the minimum singular value of the matrix to undergo inversion via the QSVT. Any singular below the black line is outside the region of where the approximation of $1/x$ is well defined.}
        \label{fig:Green_U8_K10}
\end{figure}

Figures \ref{fig:Green_U2_K10}, \ref{fig:Green_U5_K10} and \ref{fig:Green_U8_K10} give the spectral plots for the two-site single particle Anderson model at different $U, V$ values.

% \begin{figure*}[t]
%     \centering
%     \includegraphics[width=0.75\textwidth]{results/full_G_K10/Anderson_U4_V0_K10.png}
%     \caption{Spectral function (equation \ref{eq:spectral_F}) of single-particle Anderson model for $U=4$ and $V=0.745$. The top plot shows the spectral function being calculated via exact diagonalization (red and orange) and via quantum signal processing (blue) with $k=10$. The middle plot shows in the absolute error $\Delta G(\omega) = |G_{exp}(\omega) - G_{true}(\omega) |$. The bottom plot shows the minimum singular value of the matrix to undergo inversion via the QSVT. Any singular below the black line is outside the region of where the approximation of $1/x$ is defined.}
%         \label{fig:Green_U4_K10}
% \end{figure*}

\clearpage
\subsection{$k=50$ results}
\begin{figure}[h!]
    \centering
    \includegraphics[width=0.75\textwidth]{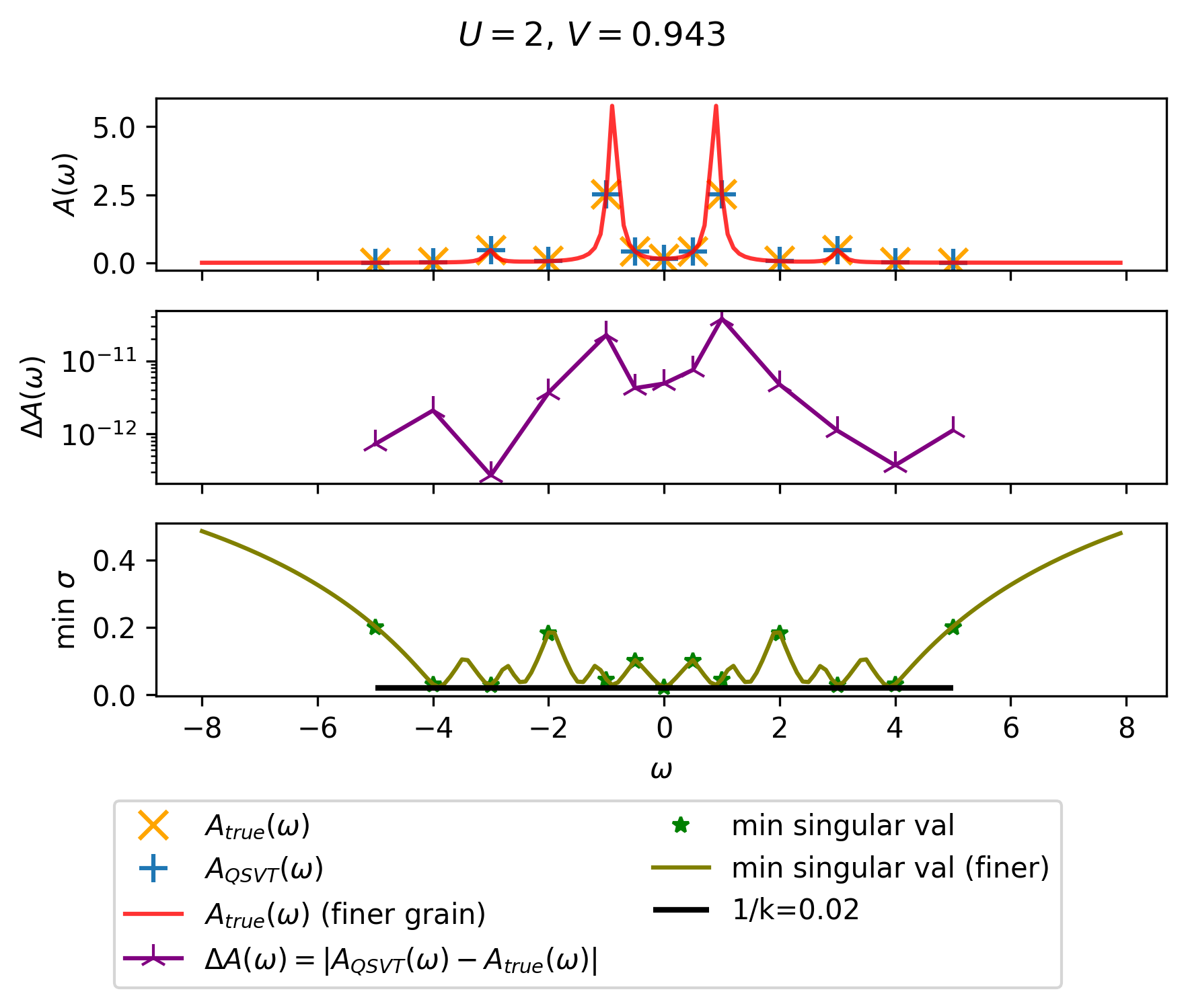}
    \caption{(top) Plot of spectral function  (equation \ref{eq:spectral_F}) of single-particle Anderson model for $U=2$ and $V=0.943$. The red line and orange points shows the spectral function function being calculated via exact diagonalization. The blue points show the spectral function calculated via the quantum singular value transform with $k=50$. (middle) Plot shows absolute error of spectral function $\Delta A(\omega) = |A_{QSVT}(\omega) - A_{true}(\omega) |$. (bottom) Plot shows the minimum singular value of the matrix to undergo inversion via the QSVT. Any singular below the black line is outside the region of where the approximation of $1/x$ is well defined.}
        \label{fig:Green_U2_K50}
\end{figure}

\begin{figure}[h!]
    \centering
    \includegraphics[width=0.75\textwidth]{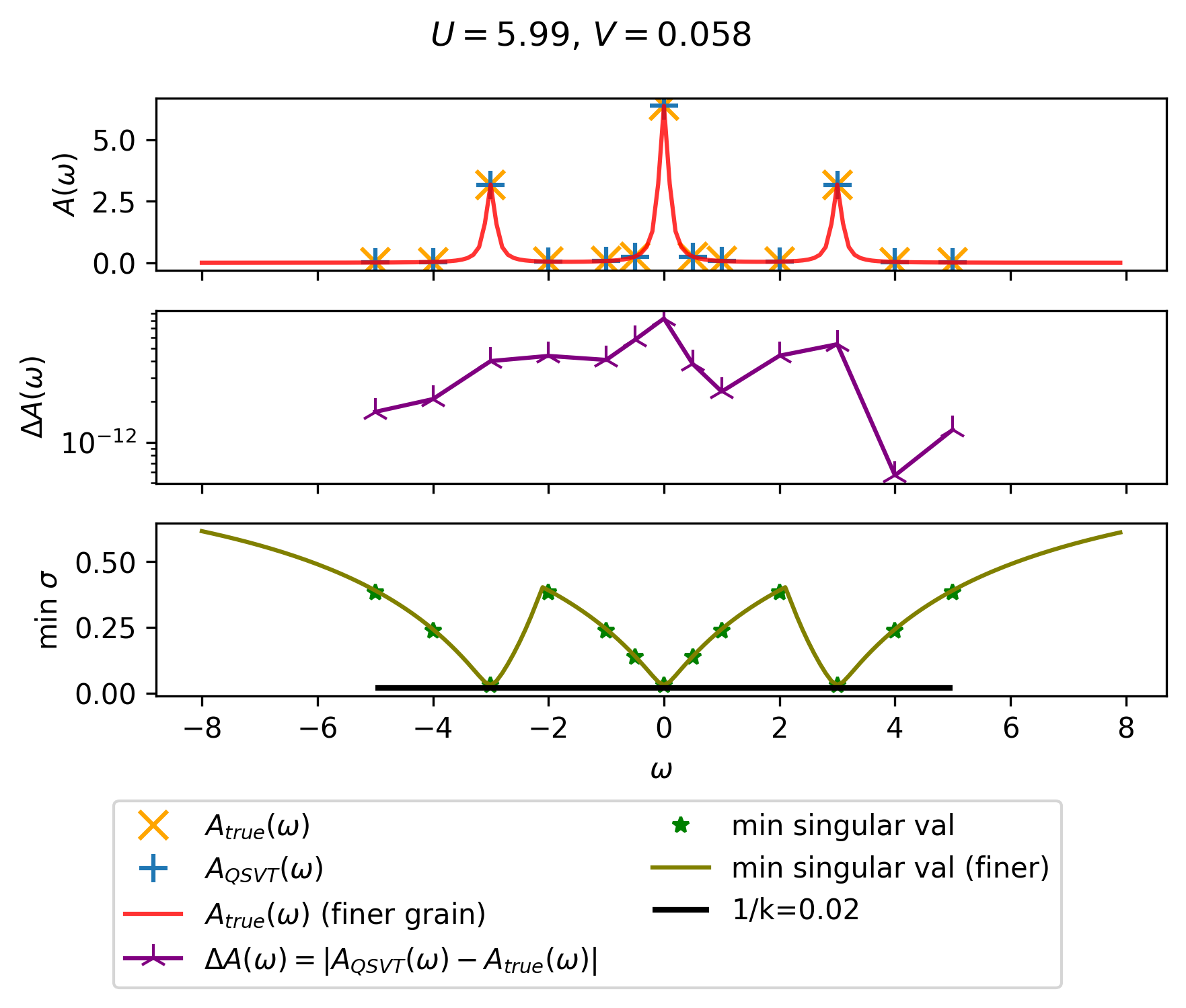}
    \caption{(top) Plot of spectral function  (equation \ref{eq:spectral_F}) of single-particle Anderson model for $U=5.99$ and $V=0.058$. The red line and orange points shows the spectral function function being calculated via exact diagonalization. The blue points show the spectral function calculated via the quantum singular value transform with $k=50$. (middle) Plot shows absolute error of spectral function $\Delta A(\omega) = |A_{QSVT}(\omega) - A_{true}(\omega) |$. (bottom) Plot shows the minimum singular value of the matrix to undergo inversion via the QSVT. Any singular below the black line is outside the region of where the approximation of $1/x$ is well defined.}
        \label{fig:Green_U5_K50}
\end{figure}

\begin{figure}[h!]
    \centering
    \includegraphics[width=0.75\textwidth]{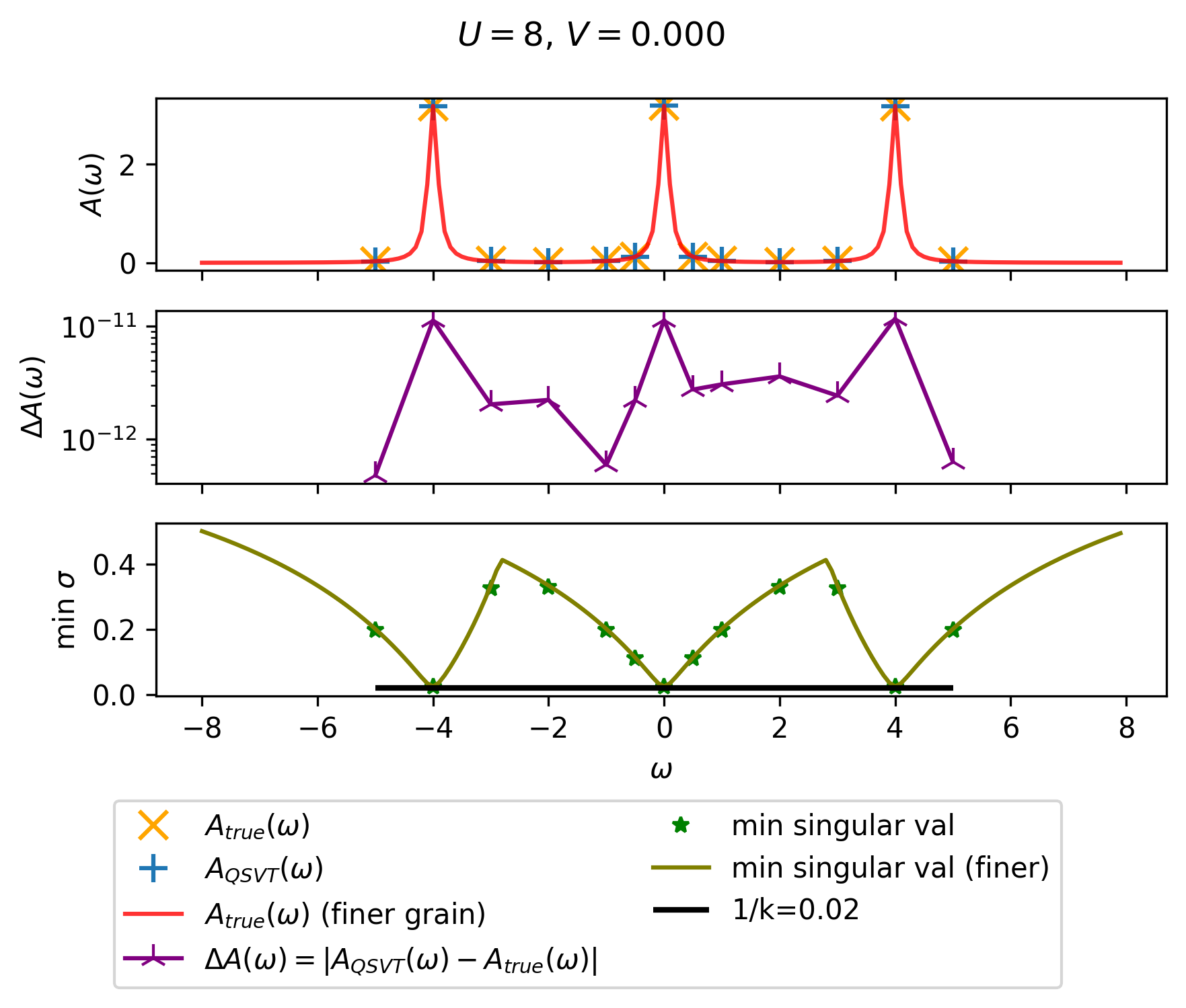}
    \caption{(top) Plot of spectral function  (equation \ref{eq:spectral_F}) of single-particle Anderson model for $U=8$ and $V=0$. The red line and orange points shows the spectral function function being calculated via exact diagonalization. The blue points show the spectral function calculated via the quantum singular value transform with $k=50$. (middle) Plot shows absolute error of spectral function $\Delta A(\omega) = |A_{QSVT}(\omega) - A_{true}(\omega) |$. (bottom) Plot shows the minimum singular value of the matrix to undergo inversion via the QSVT. Any singular below the black line is outside the region of where the approximation of $1/x$ is well defined.}
        \label{fig:Green_U8_K50}
\end{figure}

Figures \ref{fig:Green_U2_K50}, \ref{fig:Green_U5_K50} and \ref{fig:Green_U8_K50} give the spectral plots for the two-site single particle Anderson model at different $U, V$ values.  

% \begin{figure*}[t]
%     \centering
%     \includegraphics[width=0.75\textwidth]{results/full_G_K50/Anderson_U4_V0_K50.png}
%     \caption{Spectral function (equation \ref{eq:spectral_F}) of single-particle Anderson model for $U=4$ and $V=0.745$. The top plot shows the spectral function being calculated via exact diagonalization (red and orange) and via quantum signal processing (blue) with $k=10$. The middle plot shows in the absolute error $\Delta G(\omega) = |G_{exp}(\omega) - G_{true}(\omega) |$. The bottom plot shows the minimum singular value of the matrix to undergo inversion via the QSVT. Any singular below the black line is outside the region of where the approximation of $1/x$ is defined.}
%         \label{fig:Green_U4_K10}
% \end{figure*}

\clearpage
\section{Mott Phase transition} \label{sec:mott_appendix}
\begin{figure}[h!]
    \centering
    \includegraphics[width=0.5\textwidth]{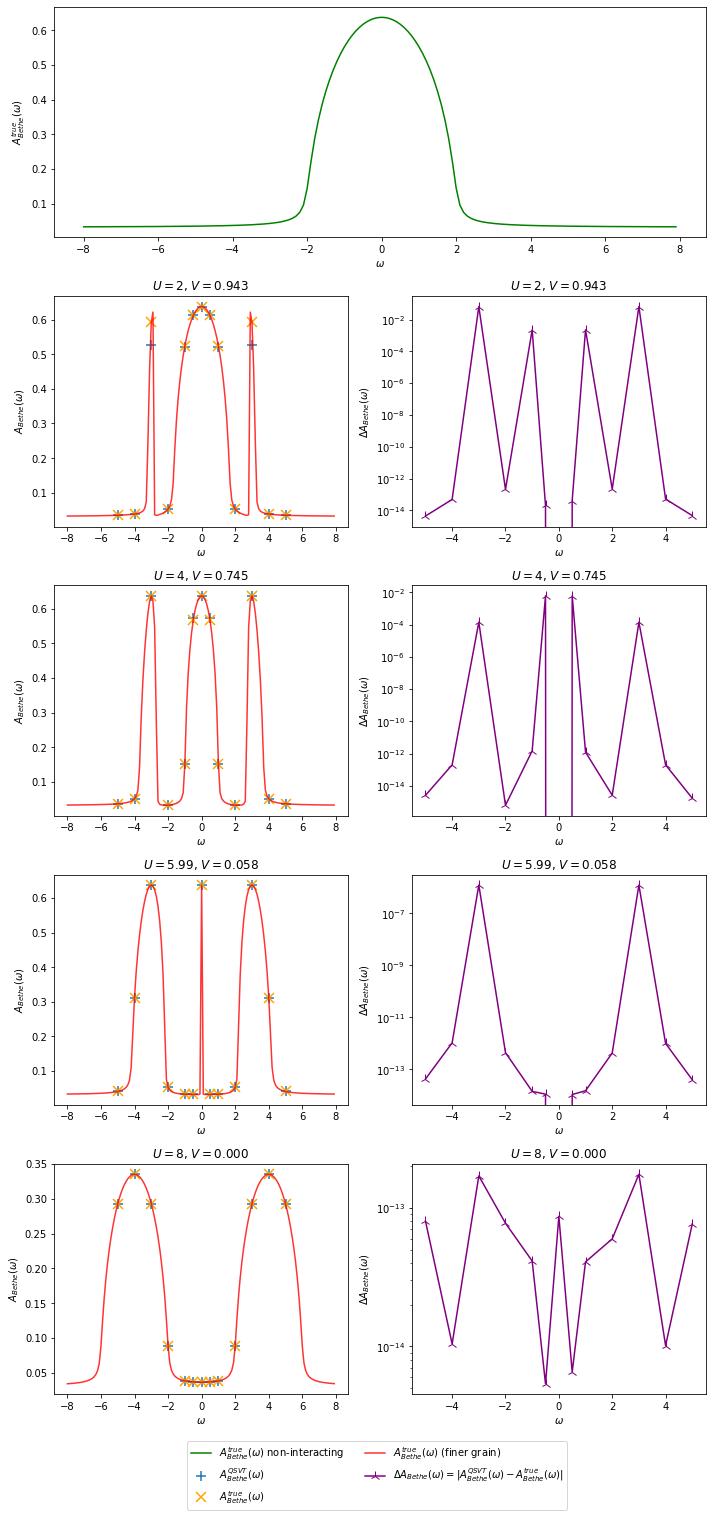}
    \caption{Metal to insulator Mott transition of the two site Anderson impurity model. The left plots show the density of states on the Bethe lattice and the right plots are absolute error in the density of states with respect to the exact solution (obtained via classical matrix inversion). The first plot gives the non-interacting system, followed by $(U,V)$ combinations of $(2,0.943), (4,0.745), (5.99, 0.058), (8, 0)$. Note the free density of states on the Bethe lattice with infinite coordination is  $\rho_{0}(x) = \frac{1}{2\pi t^{2}} \sqrt{4t^{2}- x^{2}}$ (in this work $t=1$) and the interacting density of states is $\rho(\omega)=\rho_{0}(\omega + \mu - \Sigma[\omega])$. The non-interacting Green's function is defined as $G_{imp}^{0}(\omega) = (\omega + i \delta -\epsilon_{\alpha} + \mu - \frac{|V|^{2}}{\omega+i \delta})^{-1}$ (in this work $\epsilon_{\alpha}=0$). For the final plot $(8, 0)$, particle hole symmetry was broken and the spin up and down parts of $\Sigma[\omega]$ where treated separately and combined. The yellow data points show the result obtained from a noise free QSVT simulation, where $k=10$.}
        \label{fig:mott_k10}
\end{figure}

\begin{figure}[h!]
    \centering
    \includegraphics[width=0.5\textwidth]{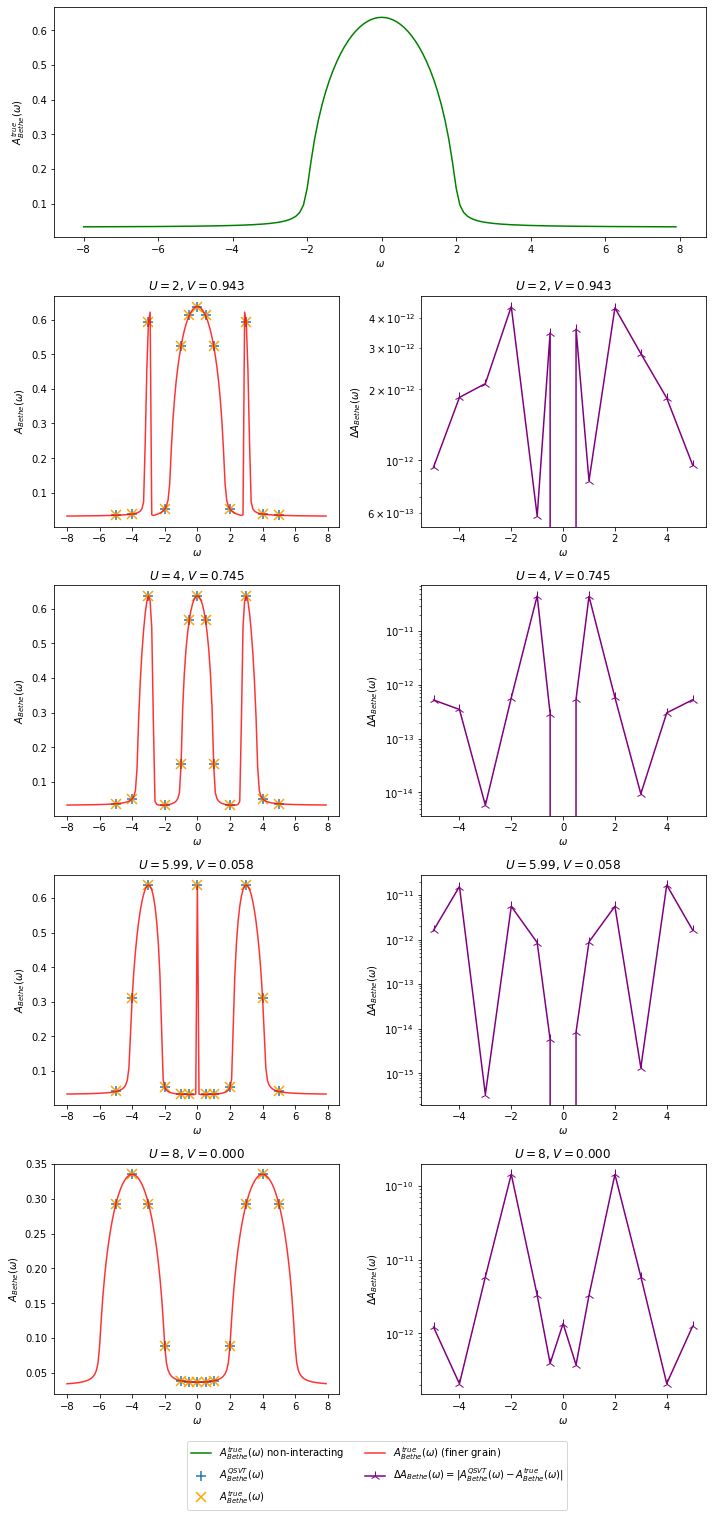}
    \caption{Metal to insulator Mott transition of the two site Anderson impurity model. The left plots show the density of states on the Bethe lattice and the right plots are absolute error in the density of states with respect to the exact solution (obtained via classical matrix inversion). The first plot gives the non-interacting system, followed by $(U,V)$ combinations of $(2,0.943), (4,0.745), (5.99, 0.058), (8, 0)$. Note the free density of states on the Bethe lattice with infinite coordination is  $\rho_{0}(x) = \frac{1}{2\pi t^{2}} \sqrt{4t^{2}- x^{2}}$ (in this work $t=1$) and the interacting density of states is $\rho(\omega)=\rho_{0}(\omega + \mu - \Sigma[\omega]$. The non-interacting Green's function is defined as $G_{imp}^{0}(\omega) = (\omega + i \delta -\epsilon_{\alpha} + \mu - \frac{|V|^{2}}{\omega+i \delta})^{-1}$ (in this work $\epsilon_{\alpha}=0$). For the final plot $(8, 0)$, particle hole symmetry was broken and the spin up and down parts of $\Sigma[\omega])$ where treated separately and combined. . The yellow data points show the result obtained from a noise free QSVT simulation, where $k=50$.}
        \label{fig:mott_k50_with_error}
\end{figure}

This section summaries the Metal to insulator Mott phase transition results for different polynomial approximtions of the inverse funciton. The results are given in Figures \ref{fig:mott_k10} and \ref{fig:mott_k50_with_error}. For $k=10$ it should be noted that the error in the spectral function changes by many orders of magnitude compared. Whereas, the $k=50$ result remains around the pico level of accuracy at all points.

\end{document}